\documentclass[
  preprint,                
  amsmath,amssymb,
  aps,prfluids,            
  superscriptaddress,
  longbibliography
]{revtex4-2}

\usepackage{graphicx}
\usepackage{epstopdf}
\usepackage{bm}
\usepackage{color}
\usepackage{setspace}

\graphicspath{{./figures/}}

\newcommand{\Rey}{\mathrm{Re}}
\newcommand{\Ek}{\mathrm{E}}
\newcommand{\dd}{\,\mathrm{d}}
\newcommand{\uvec}{\bm{u}}
\newcommand{\xvec}{\bm{x}}
\newcommand{\nvec}{\bm{n}}
\newcommand{\Vvec}{\bm{V}}
\newcommand{\Gvec}{\bm{G}}
\newcommand{\lamvec}{\bm{\lambda}}

\begin{document}
\singlespacing

\title{Two-dimensional simulations of hydrodynamic spin coupling in a
two-rotor corral}

\author{Tsorng-Whay Pan}
\email{tpan@central.uh.edu}
\affiliation{Department of Mathematics, University of Houston,
Houston, Texas 77204, USA}

\author{Jiwen He}
\email{jhe4@central.uh.edu}
\affiliation{Department of Mathematics, University of Houston,
Houston, Texas 77204, USA}

\date{\today}

\begin{abstract}
We study hydrodynamic spin coupling in a two-rotor corral using direct
numerical simulation (DNS) of the strictly two-dimensional incompressible
viscous fluid flow modeled by the Navier--Stokes equations. An
active rotor is driven at angular velocity $\Omega$, and a nearby torque-free
passive rotor selects an angular velocity $\omega$ through hydrodynamic torque
balance. The signed gear ratio $\Gamma=\omega/\Omega$ distinguishes corotation
from counterrotation, with Reynolds number $\Rey=|\Omega|r^2/\nu$. Motivated by a
recent quasi-two-dimensional experiment~\cite{SmithRistrophZhang2026}, we use a
distributed Lagrange
multiplier/\allowbreak{}fictitious domain method to compute planar phase
diagrams of $\Gamma(G,\Rey)$ at corral sizes $C=3$, $4.5$, and $6$. The planar
model recovers the benchmark gap route at $\Rey=20$: an intermediate
counterrotation band, a wide-gap transition to corotation, gear-ratio
magnitudes of order $10^{-2}$, and the observed sequence of vortex
attachment, detachment, and merger. It also produces a reentrant-like gap
structure with a small-gap corotation region whose relation to the experimental
close-range geometric state remains unresolved. The main discrepancy is the
high-$\Rey$ boundary. At the experimental mid-gap transect $G\approx0.3$,
the planar gear ratio approaches zero from the counterrotating side but does
not cross through $\Rey=400$; at the narrower gap $G=0.22$, by contrast,
the planar terminal spin reverses near $\Rey=44$. Wall-traction diagnostics
show that this crossing is not the experimental shear-competition mechanism:
the gap-facing counterrotating arc narrows but does not collapse or deflect as
in the experiment, and the reversal at $G=0.22$ occurs by redistribution of
the integrated planar torque. The strictly planar model therefore captures the
broad gap-route architecture and the existence of a Reynolds-driven spin
boundary, but displaces that boundary in gap and alters its surface-stress
mechanism. The remaining mismatch points to finite-depth secondary motion,
end-wall stresses, and apparatus geometry as plausible contributors to the
experimental shear balance.
\end{abstract}

\maketitle


\section{Introduction}

Flow-mediated interactions between rigid bodies occur throughout soft,
biological, and engineering flows. Translation-dominated examples
include sedimentation, swarming, and flocking, whereas
rotation-dominated examples include spinning colloids, active matter,
and turbine arrays~\cite{Ristroph2008,Newbolt2019,Soni2019,
Bililign2022,Brownstein2019,Chen2024}. Translational interactions have
been studied extensively, but rotor-scale spin coupling, the way one
body's rotation drives another body's rotation through the intervening
fluid, is less well understood. Classical problems isolate parts of
this physics: vortex pairs orbit, leapfrog, and
merge~\cite{Kida1994,Dritschel1995}; Taylor--Couette flow reveals
instabilities between concentric rotating
cylinders~\cite{Taylor1923,Huisman2018}; and prescribed rotor arrays generate
vortex-mixing and cellular flows~\cite{Jana1994,Hu2021}. Recent
Stokes-flow simulations of torque-driven confined cylinder pairs have
also identified hydrodynamic bound states and wide-gap
recirculation~\cite{GuoManZhu2024}. What remains less developed at finite Reynolds
number is a controlled two-body setting in which spin coupling is
measured through the emergent response of a torque-free passive rotor,
rather than imposed through a prescribed output torque or speed.

Smith, Ristroph, and Zhang~\cite{SmithRistrophZhang2026} recently
introduced such a setting: the \emph{two-rotor problem}. An active
cylinder is driven at prescribed angular velocity $\Omega$, while a
nearby test rotor on low-friction bearings rotates freely at a speed
$\omega$ set by hydrodynamic coupling. Both rotors are confined in a
circular corral of radius $R$. The state is described by three input
parameters and one output,
\begin{equation}
  \Rey=\frac{|\Omega| r^{2}}{\nu},\qquad
  G=\frac{g}{2R-4r},\qquad
  C=\frac{R}{r},\qquad
  \Gamma=\frac{\omega}{\Omega},
  \label{eq:groups}
\end{equation}
where $r$ is the rotor radius, $\nu$ the kinematic viscosity, and $g$
the surface-to-surface interrotor gap. The normalized gap $G\in[0,1]$
places rotor--rotor contact at $G=0$ and simultaneous rotor--corral
contact at $G=1$. The gear ratio $\Gamma$ records the passive rotation
per active rotation; its sign distinguishes counterrotation
($\Gamma<0$) from corotation ($\Gamma>0$). The Reynolds number is built
from the drive rate $|\Omega|$, while the sign of $\Omega$ fixes the
sense of rotation and therefore the sign convention for
$\Gamma=\omega/\Omega$.

The experimental phase diagram is rich but structured. At moderate gap
and low $\Rey$ the passive rotor counterrotates, as expected from a
gearlike inner shear. Corotation nevertheless appears broadly in the
same parameter space. The experiment rationalized the resulting
counter-to-corotation transitions through a competition between inner
and outer shear on the passive rotor: a close-range \emph{geometric}
transition, in which the inner shear zone is pinched off; a wide-gap
\emph{topological} transition, in which an intervening recirculation
reverses the washing flow; and a high-$\Rey$ \emph{inertial}
transition, in which outward inertial flow strengthens the outer shear.
The associated measurements of the inner shear-zone length $S$, the
inner-zone deflection angle, and the inner/outer velocity balance provide the
mechanistic benchmark for any numerical model.

These observations pose a natural modeling question. The apparatus is
quasi-two-dimensional, with aspect ratio $H/r=8$, so the measured spin states
combine horizontal-plane flow with finite-depth effects and end-wall-driven
secondary motion. A strictly two-dimensional computation isolates the planar
contribution. Agreement with experiment identifies mechanisms already present in
the horizontal Navier--Stokes problem; disagreement identifies where finite
depth, end walls, or apparatus geometry shift the shear balance. The same
fluid--structure framework also makes it possible to ask later how passive
translation or passive-body shape would modify the fixed-axis circular-rotor
problem.

Here we study the fixed-axis circular-rotor problem using a distributed
Lagrange multiplier/\allowbreak{}fictitious domain (DLM/FD) method. The method
was developed for rigid particles in incompressible viscous
flow (e.g., see~\cite{Glowinski1999,Glowinski2001}), then extended to curved rotating
containers by adding a second Lagrange multiplier that enforces a prescribed
exterior flow~\cite{PanGlowinskiHou2007}. It has also reproduced the wavelength
of circular particle bands in a fully filled rotating
cylinder~\cite{HouPanGlowinski2014}. In the present problem the circular corral
is embedded in a rectangular fictitious domain, the active rotor is a rigid body
with prescribed angular velocity, and the passive rotor obeys the Euler--Newton
equations. In the main fixed-axis problem, only its spin is free and that spin
is selected by hydrodynamic torque.

To our knowledge, this is the first finite-Reynolds-number direct numerical
simulation (DNS) of a
torque-free passive rotor in the two-rotor corral. The central result is a
boundary-displacement picture. The planar model recovers the main gap-controlled
architecture of the experiment and the order of magnitude of the coupling. It
also contains a fixed-gap Reynolds-driven crossing, but that crossing occurs at
a narrower gap than in the experiment: the planar boundary reaches $G=0.22$ near
$\Rey=44$ but does not reach the experimental transect near $G=0.3$ by
$\Rey=400$. The key difference is therefore the shape, placement, and
surface-stress mechanism of the high-$\Rey$ boundary, not simply the presence or
absence of a Reynolds-driven route. Passive translation and passive-body shape
are treated as outlook directions enabled by the same formulation, rather than
as primary conclusions of the present study.

The paper is organized as follows. Section~\ref{sec:model} gives the
governing equations, DLM/FD formulation, and numerical validation.
Section~\ref{sec:phase} reports the fixed-axis circular-rotor phase
diagrams, identifies the gap-driven and Reynolds-driven boundaries, analyzes
transient passive-spin response, and compares the planar results with the
quasi-two-dimensional experiment.
Section~\ref{sec:conclusion} summarizes the
boundary-displacement picture and the remaining mechanistic tests, and outlines
extensions to passive translation and passive-body shape.

\section{Model and numerical method}
\label{sec:model}

\subsection{Governing equations}

We model the experiment in a horizontal plane.  An incompressible
Newtonian fluid of density $\rho_f$ and dynamic viscosity $\mu_f$
(kinematic viscosity $\nu=\mu_f/\rho_f$) fills a circular corral
$\mathcal{C}$ of radius $R$.  Two rigid bodies are immersed in the
fluid: an active circular rotor of radius $r$, driven at prescribed
angular velocity $\Omega$ about a fixed center, and a passive body whose
motion is determined by hydrodynamic force and torque.  In the fixed-axis
two-disk problem that forms the main comparison with experiment, the
passive body is also circular and its center is fixed; only its angular
velocity $\omega$ evolves.  The same formulation also permits a freely
translating passive disk or a noncircular passive body, which we return to only
as outlook directions.

The rotor centers lie on a diameter of the corral and are placed
symmetrically about its center.  The surface-to-surface gap $g$ therefore
ranges from rotor--rotor contact, $g=0$, to simultaneous rotor--wall
contact, $g=2R-4r$.  We use the dimensionless groups introduced in
Eq.~\eqref{eq:groups}: the normalized gap $G=g/(2R-4r)$, the corral size
$C=R/r$, the Reynolds number $\Rey=|\Omega|r^2/\nu$, and the signed gear
ratio $\Gamma=\omega/\Omega$.  The Reynolds number uses the drive rate
$|\Omega|$, while the sign of $\Omega$ fixes the imposed sense of rotation;
with this convention $\Gamma>0$ denotes corotation and $\Gamma<0$
counterrotation.

In the fluid region $\mathcal{C}\setminus\mathcal{B}(t)$, where
$\mathcal{B}(t)$ is the union of the rigid bodies, the velocity
$\uvec$ and pressure $p$ satisfy
\begin{align}
  \rho_f\!\left[\frac{\partial\uvec}{\partial t}
    +(\uvec\!\cdot\!\nabla)\uvec\right]
    &= -\nabla p+\mu_f\,\Delta\uvec,
    \label{eq:ns-mom}\\[2pt]
  \nabla\!\cdot\!\uvec &= 0 .
    \label{eq:ns-div}
\end{align}
The initial velocity is divergence free, the corral wall is stationary
($\uvec=\bm 0$ on $\partial\mathcal{C}$), and each body surface satisfies
no slip with the corresponding rigid-body velocity.  Gravity, present in
the sedimentation problems for which the numerical method was originally
developed~\cite{PanGlowinskiHou2007,HouPanGlowinski2014}, is absent from
this planar spin-coupling problem.

For a passive rigid body with center of mass $\Gvec$, translational
velocity $\Vvec$, angular velocity $\omega$, mass $M_p$, and moment of
inertia $I_p$, the Euler--Newton equations are
\begin{align}
  \frac{\dd\Gvec}{\dd t} &= \Vvec, &
  M_p\frac{\dd\Vvec}{\dd t} &= \bm{F}_H, &
  I_p\frac{\dd\omega}{\dd t} &= T_H .
  \label{eq:euler-newton}
\end{align}
Here $\bm F_H$ and $T_H$ are obtained from the fluid stress
$\bm{\sigma}=\mu_f(\nabla\uvec+\nabla\uvec^{\!\top})-p\bm{I}$:
\begin{equation}
  \bm{F}_H=\oint_{\partial\mathcal{B}}\!\bm{\sigma}\,\nvec\,\dd s,
  \qquad
  T_H=\oint_{\partial\mathcal{B}}\!\bigl[(\xvec-\Gvec)\times
        \bm{\sigma}\,\nvec\bigr]\,\dd s .
  \label{eq:FT}
\end{equation}
The cross product in Eq.~\eqref{eq:FT} denotes the scalar out-of-plane
torque,
\[
(\xvec-\Gvec)\times\bm{\sigma}\nvec
 =(x_1-G_1)(\bm{\sigma}\nvec)_2-(x_2-G_2)(\bm{\sigma}\nvec)_1 .
\]
The normal $\nvec$ points outward from the body into the fluid, following
Refs.~\cite{PanGlowinskiHou2007,HouPanGlowinski2014}.  For
the active rotor we replace Eq.~\eqref{eq:euler-newton} by the prescribed
constraints $\Vvec=\bm 0$ and $\omega=\Omega$.  For the fixed-axis
passive disk in Sec.~\ref{sec:phase}, we impose $\Vvec=\bm 0$ and retain
only the angular equation.

\subsection{Distributed Lagrange multiplier/\allowbreak{}fictitious domain
formulation}
\label{sec:dlm}

The coupled problem is solved by a distributed Lagrange multiplier/\allowbreak{}fictitious
domain (DLM/FD) method
following Refs.~\cite{Glowinski1999,Glowinski2001,PanGlowinskiHou2007}.
The circular corral $\mathcal{C}$ is embedded in a slightly larger square
computational domain $\mathcal{D}$, and the annular exterior of the corral
inside this square is denoted
$\mathcal{A}=\mathcal{D}\setminus\mathcal{C}$.  The velocity is computed
on a fixed Cartesian mesh over all of $\mathcal{D}$.  This choice avoids
body-fitted remeshing as the passive body moves or rotates, while
Lagrange multipliers enforce the physical constraints.

The body multiplier $\lamvec$, supported inside each rigid body, enforces
rigid-body motion,
\begin{equation}
  \uvec(\xvec,t)=\Vvec(t)+\omega(t)\times\bigl[\xvec-\Gvec(t)\bigr],
  \qquad \xvec\in\mathcal{B}(t).
  \label{eq:rigid}
\end{equation}
In two dimensions $\omega$ is the out-of-plane scalar angular velocity and
\[
\omega\times(\xvec-\Gvec)=
\omega\,\bigl(-(x_2-G_2),\,x_1-G_1\bigr).
\]
A second multiplier, $\lamvec_{\mathcal A}$, is supported on
$\mathcal A$ and imposes the prescribed exterior velocity; in the present
stationary-corral problem this velocity is zero, which enforces no slip
on the curved boundary $\partial\mathcal C$ while retaining a rectangular
grid~\cite{PanGlowinskiHou2007}.  This corral multiplier is absent from
the earlier rectangular-domain formulations
\cite{Glowinski1999,Glowinski2001}.

For completeness, we write the weak form used in the computation.  To
avoid obscuring the notation, the body terms below are written for one
representative rigid body; in the two-rotor simulations the corresponding
constraints are applied to the active and passive bodies, with the active
translation and spin prescribed.  For almost every $t>0$, find
$\uvec(t)$, $p(t)\in L^2_0(\mathcal D)$, $\Vvec(t)$, $\Gvec(t)$,
$\omega(t)$, $\lamvec(t)$, and $\lamvec_{\mathcal A}(t)$ such that, for
all admissible test functions $\bm v$, $\bm Y$, and scalar angular test
function $\chi$,
\begin{align}
  &\rho_f\!\int_{\mathcal{D}}\!\Bigl[\frac{\partial\uvec}{\partial t}
      +(\uvec\!\cdot\!\nabla)\uvec\Bigr]\!\cdot\!\bm{v}\dd\xvec
   -\!\int_{\mathcal{D}}\! p\,\nabla\!\cdot\!\bm{v}\dd\xvec
   +\mu_f\!\int_{\mathcal{D}}\!\nabla\uvec\!:\!\nabla\bm{v}\dd\xvec
  \nonumber\\
  &\quad-\bigl\langle\lamvec,\,\bm{v}-\bm{Y}
      -\chi\times\overrightarrow{\Gvec\xvec}\bigr\rangle
   -\bigl\langle\lamvec_{\mathcal{A}},\,\bm{v}\bigr\rangle
   +\Bigl(1-\tfrac{\rho_f}{\rho_s}\Bigr)
      \Bigl[M_p\frac{\dd\Vvec}{\dd t}\!\cdot\!\bm{Y}
            +I_p\frac{\dd\omega}{\dd t}\chi\Bigr]=0,
  \label{eq:weak-mom}
\end{align}
together with
\begin{align}
  &\int_{\mathcal{D}}q\,\nabla\!\cdot\!\uvec\dd\xvec=0,
  \label{eq:weak-div}\\
  &\bigl\langle\bm{\mu},\,\uvec-\Vvec
        -\omega\times\overrightarrow{\Gvec\xvec}\bigr\rangle=0,
  \label{eq:weak-rigid}\\
  &\bigl\langle\bm{\mu}_{\mathcal{A}},\,\uvec-\bm{g}_0\bigr\rangle=0.
  \label{eq:weak-corral}
\end{align}
Here $\bm g_0$ is the prescribed velocity outside the corral, zero for the
stationary corral considered here.  The factor
$(1-\rho_f/\rho_s)$ multiplying the rigid-body inertia is the standard
fictitious-domain correction: because the fluid equations are solved
inside the body as well as outside it, the explicit solid inertia is
reduced by the inertia of the displaced fluid.  In the neutrally buoyant
case $\rho_s/\rho_f=1$ used in the production runs, this term vanishes and
the rigid velocity is recovered through the projection scheme of
Refs.~\cite{PanGlowinski2002,PanGlowinski2005}.  The terminal angular velocity reported
below is a zero-torque state and, when that state is steady and unique,
does not depend on the density ratio; inertia affects only the relaxation
toward it.  The multiplier spaces are described in
Ref.~\cite{PanGlowinskiHou2007}.

\subsection{Operator splitting and discretization}

We advance the coupled equations with the first-order Lie
operator-splitting scheme~\cite{Chorin1978} used in
Refs.~\cite{PanGlowinskiHou2007,HouPanGlowinski2014}.  Each time step
$\Delta t$ is split into five subproblems: incompressibility; nonlinear
advection; diffusion together with enforcement of the exterior-corral
constraint through $\lamvec_{\mathcal A}$; rigid-body translation; and the
rigid-motion projection through $\lamvec$.  The
quasi-Stokes problem in the incompressibility step is solved by an
Uzawa/preconditioned conjugate-gradient method with fast elliptic
preconditioners~\cite{Glowinski2003}.  The advection step is advanced by
the wave-like-equation method of Ref.~\cite{DeanGlowinski1997}, and the
saddle-point problems associated with the two multiplier projections are
solved by conjugate-gradient iteration~\cite{Glowinski2001}.

The velocity uses $P_1$-iso-$P_2$ finite elements on a fixed Cartesian
mesh, and the pressure uses $P_1$ elements on the coarser mesh
$h_p=2h_v$~\cite{Glowinski2003,PanGlowinskiHou2007,HouPanGlowinski2014}.
The body multiplier $\lamvec$ is represented by collocation at velocity
nodes inside each body together with selected boundary points.  The corral
multiplier $\lamvec_{\mathcal A}$ is represented analogously in
$\mathcal A$ and on the corral surface.  Because the Eulerian grid is
fixed, the collocation sets attached to the moving bodies are updated each
time step from the current center, velocity, and angular velocity.

\subsection{Passive spin and terminal criterion}
\label{sec:terminal}

All simulations are time dependent.  The fluid and rigid-body equations
are advanced from rest, while the active rotor is set to its prescribed
angular velocity $\Omega$ at $t=0$.  In the fixed-axis calculations the
passive rotor relaxes to a steady zero-torque spin.  We identify terminal
states by a stationarity criterion on the computed velocity field,
$\|\delta\uvec\|_{L^\infty}<10^{-10}$ between successive time steps, and then
verify the torque balance with the
nondimensional residual $\mathcal{R}_T=|T_H|/(\mu_f|\Omega|r^2)$.  At the
reported terminal states the hydrodynamic torque computed from the surface
shear satisfies $T_H\approx0$, as required for a torque-free passive rotor.
Across the conditions discussed below, including the $G=0.30$ transect to
$\Rey=400$, the monitored passive spin settles to a constant value.  We
therefore report $\Gamma=\omega/\Omega$ at the terminal state rather than as a
time average.  This procedure is the numerical counterpart of the experimental
protocol, in which the air-bearing test rotor is allowed to equilibrate at each
imposed drive before its mean spin is recorded.

The inertia of the passive body controls only the approach to the terminal
state when that terminal state is steady and unique.  In particular, the
zero-torque condition fixes the final angular velocity independently of
$\rho_s/\rho_f$ and $I_p$; for $\rho_s/\rho_f\ne1$, these quantities would
set relaxation times.  In the neutrally buoyant simulations reported here,
the spin is obtained through the projection described above and $I_p$ does
not enter the evolution.

The physical and numerical parameters used in the production runs are
summarized in Table~\ref{tab:params}.

\begin{table}[t]
  \centering
  \caption{Physical and numerical parameters.  Lengths are nondimensional
  simulation units with corral radius $R=0.5$.  The square fictitious domain
  is the unit square padded by two velocity mesh spacings,
  $\mathcal{D}=(-2h_v,\,1+2h_v)^2$.}
  \label{tab:params}
  \begin{ruledtabular}
  \begin{tabular}{lcl}
    Quantity & Symbol & Value \\
    \hline
    Corral radius             & $R$                   & $0.5$ \\
    Corral size               & $C=R/r$               & $3,\ 4.5,\ 6$ \\
    Rotor radius              & $r=R/C$               & $0.167,\ 0.111,\ 0.083$ \\
    Reynolds number           & $\Rey=|\Omega| r^2/\nu$ & $10$ to $100$ ($G=0.3$, $C=4.5$ transect to $400$) \\
    Normalized gap            & $G$                   & $0.1$ to $0.8$ \\
    Kinematic viscosity       & $\nu$                 & $0.01$ \\
    Fluid density             & $\rho_f$              & $1$ \\
    Velocity mesh size        & $h_v$                 & $1/512$ \\
    Pressure mesh size        & $h_p=2h_v$            & $1/256$ \\
    Time step                 & $\Delta t$            & $10^{-3}$ \\
    Solid-to-fluid density    & $\rho_s/\rho_f$       & $1$ \\
    Passive moment of inertia & $I_p$                 & $\tfrac{\pi}{2}\rho_s r^4$ ($\rho_s/\rho_f\neq1$ only) \\
    Terminal tolerance        & $\|\delta\uvec\|_{L^\infty}$ & $<10^{-10}$ (absolute) \\
  \end{tabular}
  \end{ruledtabular}
\end{table}

\subsection{Validation and resolution}
\label{sec:validation}

The numerical checks target the ingredients that determine the terminal gear
ratio: enforcement of the curved no-slip and no-penetration constraints,
formation of recirculating Couette flow, evaluation of hydrodynamic torque, and
resolution of the gaps and shear zones that enter the phase diagrams. We
therefore separate numerical validation from physical benchmarking. Numerical
validation is carried out against an exact Stokes solution and by resolution
checks of torque-selected quantities; the comparison with the experiment is the
physical benchmark of Sec.~\ref{sec:phase}.

The analytical benchmark is eccentric Couette flow between two circular
cylinders, for which Wannier obtained a closed-form Stokes
solution~\cite{Wannier1950}. This problem is the one-rotor limit of the
present geometry: the passive rotor is removed, the active rotor is placed
off center in a stationary circular corral, and the inner cylinder is driven
with surface speed $|\Omega|r$. In the Stokes limit the stream function
satisfies $\nabla^2\nabla^2\psi=0$. We evaluate Wannier's stream function,
with coefficients fixed by no penetration and prescribed tangential velocity
on both circles, and compare it with the DLM/FD solution computed with the
nonlinear advection step omitted.

The validation geometry has rotor radius $r=0.125$ and corral radius
$R=0.5$, so $C=R/r=4$, intermediate between the main corral sizes studied
below. The three eccentricities are chosen to bracket the rotor offsets in
the two-rotor sweep. For two rotors placed symmetrically about the corral
center, a rotor at normalized gap $G$ sits with its center a distance
$s=r+g/2$ from the corral center and has offset
$\epsilon=\frac{s}{R-r}=\frac{1+G(C-2)}{C-1}$.
At $C=4.5$, the production range $G=0.1$ to $0.8$ corresponds to
$\epsilon\simeq0.36$ to $0.86$; including the slightly tighter case
$G=0.05$ gives $\epsilon\simeq0.32$. The benchmark eccentricities
$\epsilon=1/3$, $1/2$, and $0.85$ therefore sample, respectively, a
narrow-gap-dominated circulation, the onset of a wide-gap recirculation eddy,
and a high-eccentricity state in which that eddy expands toward the corral
center. The validation meshes $h=1/128$, $1/256$, and $1/512$ provide
$16$, $32$, and $64$ cells per rotor radius.

\begin{figure}[!tbp]
  \centering
  \newcommand{\wcol}[4]{\begin{minipage}[t]{0.32\linewidth}\centering
    {\footnotesize (#4)~Eccentricity $\epsilon=#3$}\\[-7pt]
    \makebox[\linewidth][r]{\includegraphics[width=0.9\linewidth,angle=-90]{#1}\hspace*{0.01\linewidth}}\\[3pt]
    \includegraphics[width=\linewidth]{#2}\end{minipage}}
  \wcol{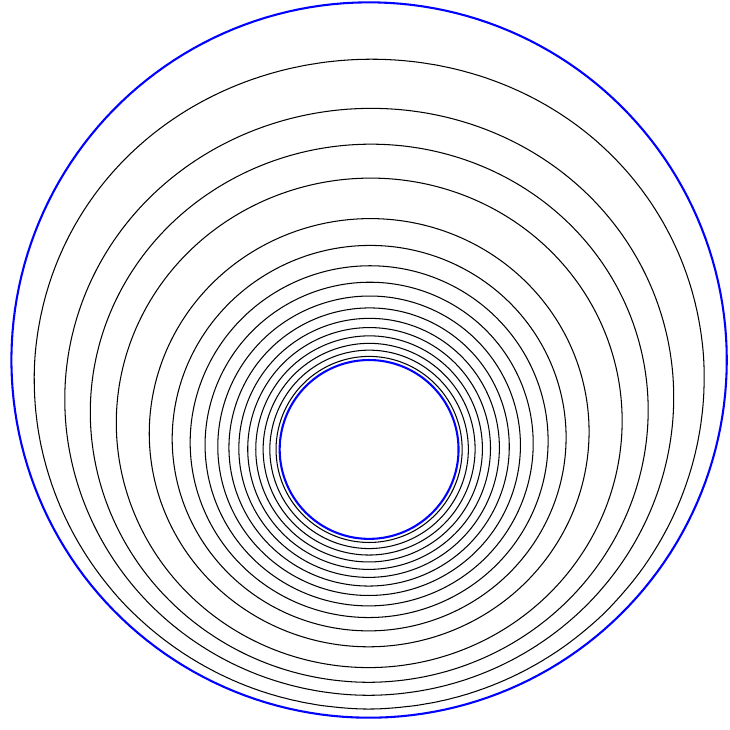}{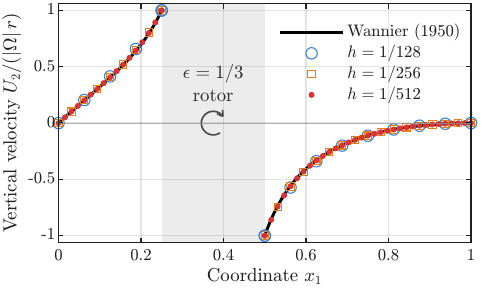}{1/3}{a}\hfill
  \wcol{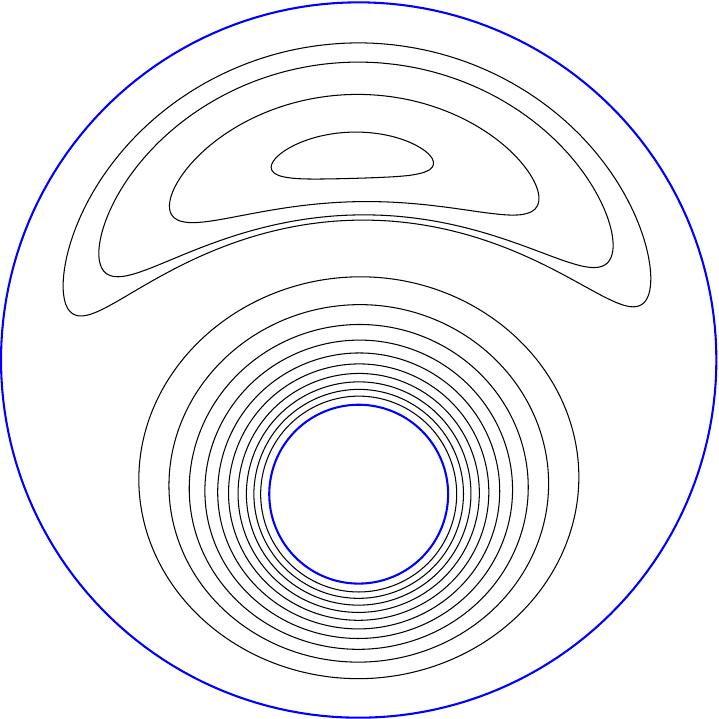}{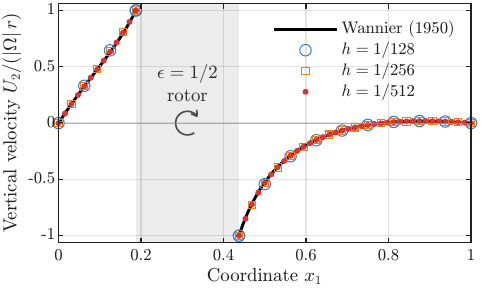}{1/2}{b}\hfill
  \wcol{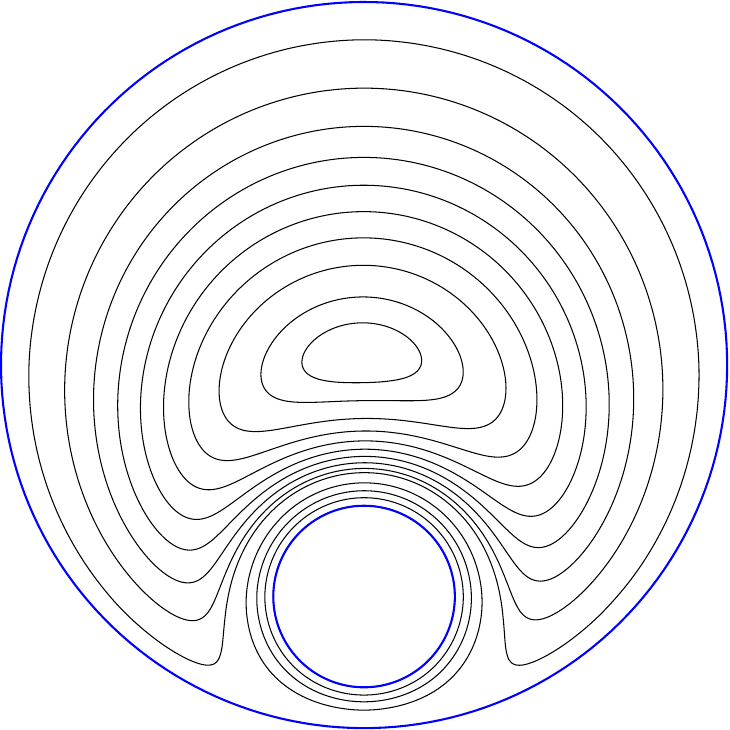}{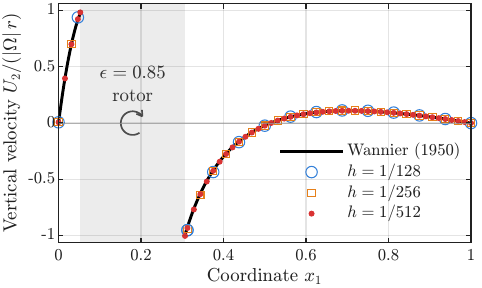}{0.85}{c}
  \caption{Validation against the Wannier eccentric Couette solution in the
  Stokes regime. Each column shows computed streamlines above the vertical
  velocity $U_2$ along the line of centers, plotted against $x_1$ and
  normalized by the rotor surface speed $|\Omega|r$. The shaded band marks the
  rotor; for the clockwise spin used here, the two points where the rotor
  surface crosses the line of centers have $U_2/(|\Omega|r)=\pm1$. The eccentricities
  $\epsilon=1/3$, $1/2$, and $0.85$ span the offset range of the two-rotor gap
  sweep, from narrow-gap-dominated circulation to wide-gap eddy onset and then
  to a large recirculation at high eccentricity. The Wannier solution (solid
  line) and DLM/FD profiles at $h=1/128$, $1/256$, and $1/512$ (markers) agree
  closely, including the sign reversal of $U_2$ caused by the wide-gap eddy;
  maximum errors are given in the text.}
  \label{fig:wannier}
\end{figure}

Figure~\ref{fig:wannier} shows the comparison. At $\epsilon=1/3$, the
streamlines are nested and crowded into the narrow gap, and the vertical
velocity $U_2$ along the line of centers decays monotonically across the wide
gap. At $\epsilon=1/2$, a separate wide-gap recirculation eddy appears,
producing a sign reversal in $U_2$. At $\epsilon=0.85$, this eddy fills much
of the wide gap and shifts toward the corral center, consistent with the
high-eccentricity flow reported for an off-center rotor in
Ref.~\cite{GuoManZhu2024}. The DLM/FD profiles track the Wannier solution at
all three eccentricities and all three resolutions, including the sign
reversal associated with the recirculation. Normalized by the rotor surface
speed $|\Omega|r$, the maximum pointwise error in $U_2$ falls from
$2.1\times10^{-2}$ at $h=1/128$ to $5.7\times10^{-3}$ at $h=1/512$ for
$\epsilon=1/3$, from $2.1\times10^{-2}$ to $5.8\times10^{-3}$ for
$\epsilon=1/2$, and from $2.7\times10^{-2}$ to $9.9\times10^{-3}$ for
$\epsilon=0.85$. The error decreases by roughly a factor of two per mesh
halving, consistent with first-order convergence in the maximum norm, and is
largest in the high-eccentricity case where the narrow gap is least resolved.
This benchmark checks both Lagrange-multiplier constraints: $\lamvec$ on the
rotor and $\lamvec_{\mathcal A}$ on the exterior corral region.

Torque provides the scalar validation most closely tied to the two-rotor
problem, because the passive gear ratio is selected by zero hydrodynamic
torque. For the Wannier benchmark, the Stokes torque on the rotating inner
cylinder is available in closed form and is used as a reference for the
computed torque. In the production calculations, the corresponding
torque-selected diagnostics are the terminal gear ratio $\Gamma$ and the
residual torque $\mathcal R_T$.

The active rotor does not introduce a Prandtl-type boundary-layer scale
$r/\sqrt{\Rey}$. In the isolated rotating-cylinder problem, the steady exterior
field is the irrotational spinner flow $u_\theta=\Omega r^2/\rho$, with
$\rho$ the distance from the rotor axis, which
matches the no-slip surface velocity and carries no vorticity outside the
cylinder. The relevant resolution requirements are therefore set by the
interrotor gap, the passive-rotor and wall shear zones, and convergence of the
torque-selected spin state. At the production resolution $h_v=1/512$, the
rotor diameter is resolved by about $171$ velocity cells at $C=3$, $114$ at
$C=4.5$, and $85$ at $C=6$. The smallest geometric scale is the interrotor gap:
at the production floor $G=0.1$, the gap contains about twenty-eight velocity
cells at $C=4.5$ and about seventeen at $C=3$. Thus the tightest small-gap
points are the least resolved and are used mainly to establish the existence of
the small-gap corotation leg; the quantitative boundary comparisons below rely
on the better-resolved interior of the sweep.


\section{Two-dimensional phase diagrams and comparison with the
quasi-two-dimensional experiment}
\label{sec:phase}

This section contains the main physical comparison. The argument is organized
around increasingly stringent tests of the planar reduction. We first ask
whether a strictly two-dimensional calculation reproduces the experimental gap
route at fixed $\Rey$, where the reported dynamics are steady and the mechanism
is primarily topological. We then examine fixed-gap Reynolds-number transects,
where the planar model contains a spin reversal but places it at a smaller gap
than the experiment. Finally, we assemble the two-dimensional phase maps,
examine the start-up response of the passive rotor, and compare the resulting
planar picture with the finite-depth, quasi-two-dimensional experiment.

\subsection{Horizontal transect: the gap route at $\Rey=20$}
\label{sec:gaproute}

\begin{figure*}[!tp]
  \centering
  \includegraphics[width=\linewidth]{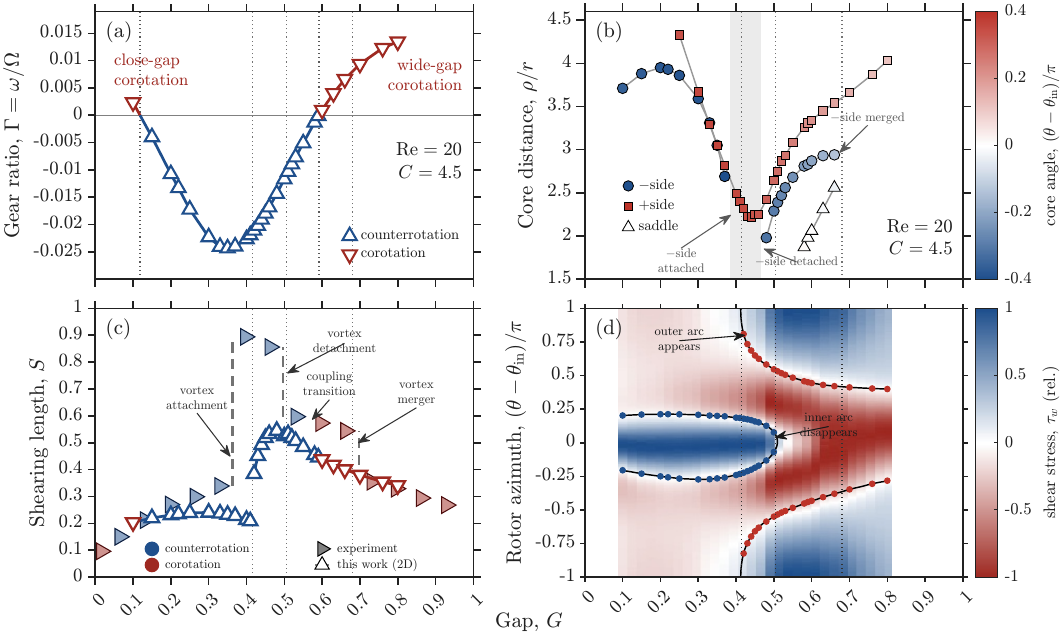}
  \caption{Gap-route diagnostics at $\Rey=20$, $C=4.5$, plotted on a common
  gap axis. (a)~Terminal gear ratio $\Gamma=\omega/\Omega$. The passive rotor
  counterrotates ($\Gamma<0$, blue up triangles) between the close-gap and
  wide-gap zero crossings at $G\approx0.12$ and $G\approx0.59$, and corotates
  ($\Gamma>0$, red down triangles) outside this interval. The wide-gap crossing
  is the planar counterpart of the experimental transition near
  $G\approx0.57$. (b)~Gap-region critical points measured from the passive
  rotor center: the $-$side vortex above the gap axis (circles), the $+$side
  vortex below it (squares), and the separating saddle (triangles). Color gives
  the azimuthal position $(\theta-\theta_{\mathrm{in}})/\pi$. The $-$side vortex
  attaches to the passive surface, detaches, and then merges with the $+$side
  vortex as $G$ increases. (c)~Inner shear-zone length $S(G)$ from the planar
  wall traction (open triangles) compared with experimental values digitized
  from Fig.~3(b) of Ref.~\cite{SmithRistrophZhang2026} (filled right
  triangles). Colors indicate the sign of the terminal coupling. Across the
  attachment window the planar traction separates into an inner arc
  $S_{\mathrm{inner}}$ and an outer bracket $S_{\mathrm{outer}}$. (d)~Normalized
  wall shear stress $\tau_w(\theta,G)$ on the passive rotor; blue shear promotes
  counterrotation and red shear promotes corotation. The zero-shear contours
  delimit the counterrotating wall arcs used in panel~(c). Dotted vertical
  lines mark the planar attachment, detachment, and merger gaps; dashed lines
  in panel~(c) mark the corresponding experimental gaps.}
  \label{fig:gaproute}
\end{figure*}

We begin with the horizontal transect $\Rey=20$ at $C=4.5$. This is the
cleanest benchmark for the gap-controlled route because it lies below the
experimental range where secondary vertical motion and long-time passive-spin
oscillations become important. Along this path the experiment reports a
reentrant sequence: close-gap corotation, a broad counterrotating band, and
wide-gap corotation. The planar calculation recovers that sequence in the
terminal gear ratio [Fig.~\ref{fig:gaproute}(a)]. The passive rotor corotates
for $G\lesssim0.12$ and again for $G\gtrsim0.59$, counterrotates between these
crossings, and reaches its strongest response, $|\Gamma|\approx0.024$, near
$G\approx0.35$. The monitored passive spin and the displayed planar fields
remain steady throughout the sweep, so this transect isolates the steady
gap-route mechanism before the Reynolds-driven boundary is considered.

Figure~\ref{fig:gaproute} collects the four diagnostics needed for that
comparison on a common gap axis. The panel order follows the logic of the
mechanism: panel~(a) gives the terminal spin response, panel~(b) tracks the
gap-region vortex centers and saddle, panel~(c) compares the computed
inner shear-zone length $S(G)$ with the experimental values digitized from
Ref.~\cite{SmithRistrophZhang2026}, and panel~(d) shows the passive-wall shear
stress $\tau_w(\theta,G)$, whose zero contours define the surface arcs used to
compute $S$.

The vortex diagnostics identify the same progression of rearrangements seen in
the experiment. A $-$side vortex above the gap axis moves onto the passive
surface between $G=0.37$ and $0.40$, close to the experimental attachment near
$G\approx0.37$. It detaches between $G=0.46$ and $0.48$, close to the
experimental $G\approx0.5$, and later merges with the $+$side vortex between
$G=0.66$ and $0.70$, close to the experimental $G\approx0.7$. The spin
transition is not locked to any single topological event: the terminal gear
ratio changes sign at $G\approx0.59$, one gap sample above the experimental
counter-to-corotation transition near $G\approx0.57$.

\begin{figure*}[!tp]
  \centering
  \newcommand{\vstream}[2]{\begin{minipage}[t]{0.19\linewidth}\centering
    {\footnotesize $G=#2$}\\[1pt]\includegraphics[width=\linewidth]{#1}\end{minipage}}
  \vstream{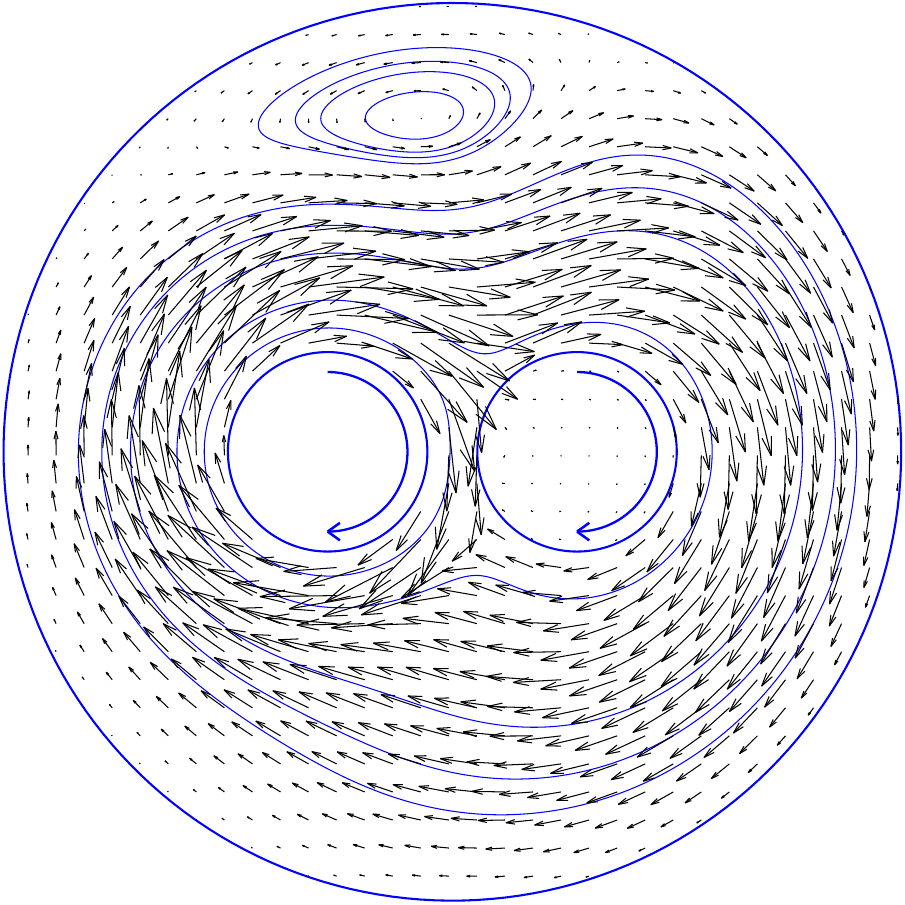}{0.10}\hfill
  \vstream{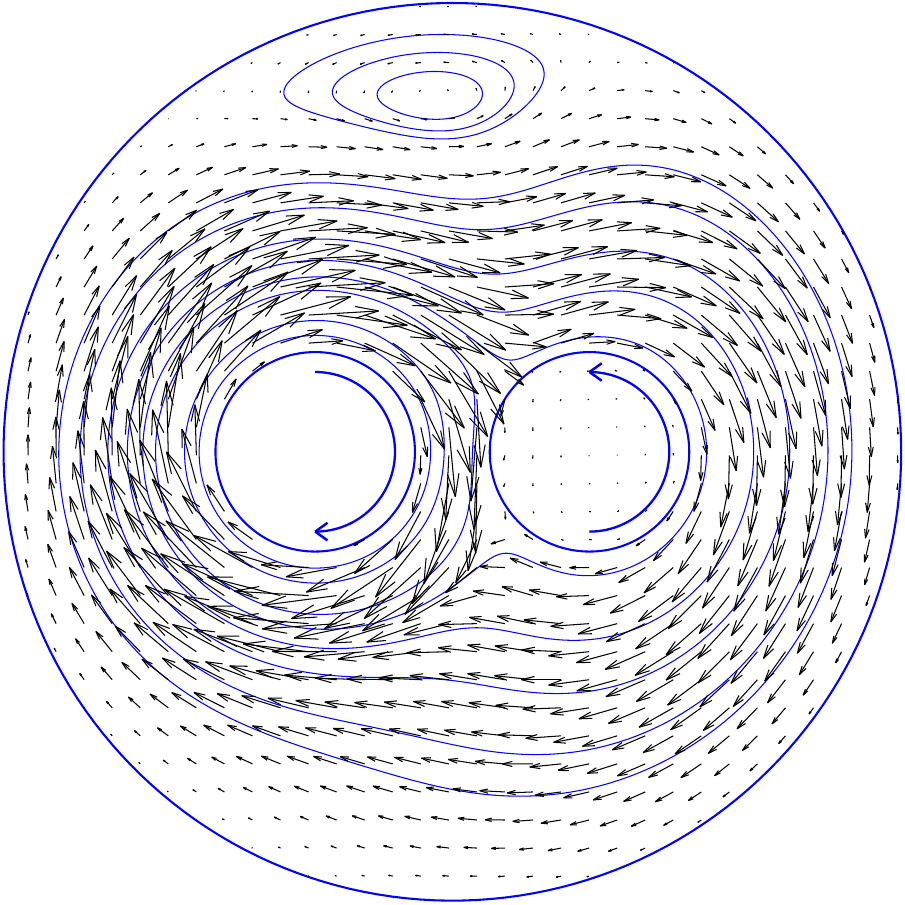}{0.15}\hfill
  \vstream{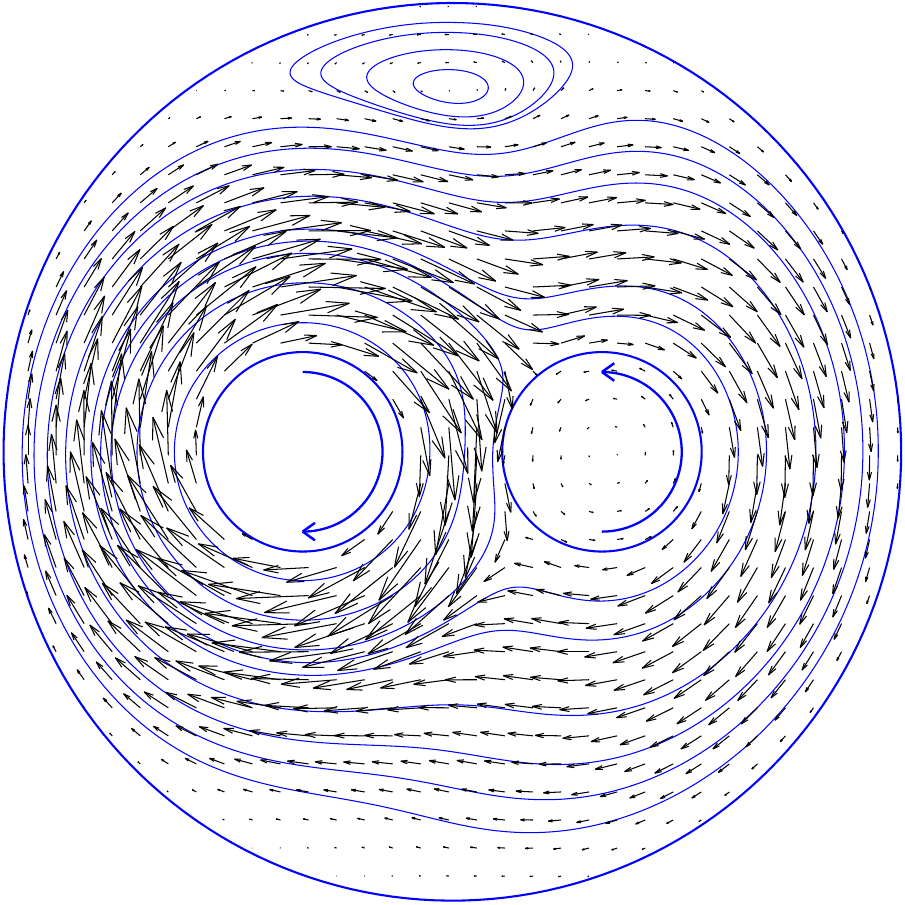}{0.20}\hfill
  \vstream{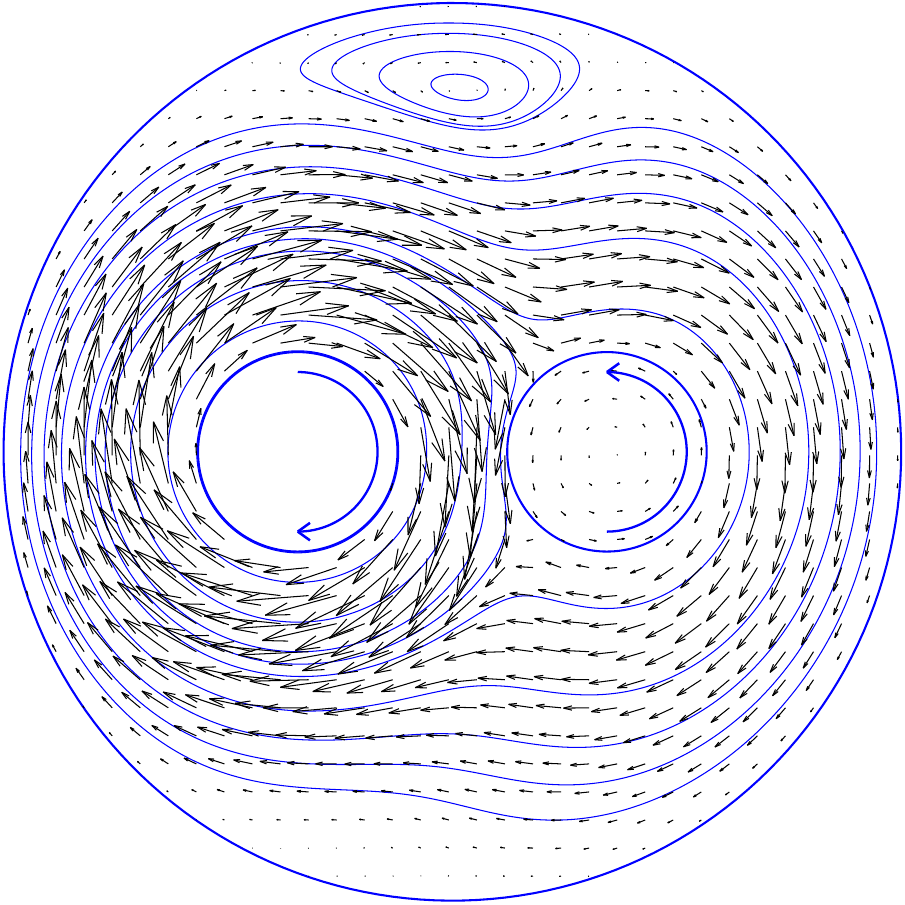}{0.22}\hfill
  \vstream{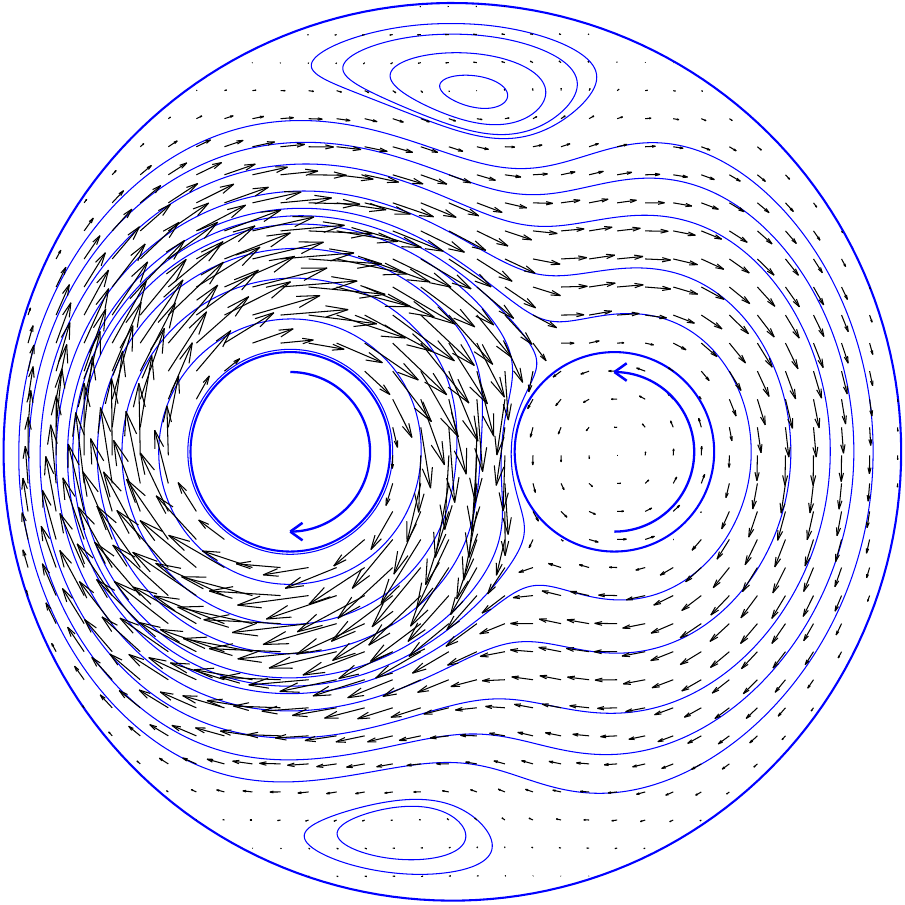}{0.25}

  \vspace{5pt}
  \vstream{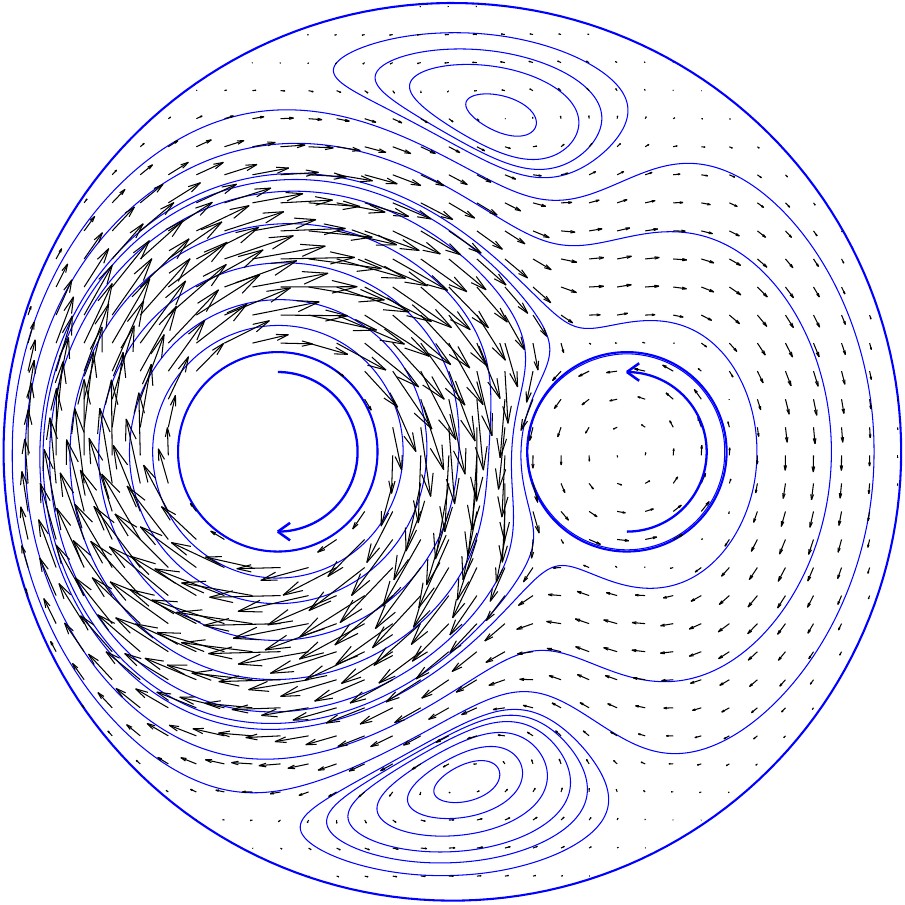}{0.30}\hfill
  \vstream{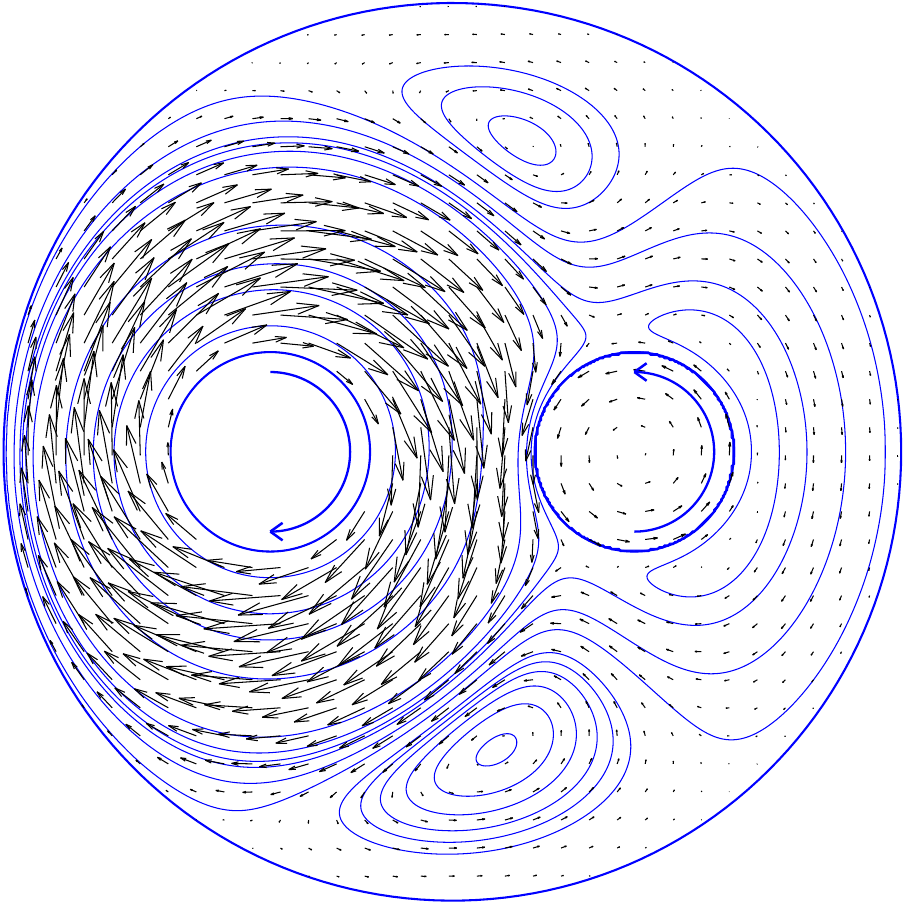}{0.33}\hfill
  \vstream{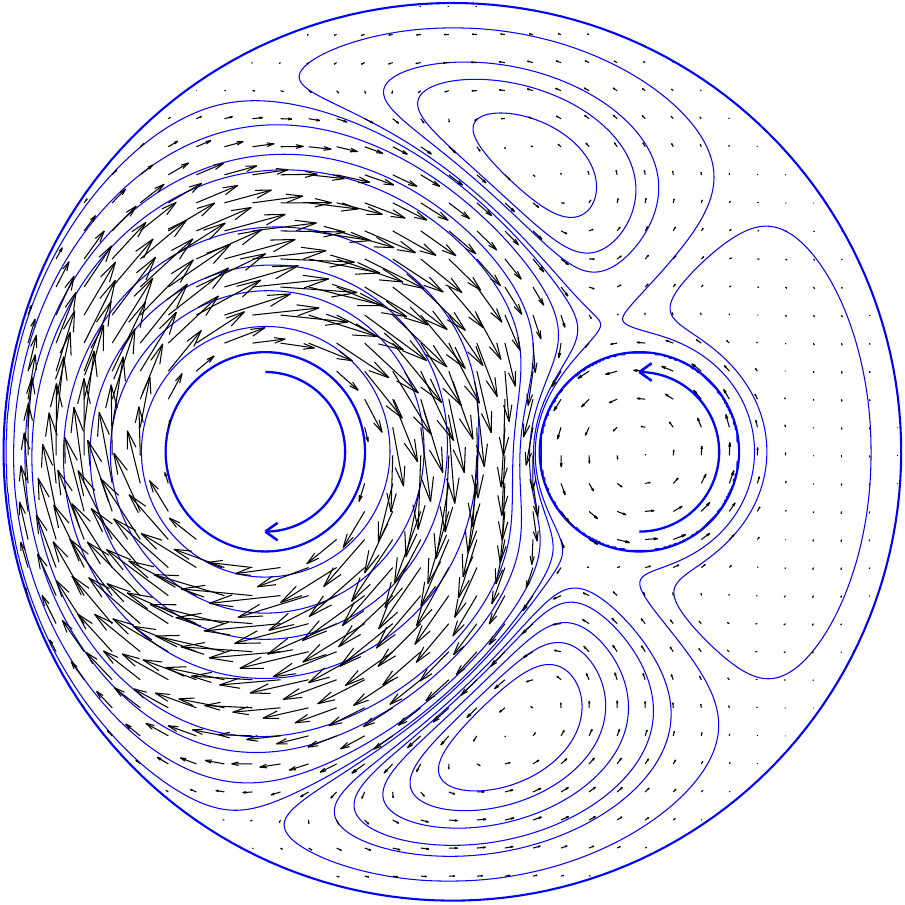}{0.35}\hfill
  \vstream{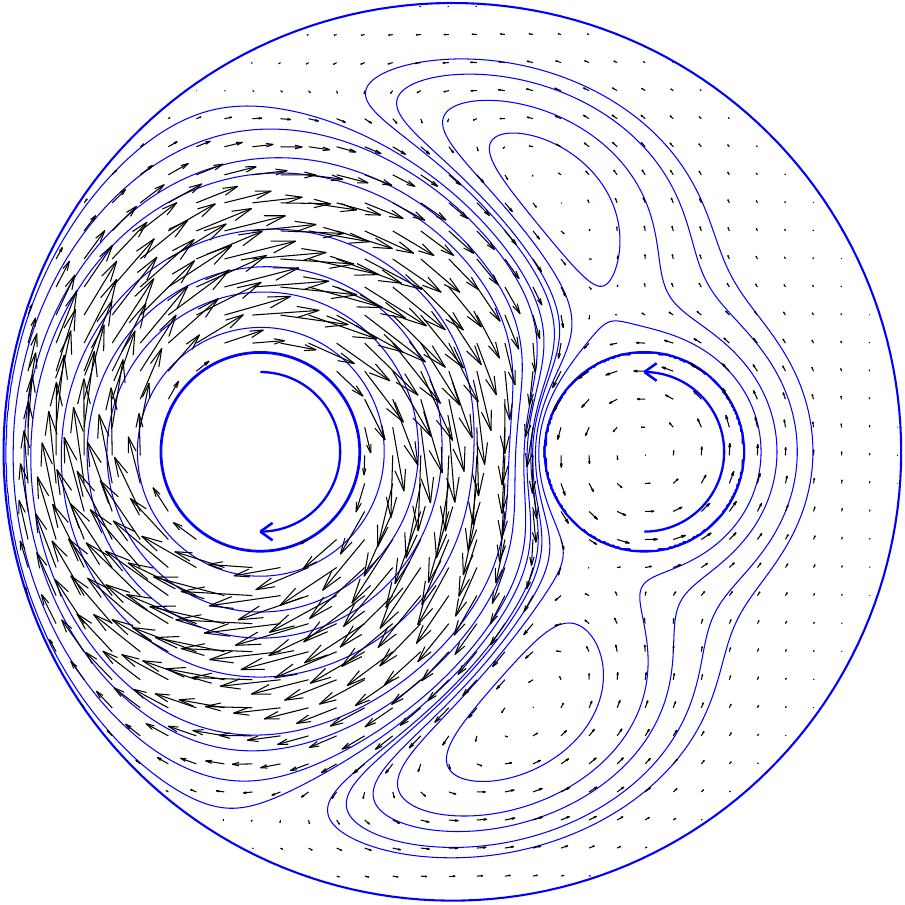}{0.37}\hfill
  \vstream{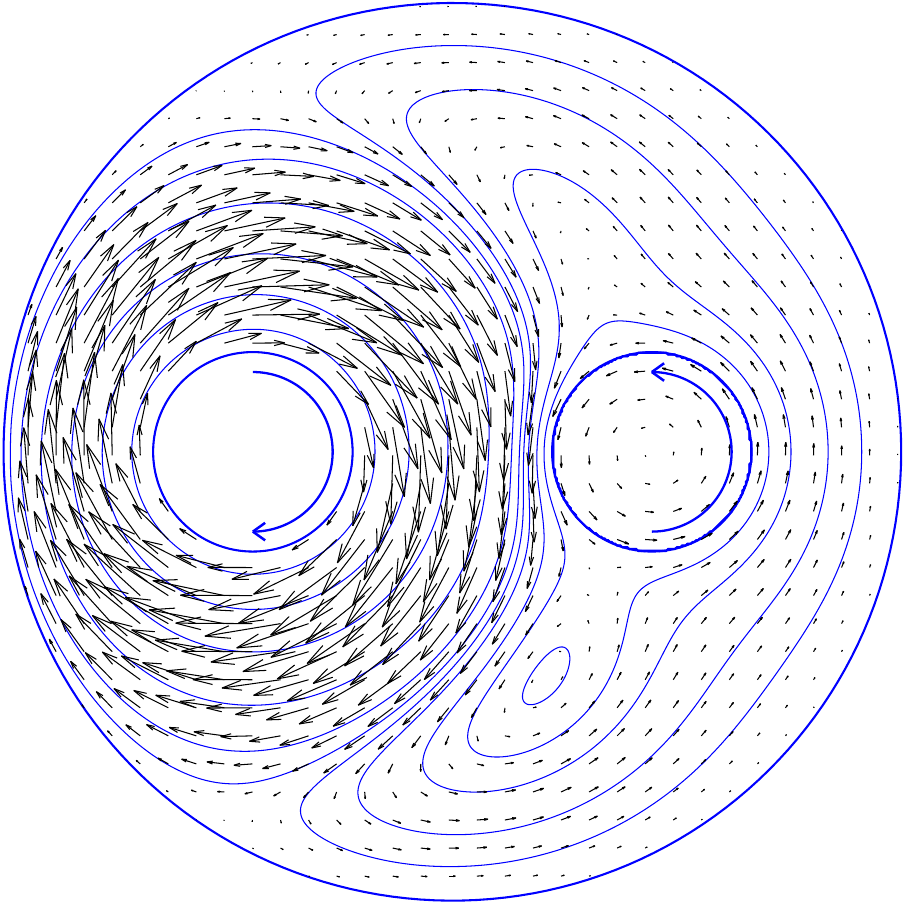}{0.40}

  \vspace{5pt}
  \vstream{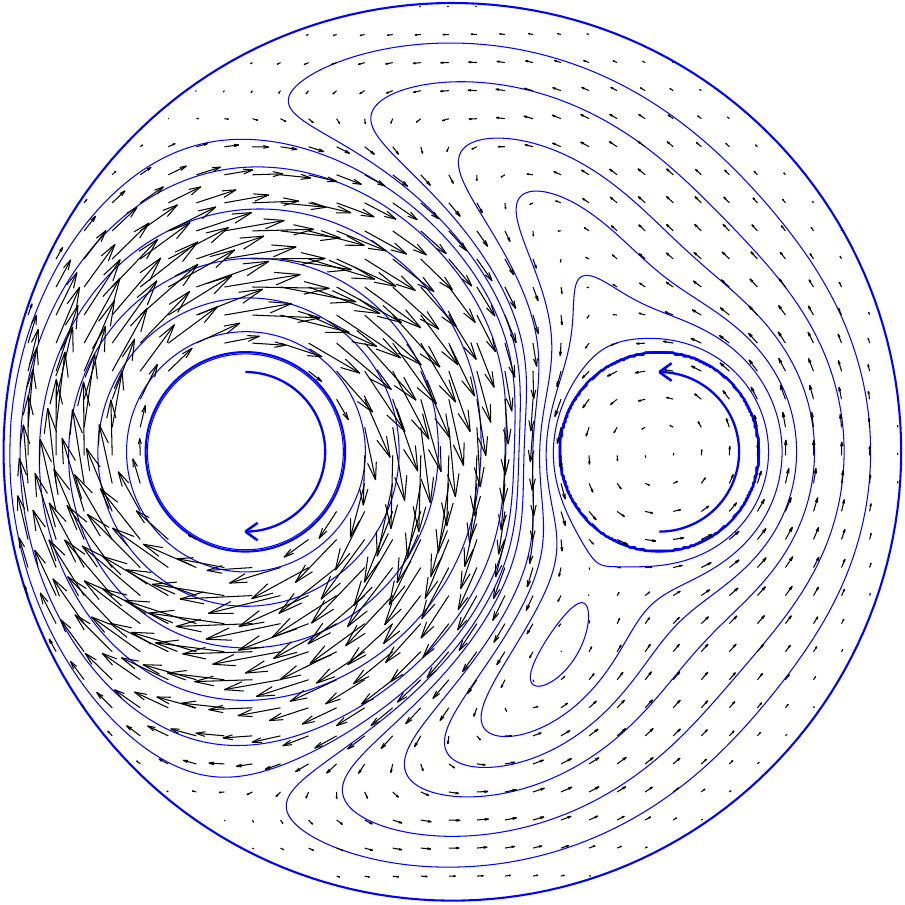}{0.43}\hfill
  \vstream{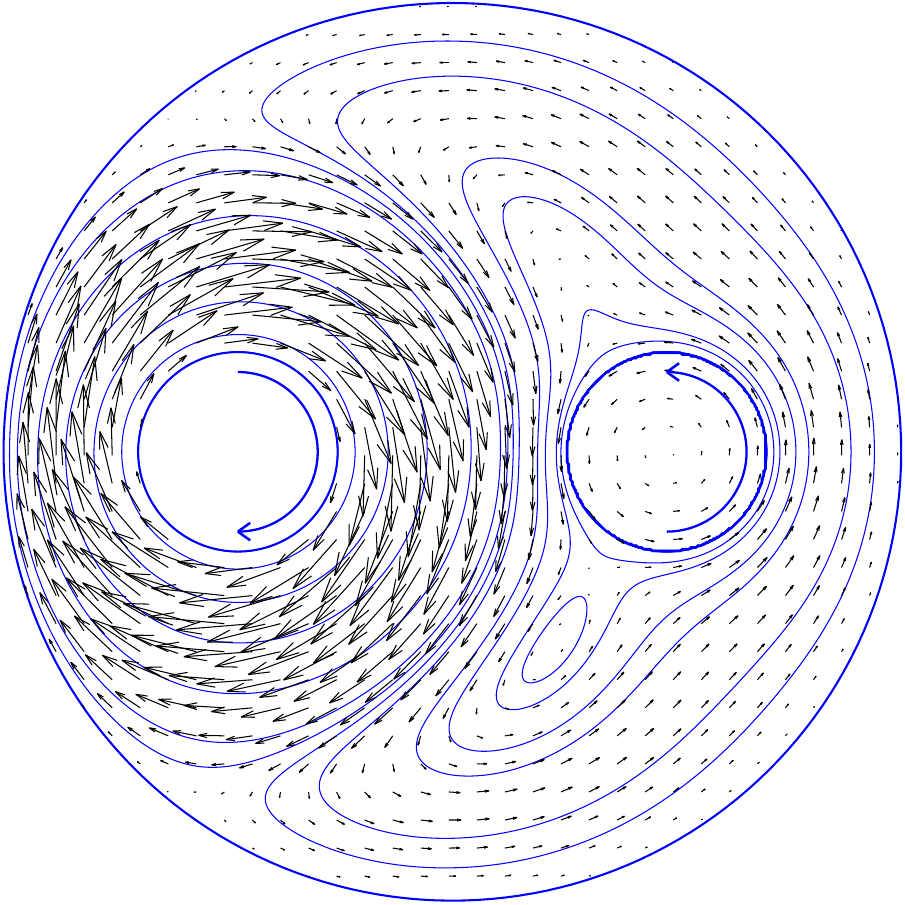}{0.46}\hfill
  \vstream{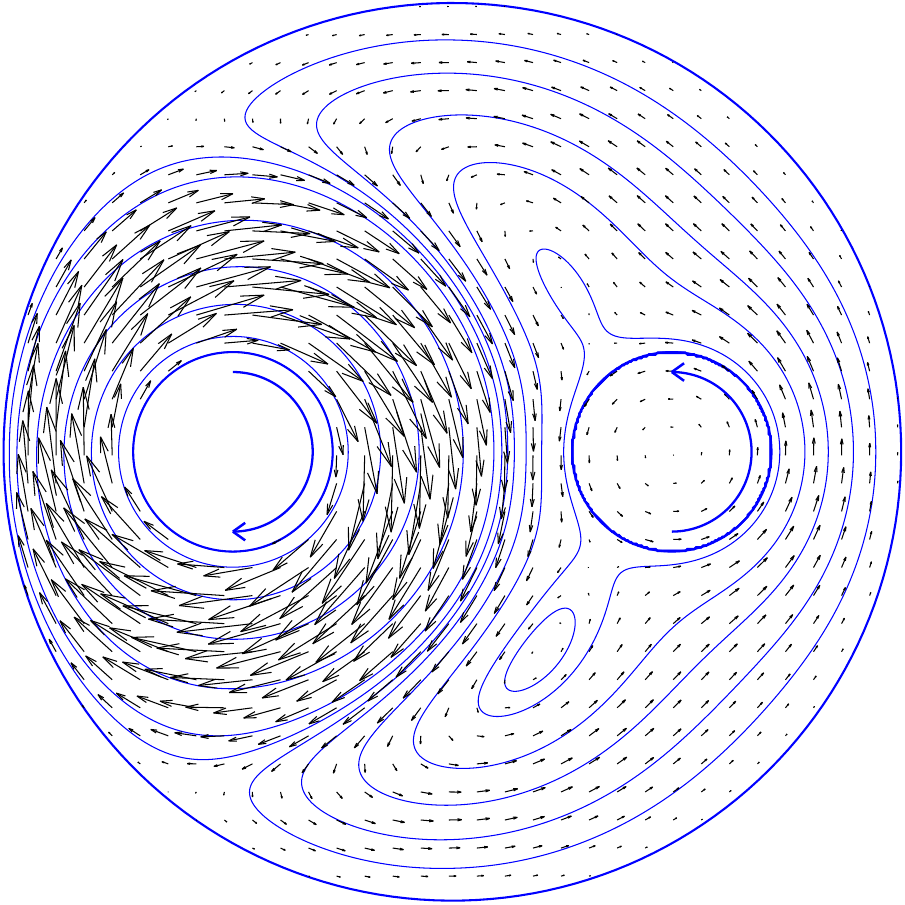}{0.48}\hfill
  \vstream{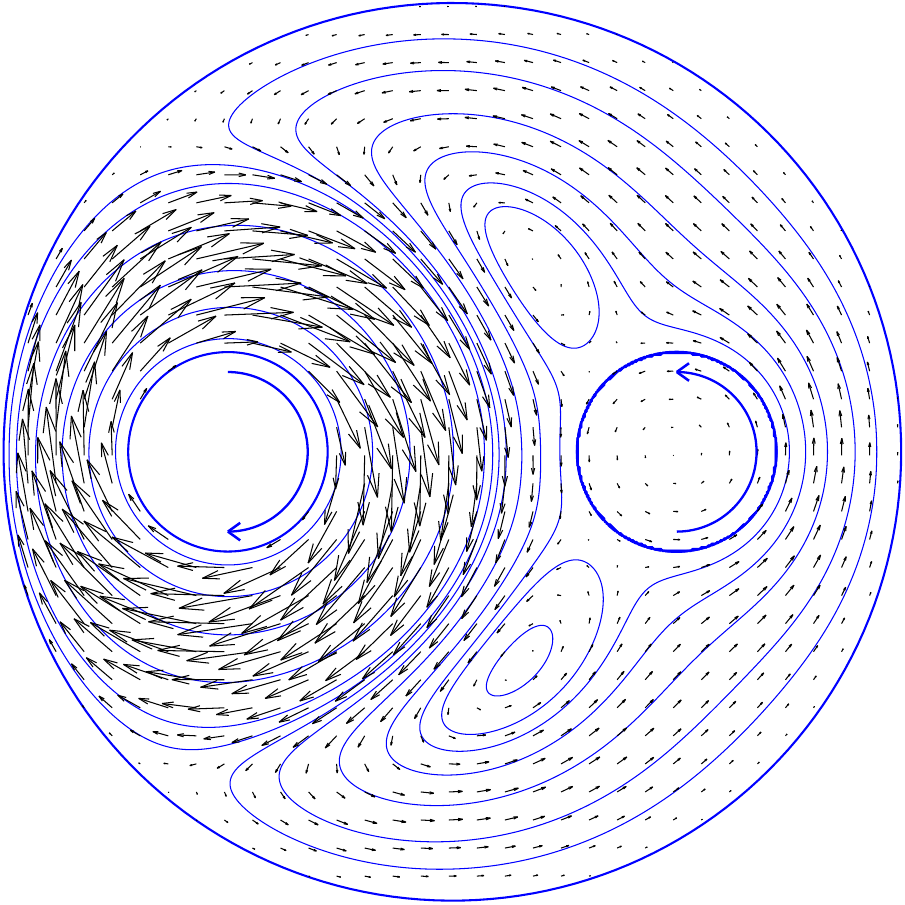}{0.50}\hfill
  \vstream{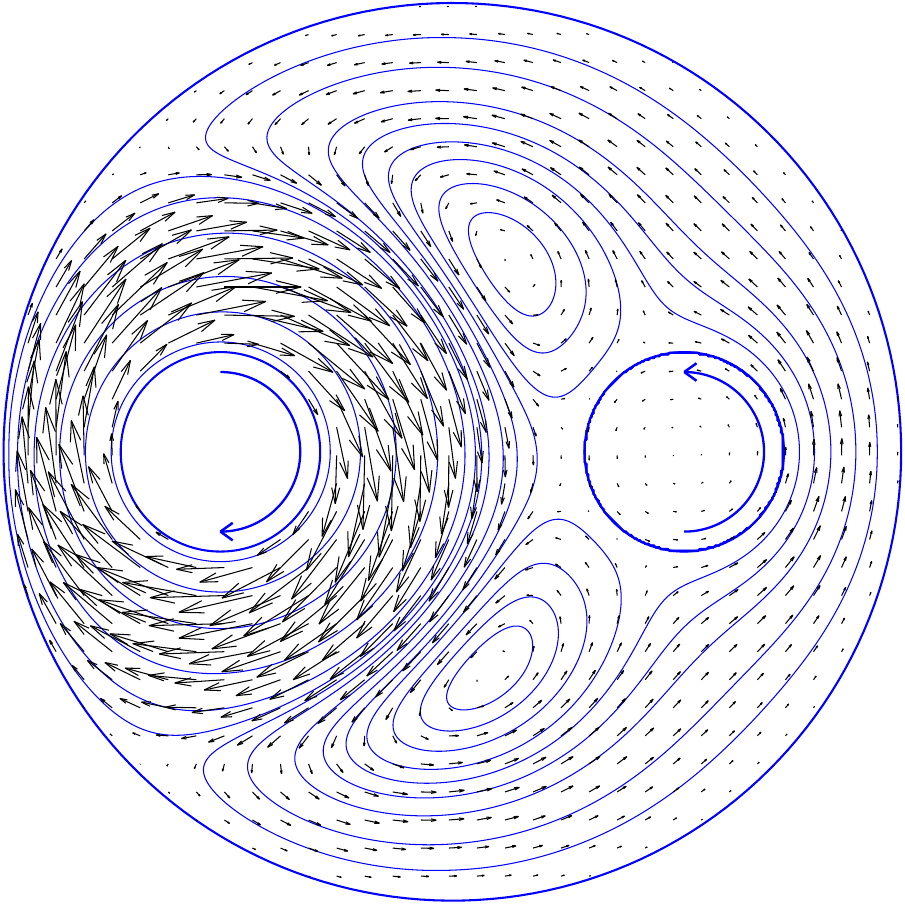}{0.53}

  \vspace{5pt}
  \vstream{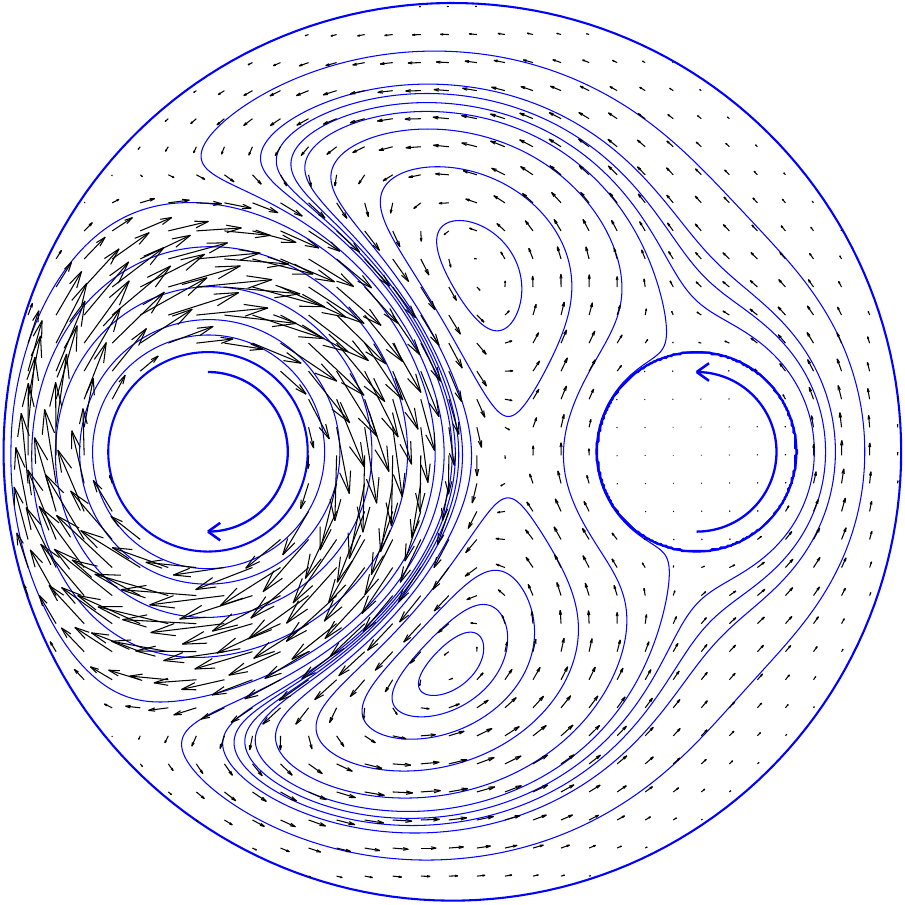}{0.58}\hfill
  \vstream{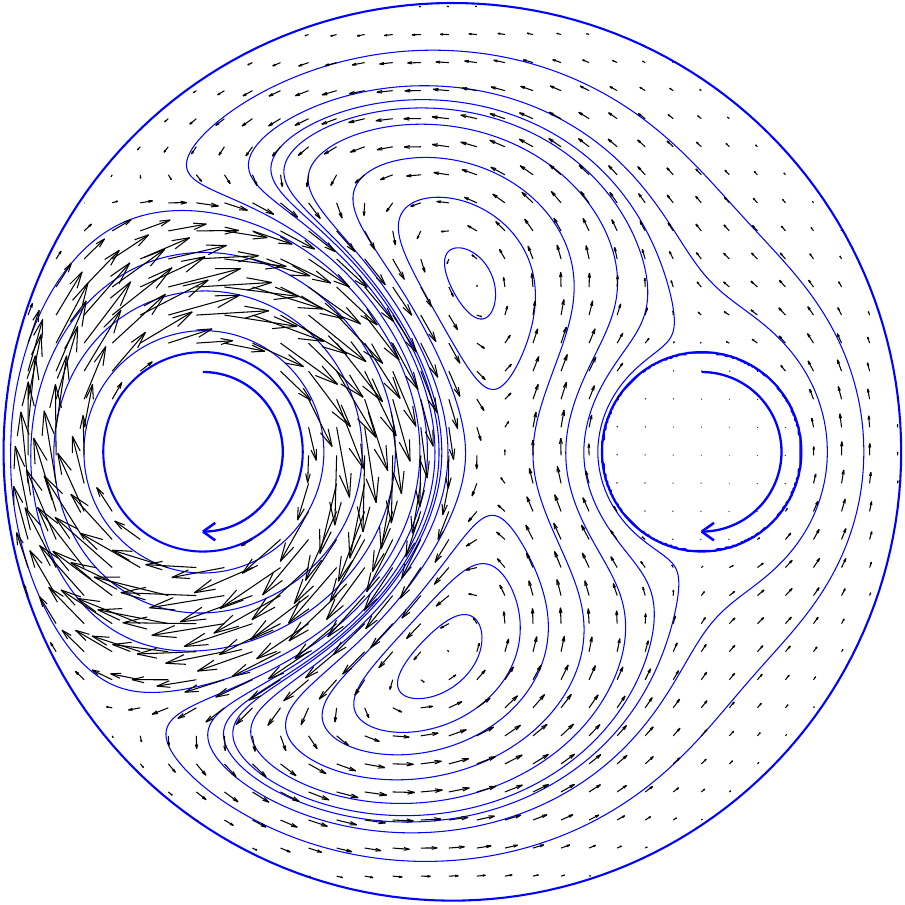}{0.60}\hfill
  \vstream{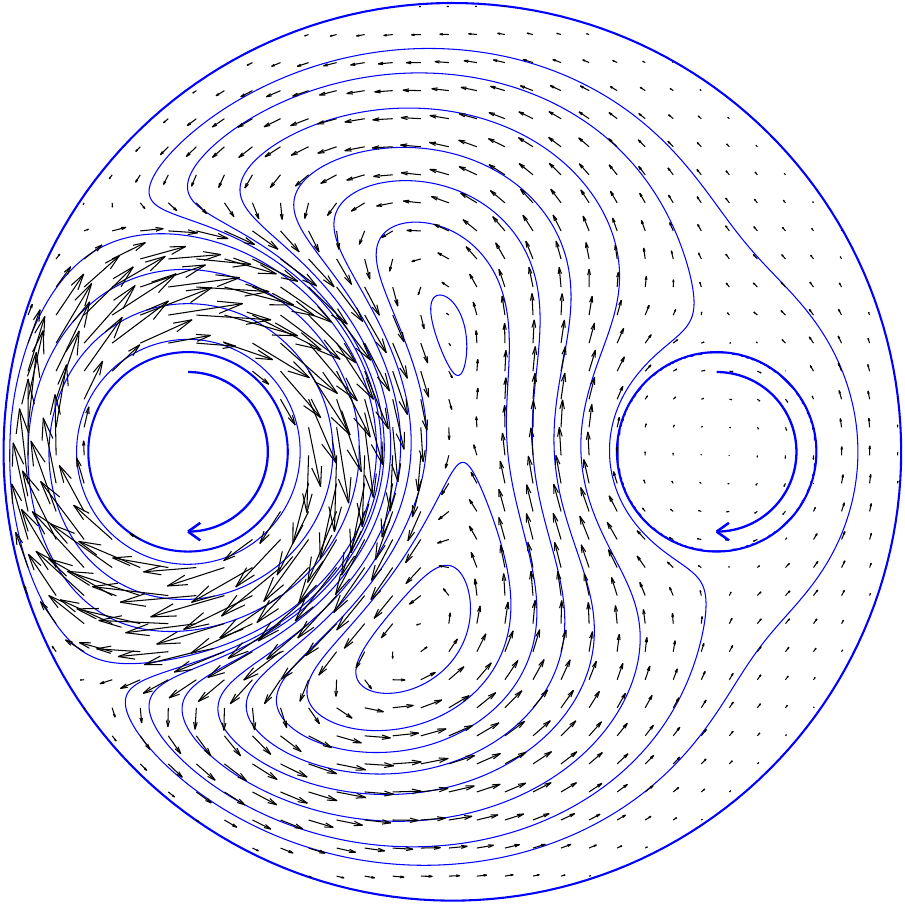}{0.66}\hfill
  \vstream{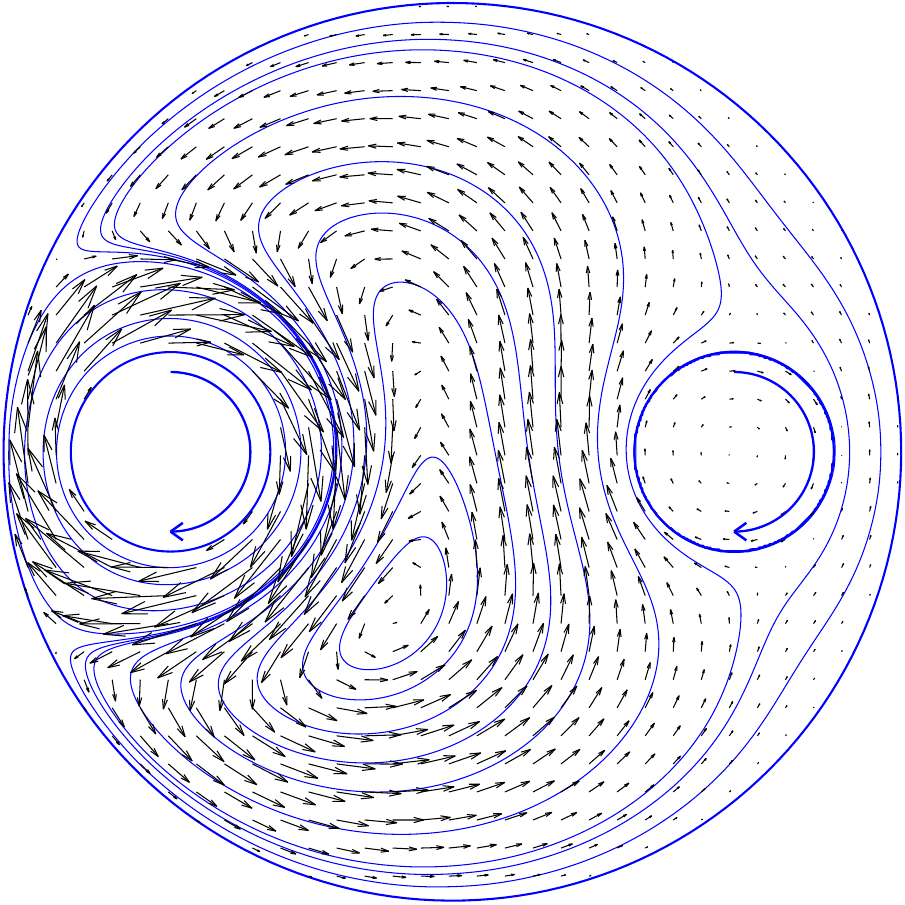}{0.73}\hfill
  \vstream{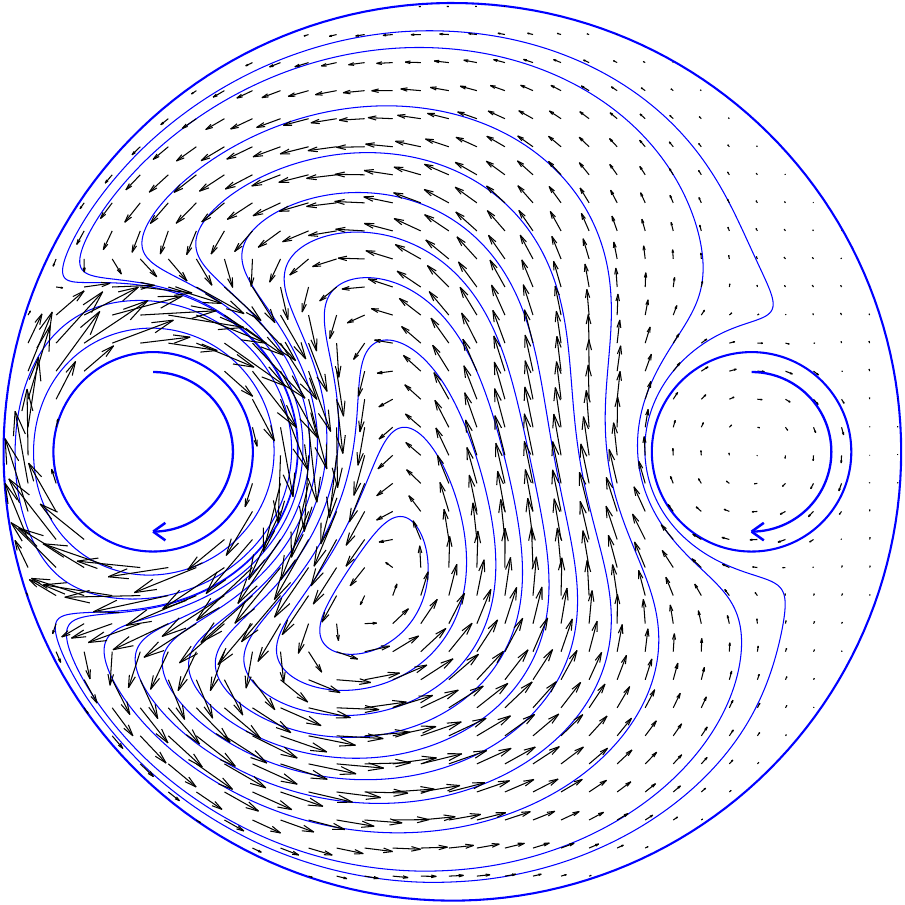}{0.80}
  \caption{Computed velocity fields with superimposed streamlines at
  $\Rey=20$, $C=4.5$ for twenty interrotor gaps from $G=0.10$ to $0.80$,
  arranged by increasing gap. The active rotor is on the left and the passive
  rotor on the right. Two corotating gap vortices organize the gap-route
  rearrangements: a $-$side center above the gap axis and a $+$side center
  below it. The $-$side is absorbed into the passive boundary layer across
  $G=0.40$ to $0.46$ and re-emerges at $G=0.48$; the pair coalesces into a
  single center near $G\approx0.68$. These planar attachment, detachment, and
  merger events are close to the experimental values $G\approx0.37$, $0.50$,
  and $0.70$.}
  \label{fig:vec}
\end{figure*}

The wall-traction map shows how this bulk sequence reaches the passive rotor.
At small and intermediate gaps, the dominant blue traction arc on the
gap-facing side promotes counterrotation. Near attachment, an additional outer
arc appears and the shear-zone diagnostic separates into two readings: a
simple inner surface arc and a larger bracket associated with the rearranging
gap vortex. As the gap widens, the inner counterrotating arc retreats, the
outer red traction expands, and the wide-gap corotating state emerges. The
streamline fields in Fig.~\ref{fig:vec} provide the corresponding physical
view: through the counterrotating band a corral-scale circulation washes the
passive rotor, while at large gap the outer return flow increasingly controls
the passive surface.

Thus the planar calculation reproduces the main steady gap-route architecture:
the reentrant spin sequence, the wide-gap transition, and the attachment,
detachment, and merger progression of the gap vortices. The agreement is
selective, however. The close-gap corotation at $G=0.10$ occupies a much wider
range than the experimental close-range geometric state near $G\approx0.02$,
and not every jump in the experimental pathline-based $S(G)$ appears as a jump
of the simple wall-traction arc. We return to this distinction in
Sec.~\ref{sec:shear}, where we separate the surface contribution from the
outer-bracket contribution to the experimental shear-zone diagnostic.


\subsection{The inner shear-zone length and its two channels}
\label{sec:shear}

The gap-route rearrangements provide a stricter test than the sign of
$\Gamma$ alone. The relevant diagnostic is the inner shear-zone length $S(G)$
on the passive rotor, defined as the fraction of the circumference over which
the wall shear promotes counterrotation. In the experiment, this length is
inferred from pathline photographs by identifying stagnation and separation
points. In the calculation, the corresponding points can be located directly as
zeros of the tangential wall traction. The computed $S(G)$ is therefore a
surface-traction counterpart of the experimental pathline-based diagnostic,
rather than an identical observable.

Figure~\ref{fig:gaproute}(b,d) puts the surface traction and the bulk vortex
topology on the same gap axis. Panel~(d) maps the normalized wall shear
$\tau_w(\theta,G)$ on the passive rotor, and panel~(b) tracks the critical
points of the gap-vortex system in the corral interior. We evaluate $S$ at
$\Rey=20$, $C=4.5$ using the fields of Fig.~\ref{fig:vec} together with
intermediate gaps. The extracted arc length changes by less than one percent
as the near-wall offset used for the traction gradient is varied from one to
eight mesh spacings. The comparison then separates two channels of the
experimental diagnostic: a surface channel carried directly by the wall
traction, and a bracket channel associated with how the experimental
stagnation--separation pair wraps around the rearranging gap vortex.

The planar surface arc reproduces part, but not all, of the experimental
$S(G)$ curve (Fig.~\ref{fig:gaproute}(c)). Before attachment, $S$ rises to about
$0.24$ near $G=0.30$, roughly $0.06$ below the experimental value, and then
falls slightly to about $0.21$ by $G=0.41$. At attachment it jumps from about
$0.21$ to about $0.38$ and then rises to about $0.53$ by $G=0.46$, whereas the
experiment jumps from about $0.35$ to about $0.85$ near $G\approx0.37$. At
larger gaps the computed surface arc decreases smoothly and reaches about
$0.34$ by $G=0.80$, matching the experimental post-merger branch to within
$0.01$ over $G=0.70$ to $0.80$.

This comparison shows that the planar surface traction captures the lower
branch, the post-merger branch, and the attachment jump. It does not, by
itself, reproduce the elevated attachment branch near $S\approx0.85$ or the
detachment branch near $S\approx0.60$. Those branches appear instead in the
outer bracket of the wall-traction map: the arc obtained by continuing to the
next pair of traction zeros after the first gap-facing counterrotation arc.
Thus the planar flow contains the same surface information, but divided
between a simple inner arc and an outer bracket that is displaced to slightly
wider gap and steepened relative to the experimental pathline reading.

The missing jumps are not jumps of the simple surface arc. The number of
counterrotating wall arcs rises from one to two at attachment and returns to
one at detachment, but $\Gamma$ varies smoothly through every rearrangement,
including the spin crossing at $G\approx0.59$ (Fig.~\ref{fig:gaproute}(a)). The
surface arc is continuous through detachment and merger, where the
experimental $S$ jumps downward. Only the attachment relocates the
passive-surface traction zeros discontinuously in the planar calculation.

The bulk vortex topology nevertheless contains the full attachment,
detachment, and merger sequence. As the gap increases, the $-$side vortex
reaches the passive surface and is absorbed into the boundary layer between
$G=0.40$ and $0.46$; it re-emerges as a free center at $G=0.48$ while the
$+$side remains free. From $G=0.58$ onward a saddle separates the two centers,
and near $G\approx0.68$ the weaker $-$side merges into the stronger $+$side as
the saddle annihilates. The planar flow therefore reproduces the experimental
bulk topology, even though the simple surface arc registers only the attachment
jump.

Taken together, the surface and bulk diagnostics decompose the experimental
$S(G)$ into two planar contributions. The surface channel carries the lower
and post-merger branches and the attachment jump. The bracket channel carries
the elevated branch associated with attachment and detachment, while the bulk
vortex topology carries the merger. This decomposition explains why the
planar calculation recovers the gap-route rearrangements and the wide-gap
coupling transition, but does not reproduce every jump in the experimental
pathline-based $S(G)$ as a jump of the wall-traction arc. Quantitatively, the three experimental jumps digitized from Fig.~3(b) of
Ref.~\cite{SmithRistrophZhang2026} (uncertain to about $0.02$) divide between
the two channels. The attachment jump, $\Delta S\approx+0.5$, is the only one
with a substantial surface contribution: the wall-traction arc accounts for
about $+0.32$, leaving a bracket remainder near $+0.18$. The detachment and
merger jumps, $\Delta S\approx-0.25$ and $-0.18$, have no wall-traction
counterpart and are carried entirely by the outer bracket and the bulk
gap-vortex topology.

The result is selective rather than uniformly positive or negative. The planar
model recovers the rearrangement sequence, places the wide-gap spin transition
within one gap sample of the experiment, matches the post-merger branch of
$S(G)$, and captures the attachment jump in the surface traction. It does not
reproduce the experimental detachment and merger jumps as jumps of the simple
wall-traction arc; those features are carried instead by the outer bracket and
the gap-vortex topology. This distinction will matter in the Reynolds-number
transects, where a terminal spin reversal can also occur without collapse of
the gap-facing shear zone.

\subsection{Vertical transects and the displaced Reynolds-driven boundary}
\label{sec:inertial}

\begin{figure*}[!tp]
  \centering
  \includegraphics[width=\linewidth]{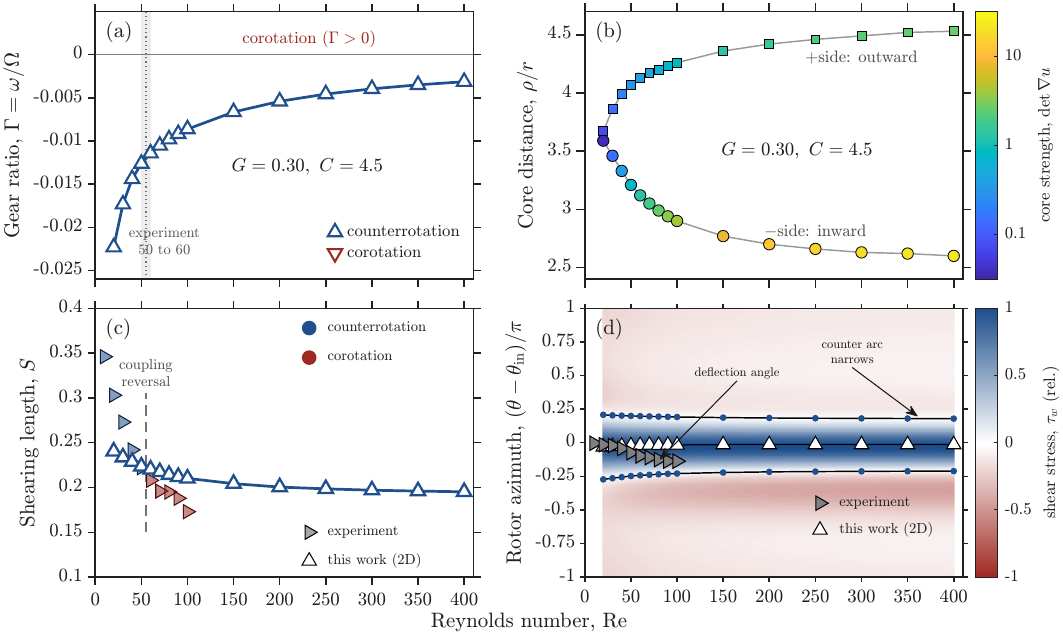}
  \caption{Reynolds-route diagnostics at the experimental gap, $G=0.30$,
  $C=4.5$. All panels share the same Reynolds axis. (a)~Terminal gear ratio
  $\Gamma=\omega/\Omega$. The planar passive rotor remains counterrotating
  ($\Gamma<0$) over the computed range; the shaded band marks the experimental
  transition window near $\Rey=55$. (b)~Distances of the $-$side (circles) and
  $+$side (squares) gap-vortex cores from the passive-rotor center, colored by
  core strength $\det\nabla u$ on a logarithmic scale. The cores separate and
  strengthen smoothly, with no creation or annihilation event. (c)~Wall-based
  shearing length $S(\Rey)$ in the planar calculation, compared with values
  digitized from Fig.~3(d) of Ref.~\cite{SmithRistrophZhang2026}. Both decrease
  with Reynolds number, but only the experiment crosses to corotation.
  (d)~Normalized wall shear stress $\tau_w$ on the passive rotor, plotted
  against surface azimuth and Reynolds number; blue shear promotes
  counterrotation and red shear promotes corotation. The gap-facing
  counterrotating arc narrows but persists. Overlaid markers compare the center
  of this planar arc (white up triangles) with the experimental deflection
  angle of the inner shearing zone (grey side triangles), digitized from
  Fig.~S8(c) of Ref.~\cite{SmithRistrophZhang2026} and entered on the same
  azimuth axis as minus the deflection angle in units of $\pi$, so its clockwise
  sense shares the gap-facing side. The experimental inner zone moves
  away from the gap axis as $\Rey$ rises, whereas the planar arc center remains
  near it.}
  \label{fig:reroute}
\end{figure*}

We next fix the rotor separation and vary the Reynolds number. This vertical
transect is the most stringent comparison because it asks whether the
experimental high-$\Rey$ counter-to-corotation route is recovered at the same
gap in a strictly planar calculation. The natural benchmark is $G=0.30$,
$C=4.5$, close to the experimental inertial transect, reported at $G=0.33$ and
rounded to $G\approx0.3$ in
Ref.~\cite{SmithRistrophZhang2026}. Experimentally, the passive rotor reverses
near $\Rey=55$ (between the measured points $\Rey=50$ and $60$). That route is
tied to three linked observations: high-$\Rey$ axial inflow and
Taylor-roll-like secondary motion in vertical-plane images, outward-spiraling
pathlines at the midplane, and a shear-competition model in which both the
inner shearing length $S(\Rey)$ and its deflection angle enter the zero-torque
balance. Figure~\ref{fig:reroute} therefore tests the planar calculation at
four levels: terminal gear ratio, gap-vortex motion, inner shear-zone length,
and the wall-traction field, including the position of its inner arc.

At the experimental gap, the planar calculation differs first at the level of
the terminal spin. Figure~\ref{fig:reroute}(a) shows that $\Gamma$ remains
negative throughout the computed range. Its magnitude decreases, from
$\Gamma\approx-0.022$ at $\Rey=20$ to $\Gamma\approx-0.003$ at $\Rey=400$, so
the transect approaches the stagnation line from the counterrotating side but
does not cross it. The smaller corral $C=3$ follows the same qualitative trend,
with larger counterrotating magnitude. Thus the planar model has a clear
Reynolds-number response at this gap, but the response weakens
counterrotation rather than producing the experimental reversal.

The bulk-flow diagnostics show continuous strengthening and displacement, not a
topological event. The two gap vortices persist throughout the sweep
[Fig.~\ref{fig:reroute}(b)]. The $-$side core moves inward toward the passive
rotor, its distance $\rho$ from the passive-rotor center decreasing from
$\rho/r\approx3.6$ to $2.6$, while the $+$side core moves outward
toward the wall, from $\rho/r\approx3.7$ to $4.5$; both strengthen with
Reynolds number, with the $-$side core circulation increasing by more than two
orders of magnitude. These changes are continuous displacements and
amplifications of existing structures, not vortex birth, merger, or loss, as
the velocity-field sequence shows directly (Fig.~\ref{fig:vecRe}).

The wall diagnostics make the mismatch sharper. In the experimental
shear-competition picture, gap-facing shear promotes counterrotation while the
outer shear over the rest of the passive-rotor perimeter promotes corotation;
the passive rotor settles where the two contributions balance. Two geometric
changes weaken the gap-facing contribution: shortening of the inner shearing
length $S$, and deflection of the inner zone away from the gap axis, which
tilts the fastest inner shear away from the line of centers. In the experiment
both changes occur. The inner zone shrinks substantially, and its center swings
to a deflection of about $0.43$ rad, or $25^\circ$, by $\Rey=100$. In the
planar calculation, only a weak version of the first trend appears. The
shearing length decreases modestly, from about $0.24$ to $0.20$
[Fig.~\ref{fig:reroute}(c)], while the wall-traction map retains a single
gap-facing counterrotating arc whose center remains close to the gap axis
[Fig.~\ref{fig:reroute}(d)]. The planar and experimental markers are not
identical observables: one is the center of a computed wall-traction arc, the
other the center of a pathline-defined inner shearing zone. The overlay should
therefore be read as a directional diagnostic, not as a pointwise validation
test. At the experimental gap, the planar flow captures inertial narrowing of
the inner shear zone, but not the accompanying angular deflection, shear-zone
collapse, or torque reversal.

The deflection angle therefore clarifies what is displaced. In the experiment,
the shortening and rotation of the inner shear zone are interpreted as
downstream consequences of centrifugal routing at the midplane: as $\Rey$
increases, pathlines around the active rotor change from closed loops to
outward spirals that send the fastest flow toward the outside of the passive
rotor. The experiment also reveals high-$\Rey$ vertical inflow and
Taylor-roll-like secondary motion, which can feed this midplane outflow while
acting mainly axially rather than directly shearing the passive rotor. At
$G=0.30$, the strictly planar calculation develops the bulk-vortex displacement
and strengthening just described, but not enough surface redirection to move the
inner arc off the gap axis or reverse the net torque. The discrepancy is
therefore not only that $S(\Rey)$ fails to collapse. The planar model also
lacks, or strongly underproduces, the deflection of the inner shear zone that is
central to the experimental high-$\Rey$ shear-balance model. Whether that
missing strength is intrinsic to strictly two-dimensional dynamics or supplied
by finite-depth axial flow and apparatus geometry is not resolved by the present
data, and remains part of the end-wall question considered in
Sec.~\ref{sec:compare}.

\begin{figure*}[!tp]
  \centering
  \includegraphics[width=\linewidth]{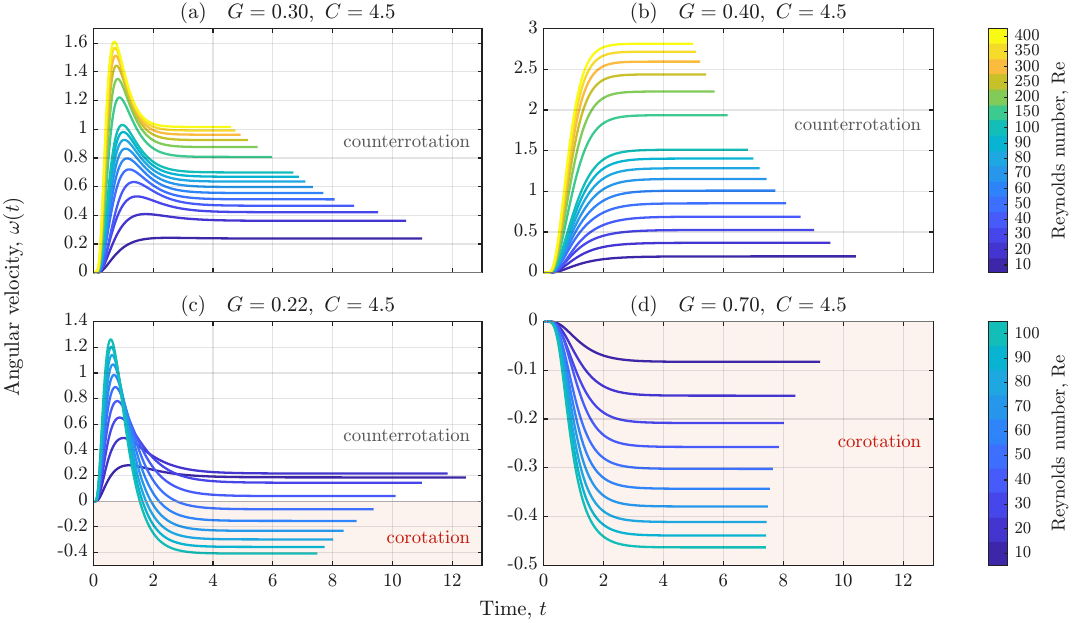}
  \caption{Passive angular-velocity histories $\omega(t)$ at $C=4.5$, colored
  by Reynolds number. The active rotor turns clockwise ($\Omega<0$), so
  $\omega>0$ denotes counterrotation and $\omega<0$ denotes corotation. Panels
  (a)~$G=0.30$ and (b)~$G=0.40$ sweep $\Rey=10$ to $400$; panels
  (c)~$G=0.22$ and (d)~$G=0.70$ sweep $\Rey=10$ to $100$. The lower color scale
  is the $\Rey=10$ to $100$ segment of the same mapping, so equal Reynolds
  numbers have equal colors in all panels. At the intermediate gaps (a,b), the
  terminal values remain positive and increase with drive, so the passive
  rotor stays counterrotating. At the narrow gap (c), the terminal value changes
  sign near $\Rey=44$ after an early counterrotating overshoot. At the wide gap
  (d), the passive rotor is corotating throughout and relaxes monotonically.
  Each panel has its own vertical scale; the time axis is common.}
  \label{fig:history}
\end{figure*}

The angular-velocity histories in Fig.~\ref{fig:history} clarify a point hidden
by the normalized gear ratio. Since $\omega=\Gamma\Omega$ and
$|\Omega|\propto\Rey$, a decrease in $|\Gamma|$ need not mean that the passive
rotor slows in dimensional angular velocity. At $G=0.30$, the terminal
$\omega$ remains positive, and therefore counterrotating, while increasing
with Reynolds number [Fig.~\ref{fig:history}(a)]. The neighboring intermediate
gap $G=0.40$ behaves similarly [Fig.~\ref{fig:history}(b)]. The approach of
$\Gamma$ toward zero at these gaps is therefore not an incipient terminal
reversal; it means that the active speed grows faster than the passive response.

\begin{figure*}[!tp]
  \centering
  \newcommand{\vstreamre}[2]{\begin{minipage}[t]{0.19\linewidth}\centering
    {\footnotesize $\Rey=#2$}\\[1pt]\includegraphics[width=\linewidth]{#1}\end{minipage}}
  \vstreamre{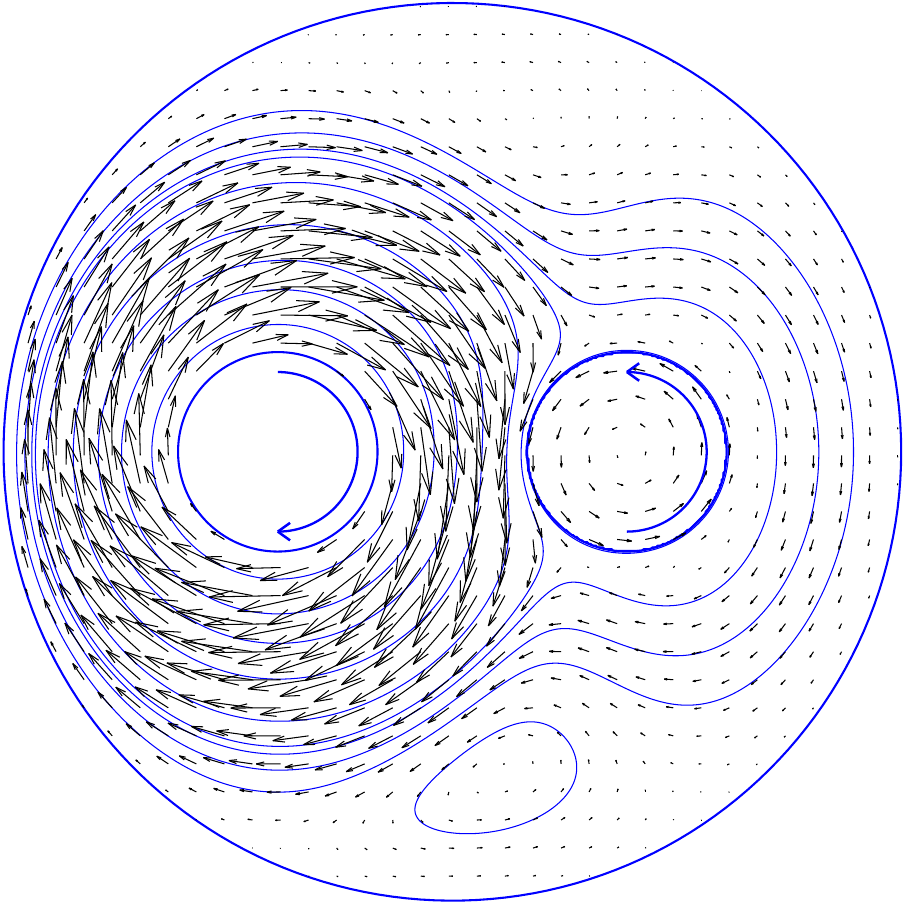}{10}\hfill
  \vstreamre{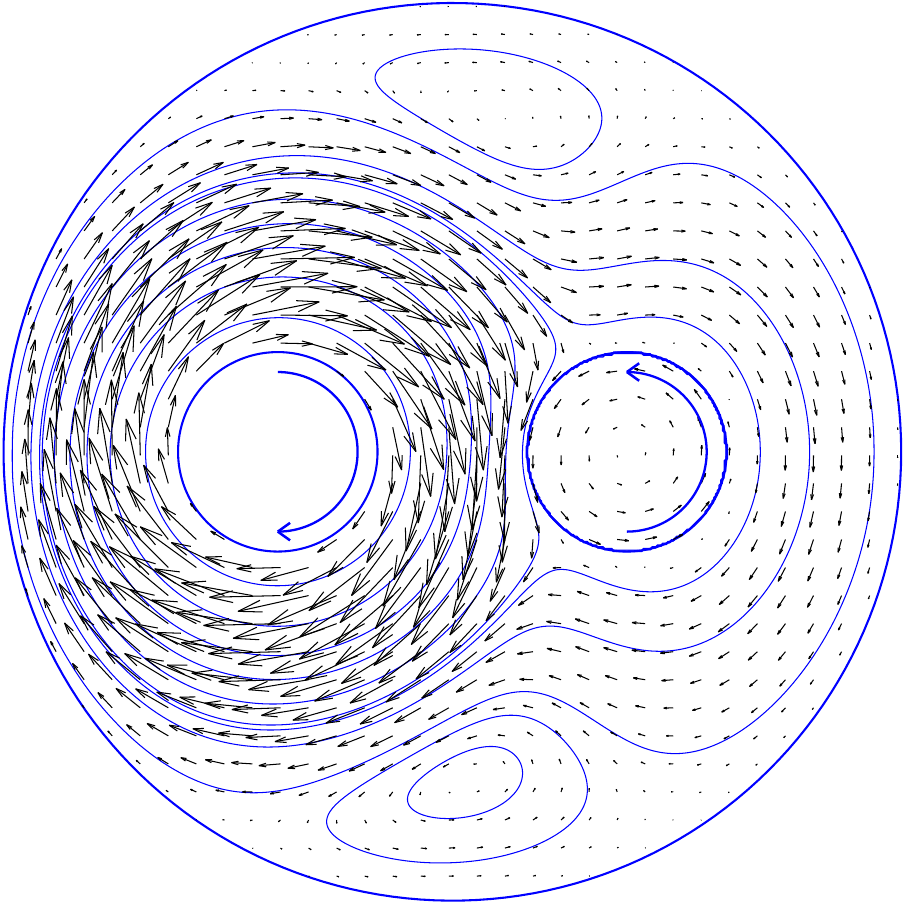}{20}\hfill
  \vstreamre{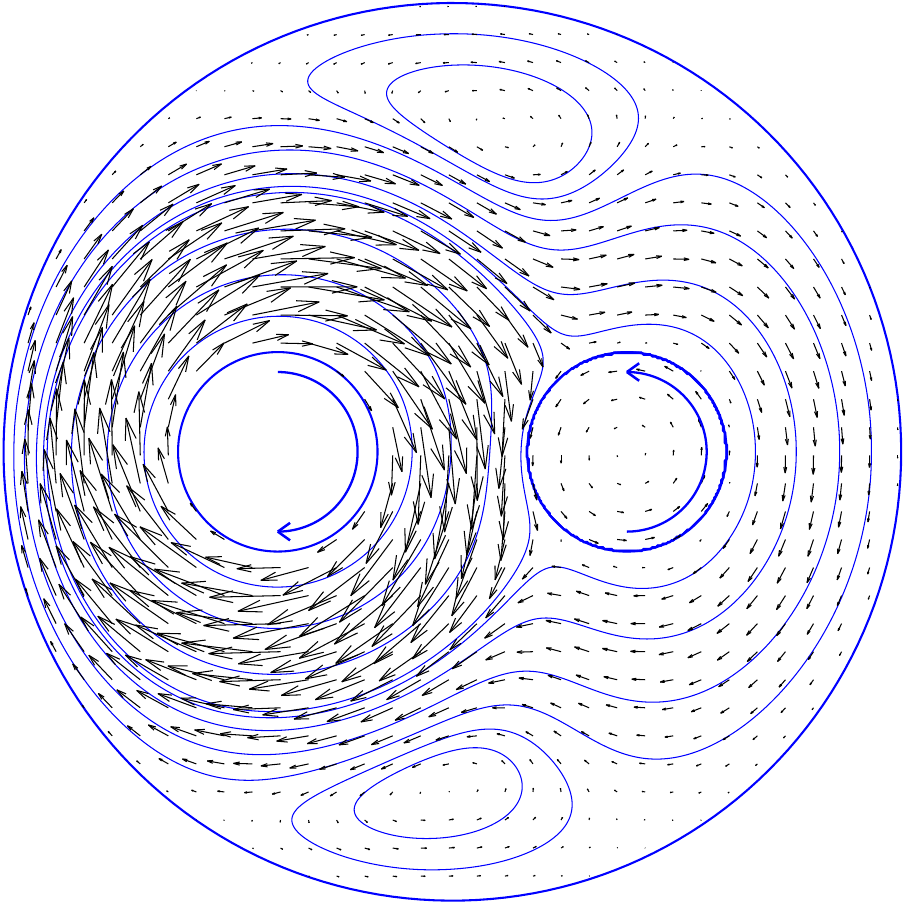}{30}\hfill
  \vstreamre{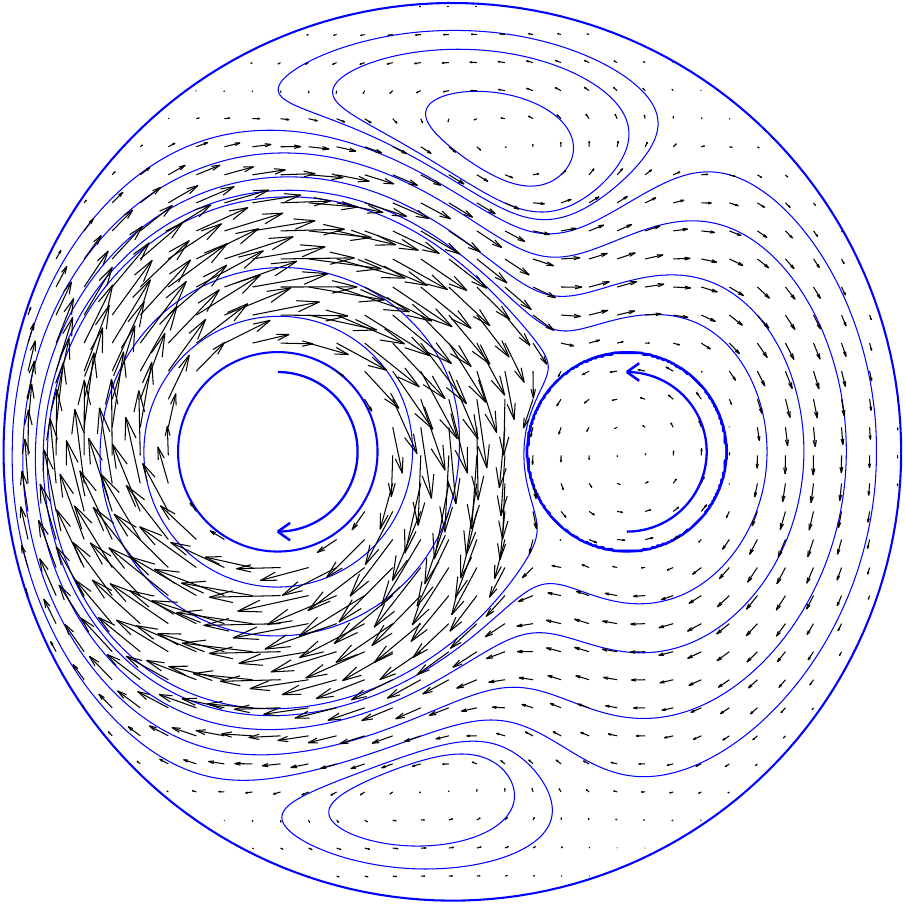}{40}\hfill
  \vstreamre{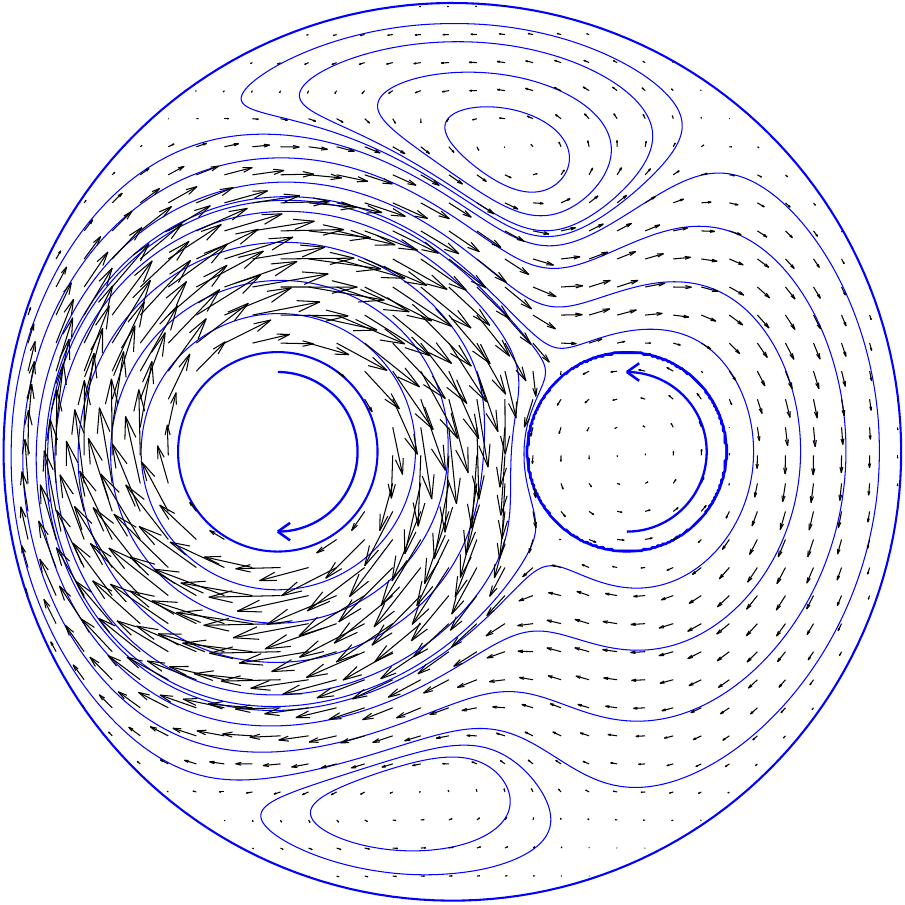}{50}

  \vspace{5pt}
  \vstreamre{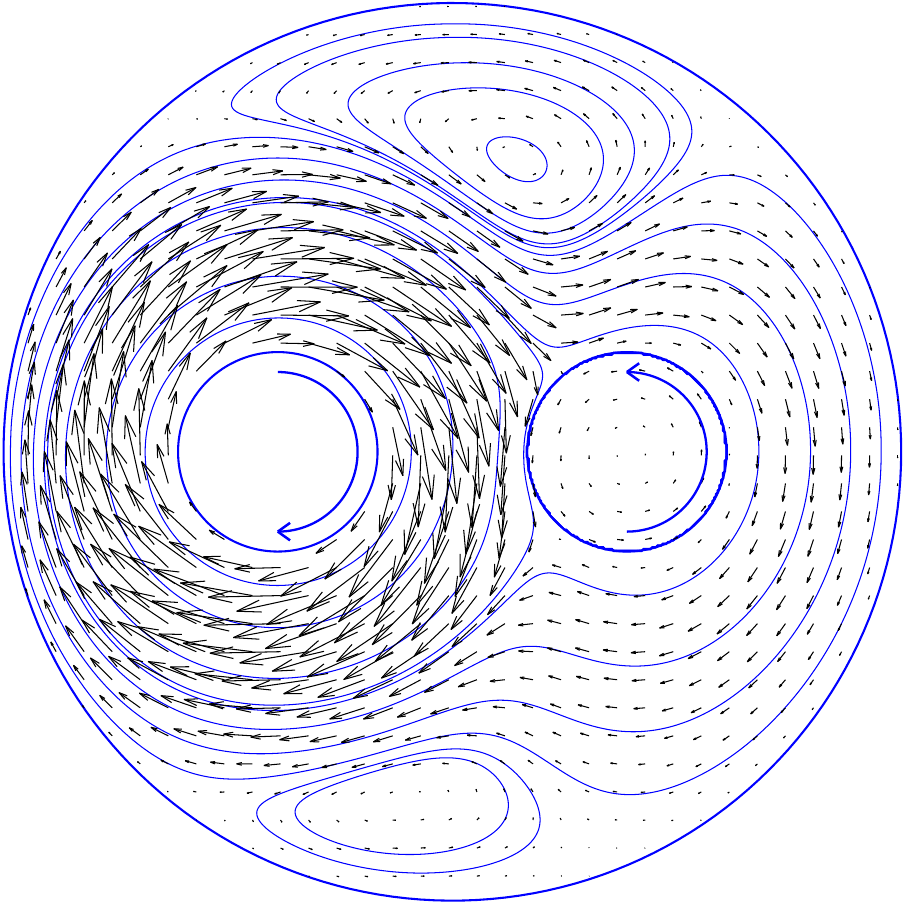}{60}\hfill
  \vstreamre{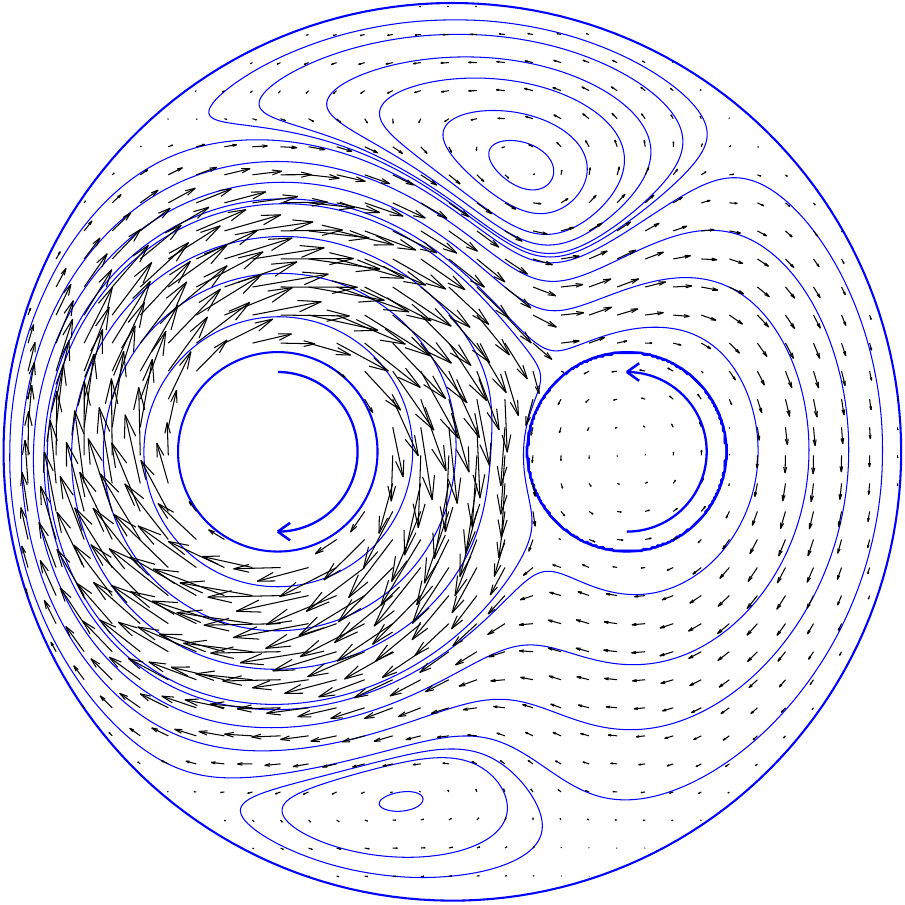}{70}\hfill
  \vstreamre{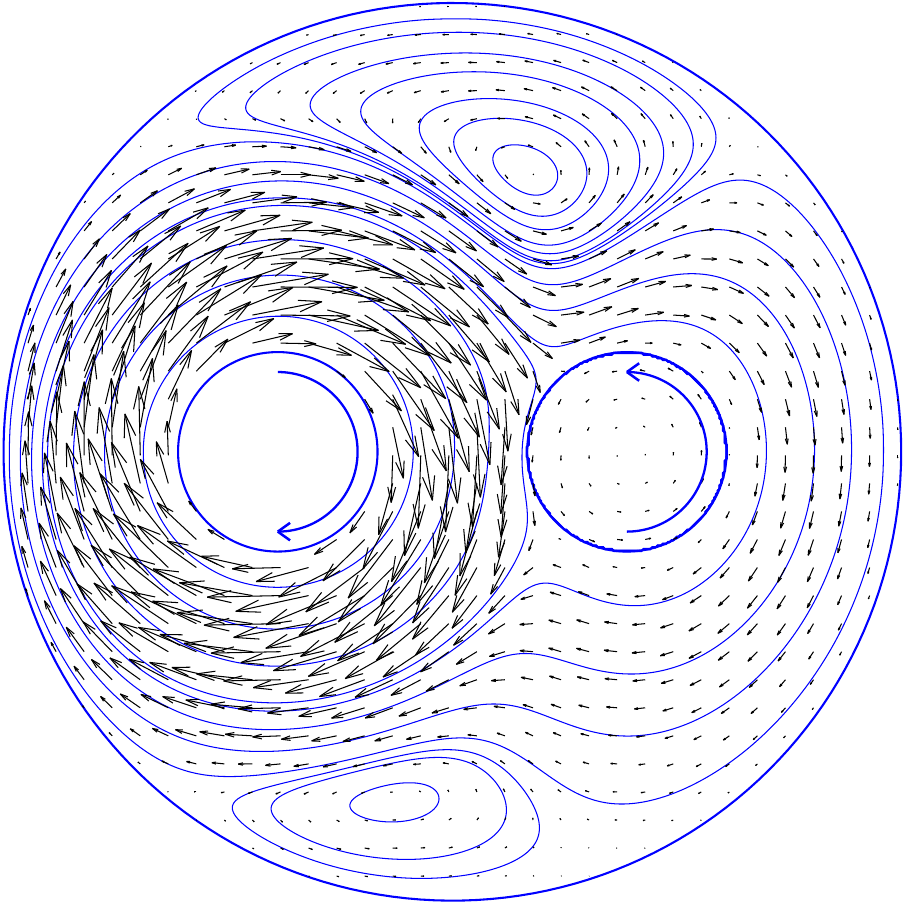}{80}\hfill
  \vstreamre{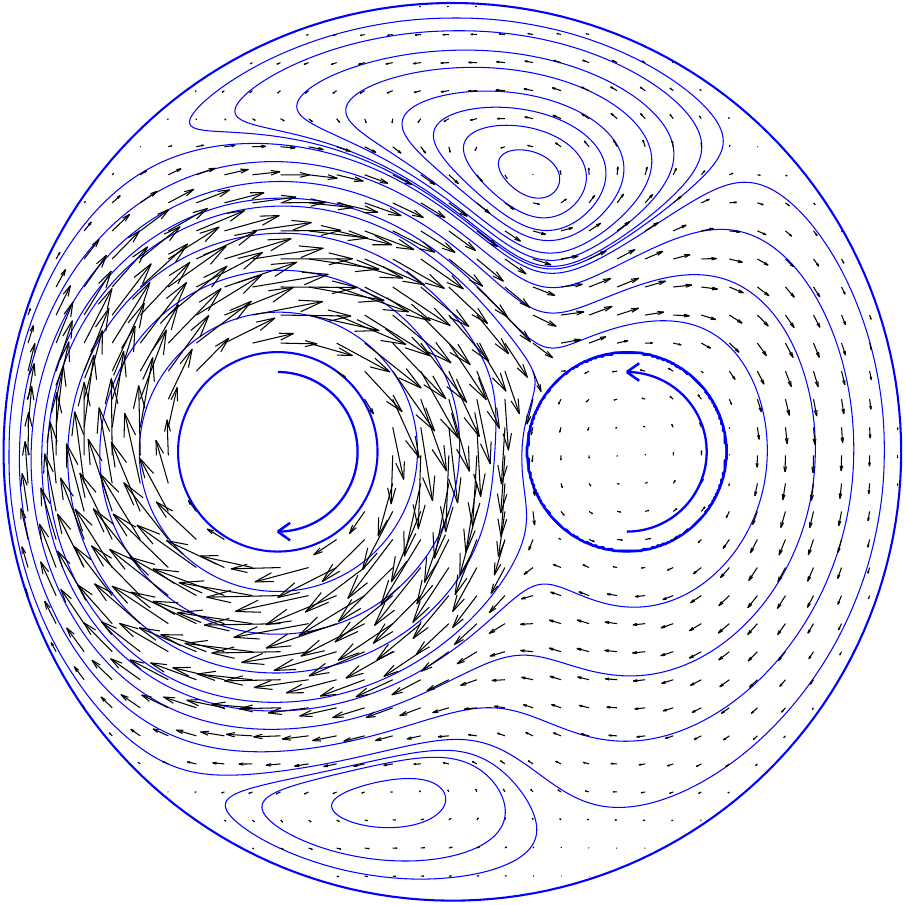}{90}\hfill
  \vstreamre{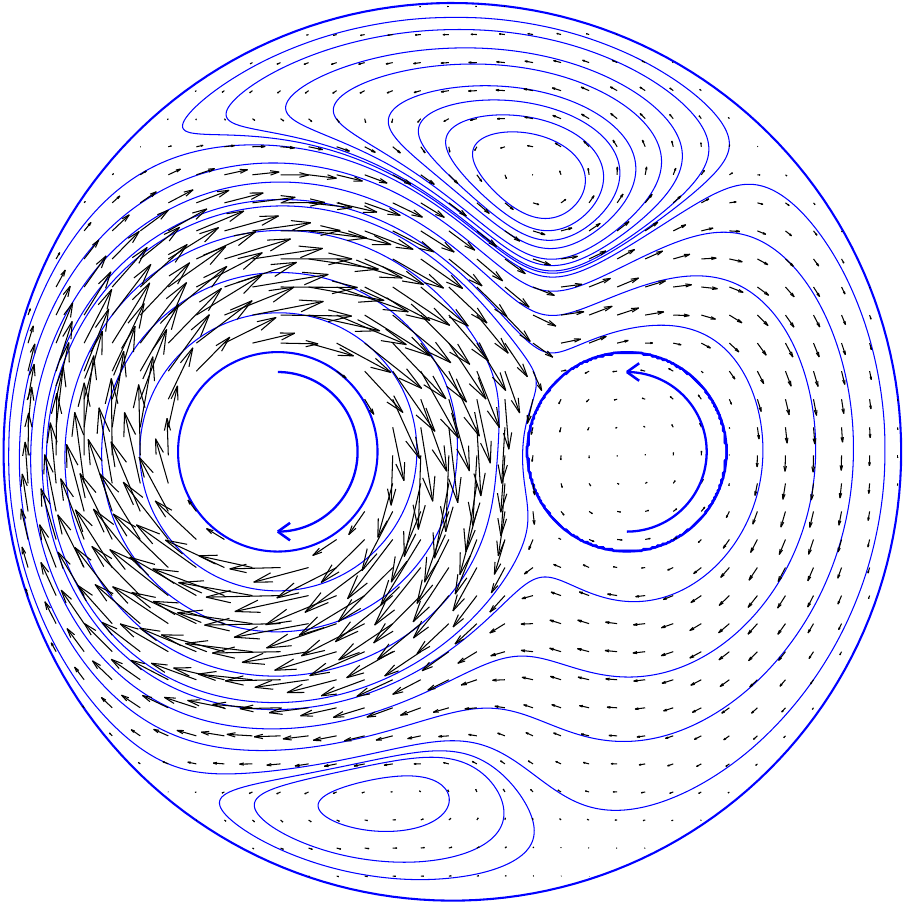}{100}
  \caption{Computed velocity fields with streamlines at $G=0.30$, $C=4.5$,
  for $\Rey=10$ to $100$. The two gap vortices persist and strengthen as
  $\Rey$ rises. No vortex is born, absorbed, or merged, and the passive rotor
  remains counterrotating throughout; the experimental counter-to-corotation
  transition near $\Rey=55$ has no planar counterpart at this gap.}
  \label{fig:vecRe}
\end{figure*}

\begin{figure*}[!tp]
  \centering
  \includegraphics[width=\linewidth]{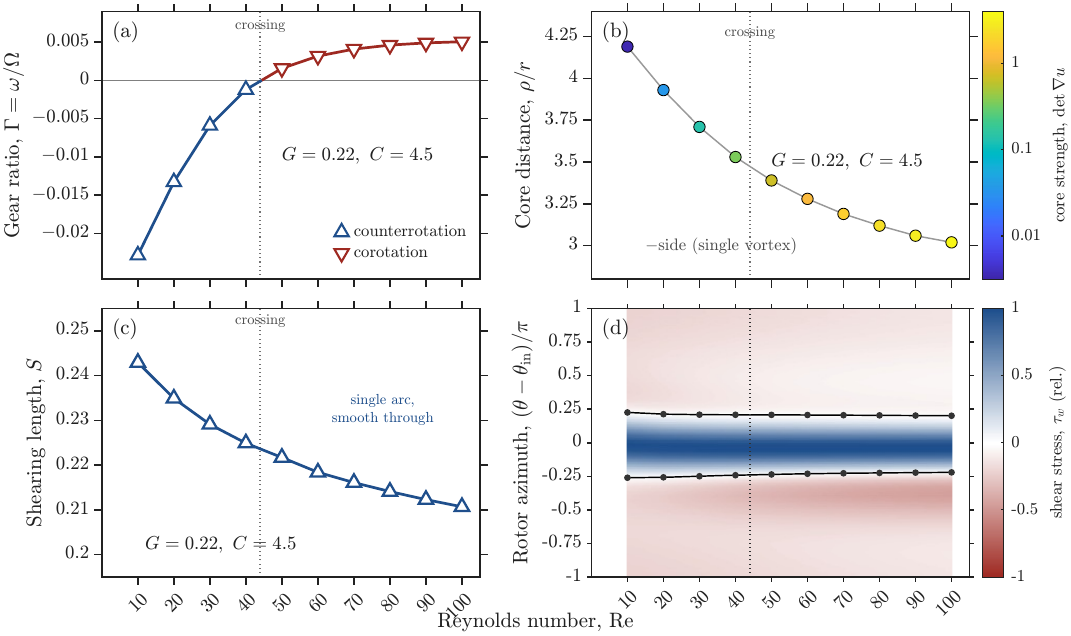}
  \caption{Reynolds-route diagnostics at the narrower gap $G=0.22$, $C=4.5$,
  where the planar transect crosses to corotation.
  All panels share the same Reynolds axis, and the dotted line marks the
  gear-ratio crossing near $\Rey=44$. (a)~Terminal gear ratio
  $\Gamma=\omega/\Omega$, which changes sign near $\Rey=44$: the passive rotor
  counterrotates below ($\Gamma<0$) and corotates above. At the experimental gap
  $G=0.30$, the corresponding planar transect approaches zero without crossing
  [Fig.~\ref{fig:reroute}(a)]. (b)~Distance of the single $-$side gap-vortex
  core from the passive-rotor center; the $+$side vortex does not form at this
  gap. The marker color is the core strength $\det\nabla u$ on a logarithmic
  scale. The core moves inward and strengthens by about three orders of
  magnitude, with no core born or annihilated across the crossing. (c)~Wall-based
  shearing length $S(\Rey)$, a single arc that drifts smoothly through
  the crossing without collapsing. (d)~Normalized wall shear stress
  $\tau_w(\theta,\Rey)$ on the passive rotor; blue shear promotes
  counterrotation and red shear promotes corotation. The gap-facing
  counterrotating arc persists through the crossing and does not change sign
  even where the terminal spin becomes corotating.}
  \label{fig:reroute22}
\end{figure*}

A terminal sign reversal does occur in the planar model, but at a narrower gap.
At $G=0.22$, the gear ratio increases monotonically with Reynolds number and
changes sign near $\Rey=44$ [Fig.~\ref{fig:reroute22}(a)]. The histories confirm
that this is a reversal of the terminal spin itself, not an artifact of
normalization: the passive rotor counterrotates below the crossing and
corotates above it [Fig.~\ref{fig:history}(c)]. We therefore call this a
fixed-gap Reynolds-driven crossing. We do not identify it with the experimental
inertial route, because it lies on the small-gap side of the planar phase map
and because the wall diagnostics show a different surface mechanism.

\begin{figure*}[!tp]
  \centering
  \newcommand{\vstreamreb}[2]{\begin{minipage}[t]{0.19\linewidth}\centering
    {\footnotesize $\Rey=#2$}\\[1pt]\includegraphics[width=\linewidth]{#1}\end{minipage}}
  \vstreamreb{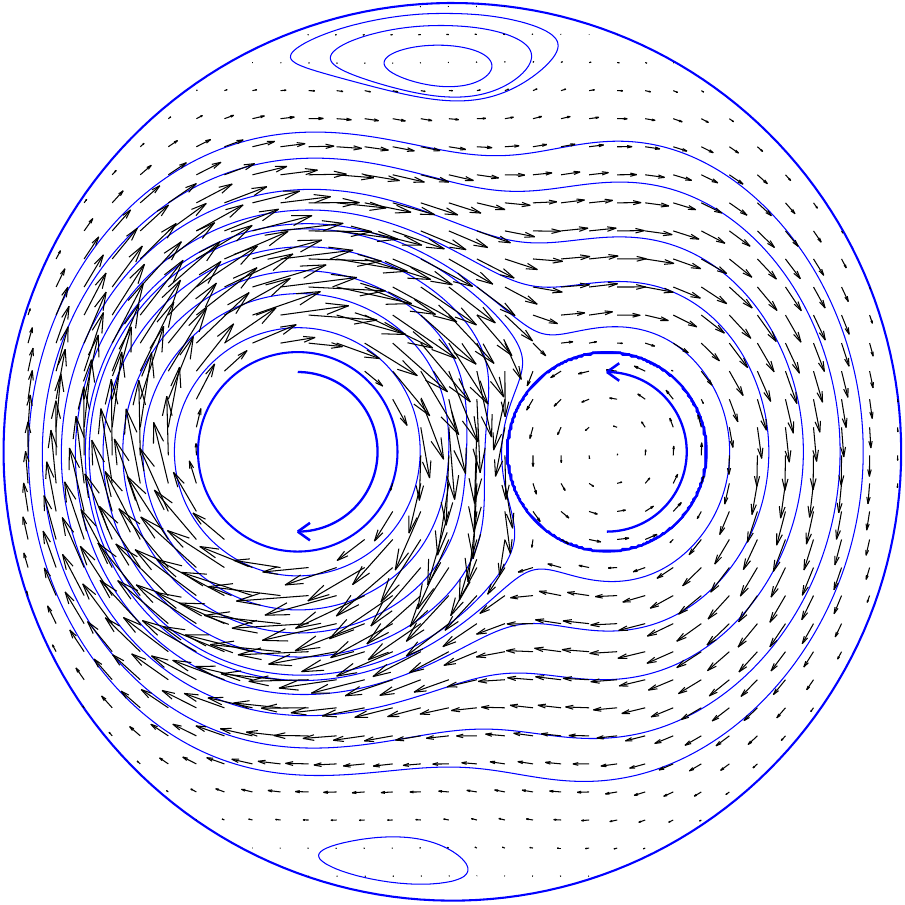}{10}\hfill
  \vstreamreb{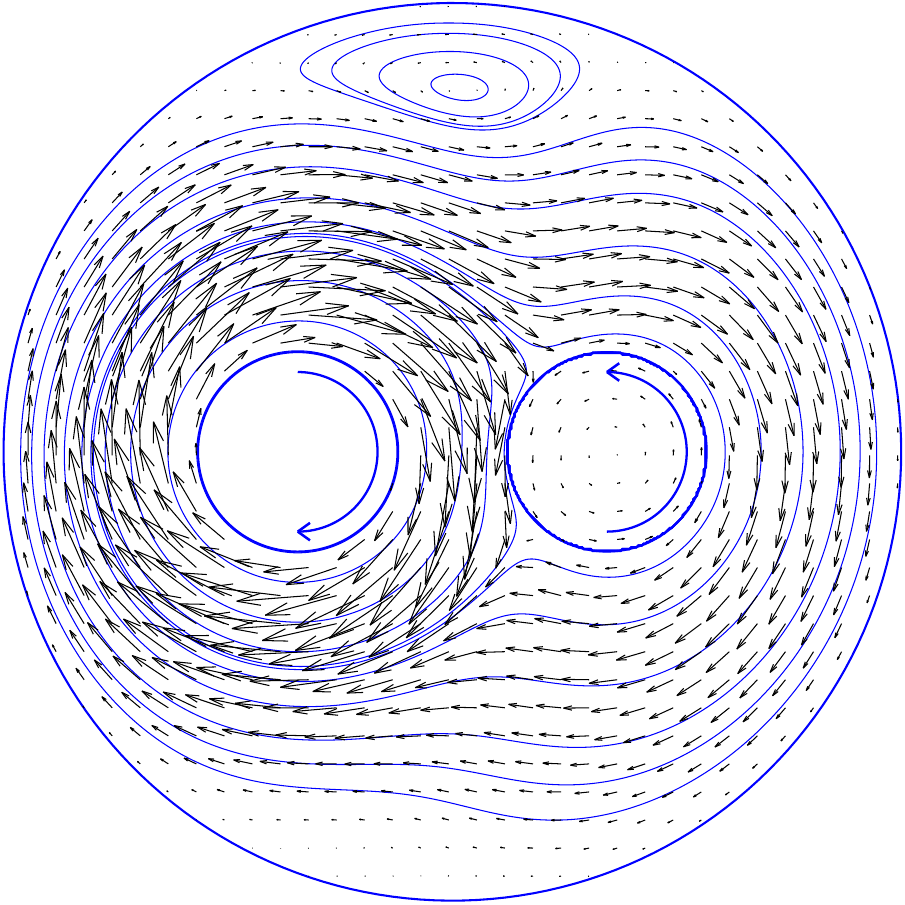}{20}\hfill
  \vstreamreb{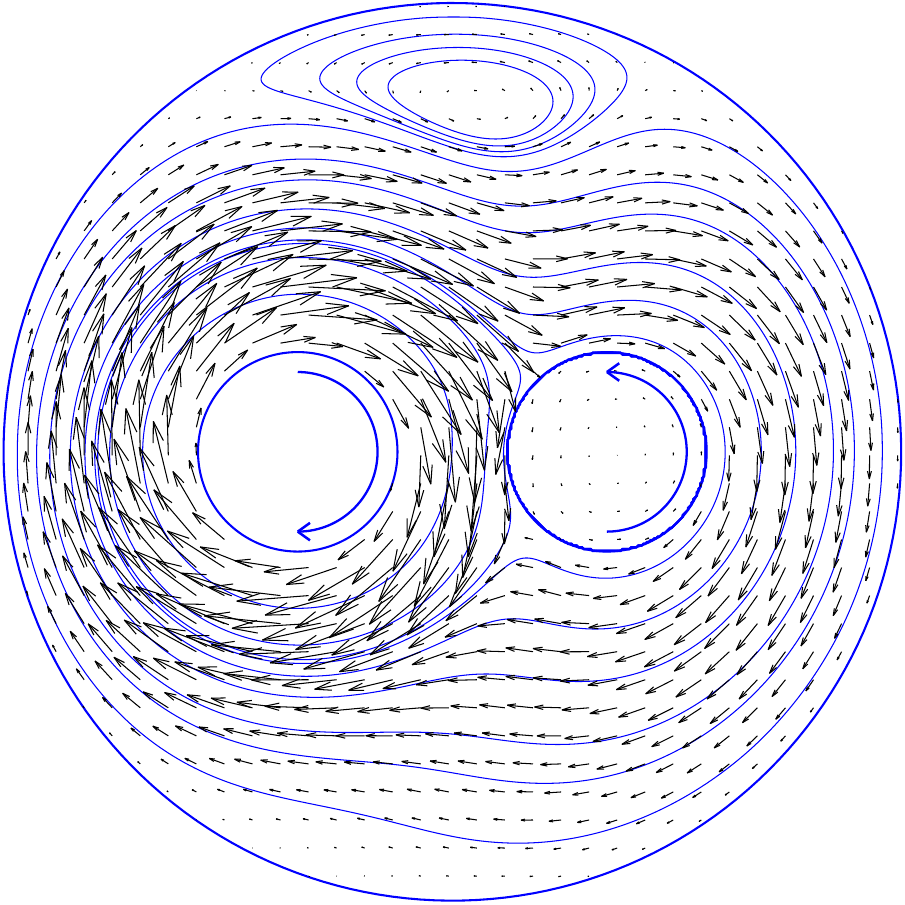}{30}\hfill
  \vstreamreb{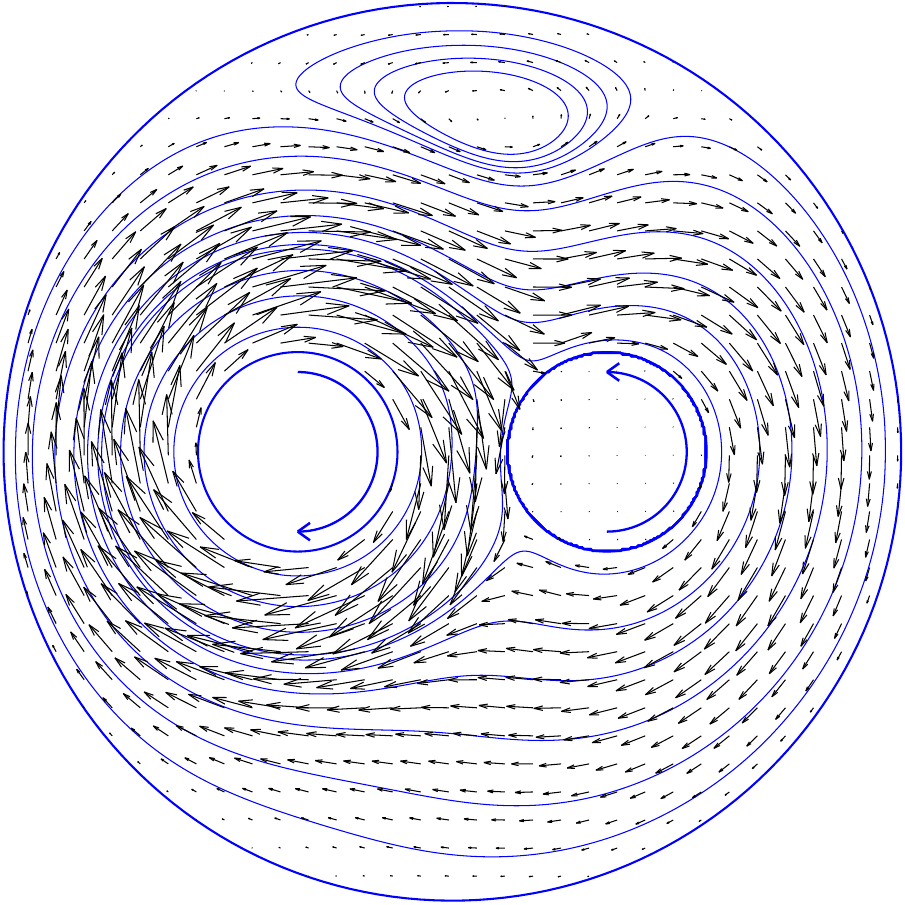}{40}\hfill
  \vstreamreb{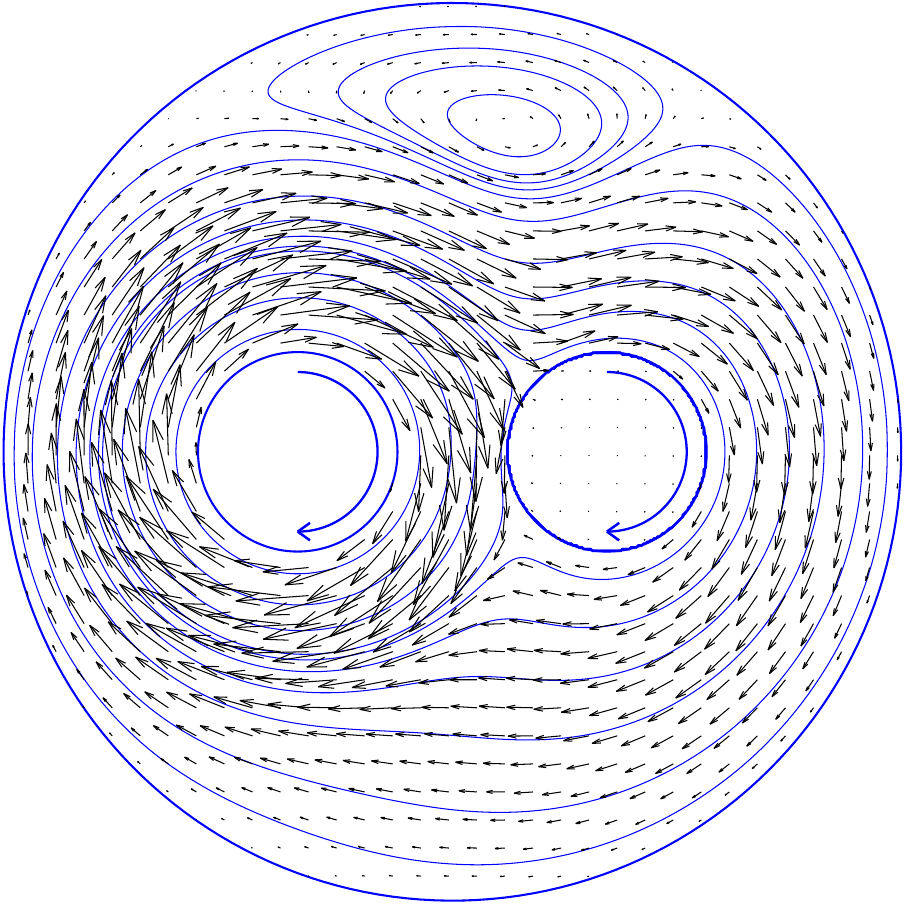}{50}

  \vspace{5pt}
  \vstreamreb{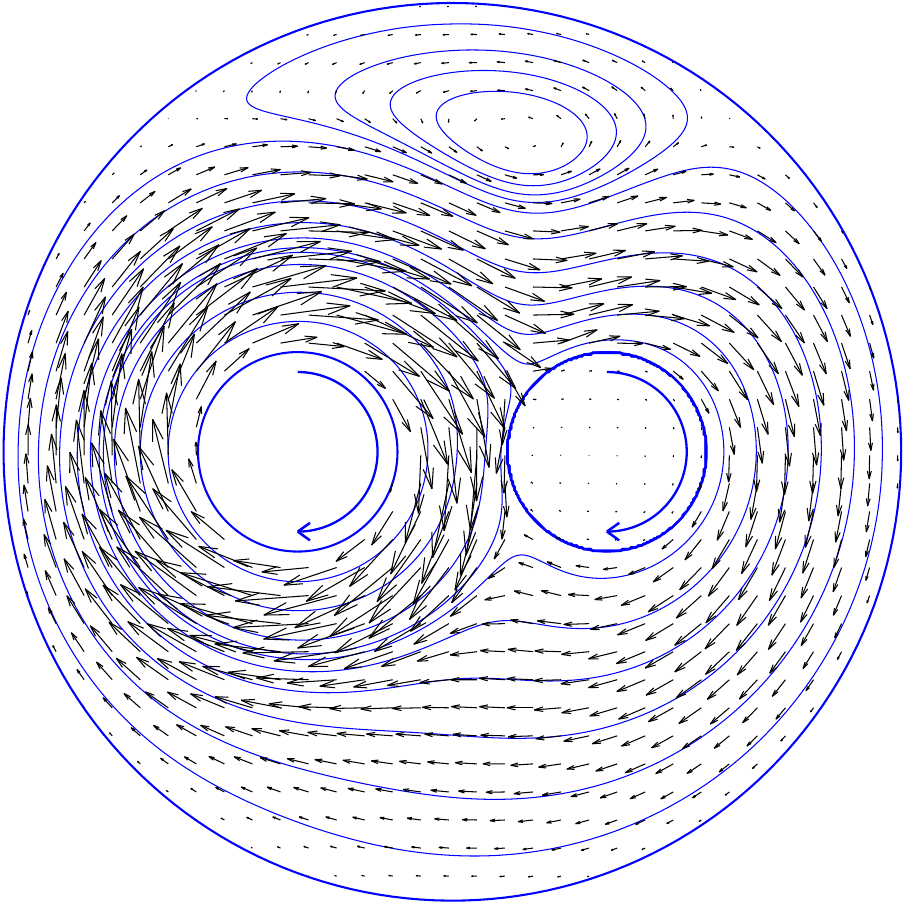}{60}\hfill
  \vstreamreb{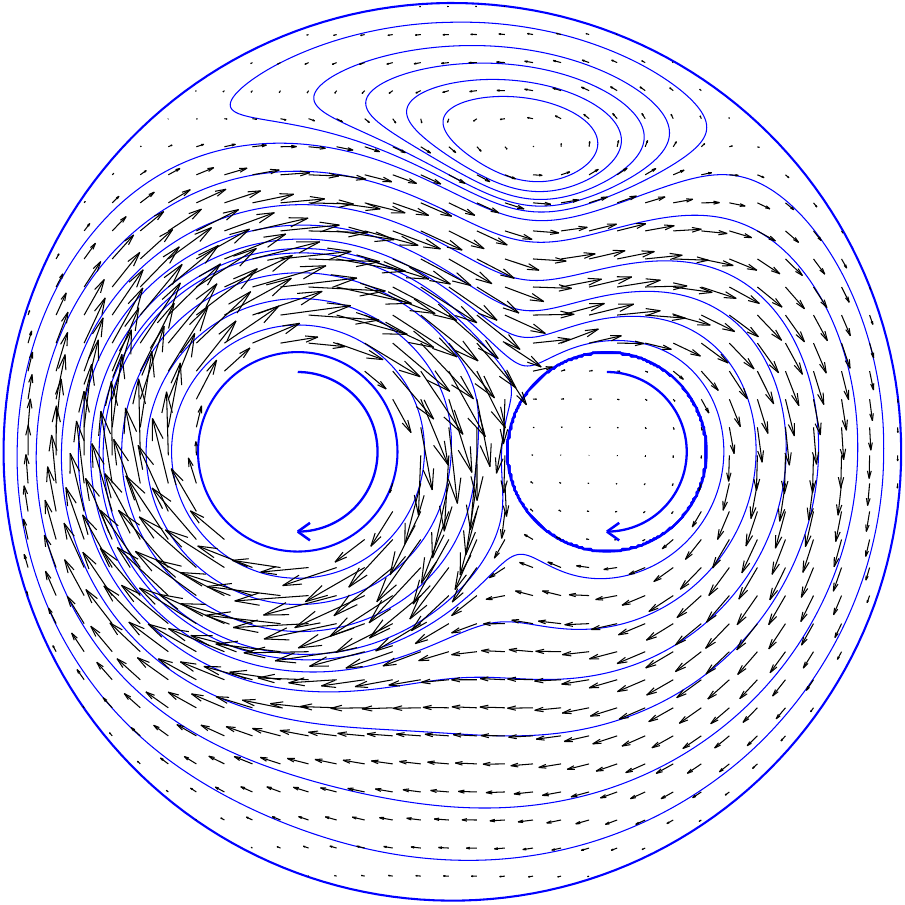}{70}\hfill
  \vstreamreb{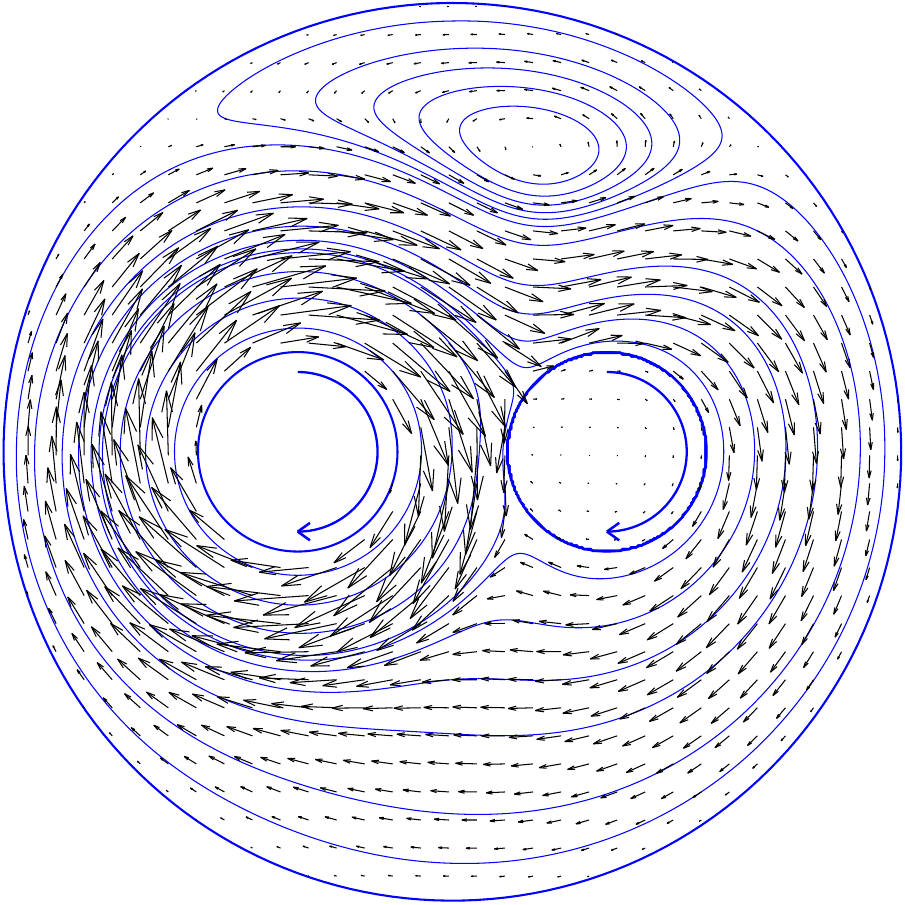}{80}\hfill
  \vstreamreb{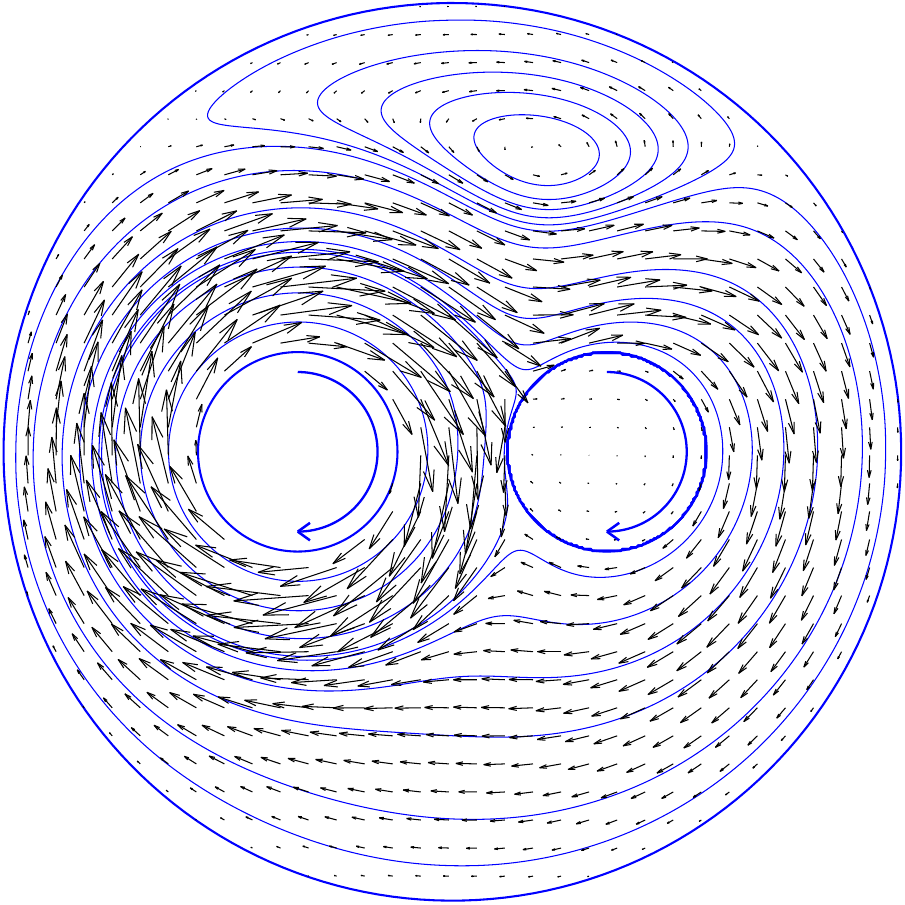}{90}\hfill
  \vstreamreb{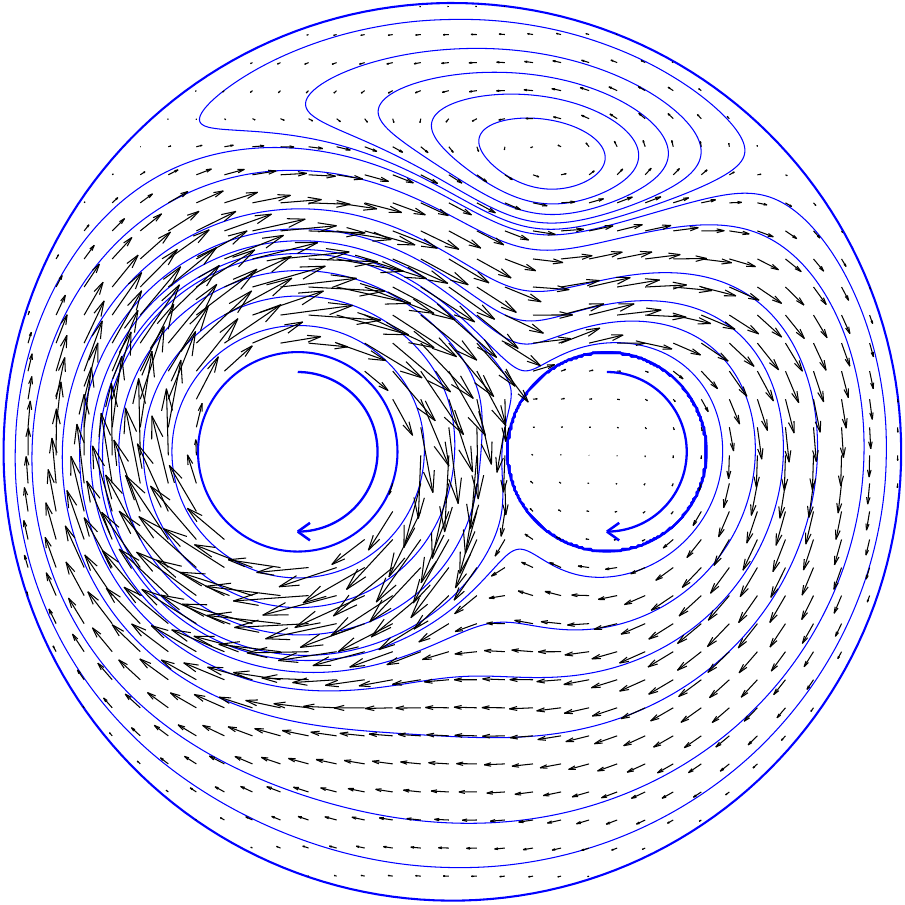}{100}
  \caption{Velocity fields with streamlines at $G=0.22$, $C=4.5$, for
  $\Rey=10$ to $100$. A single $-$side gap vortex organizes the flow at every
  Reynolds number and tightens as $\Rey$ rises; the $+$side vortex present at
  $G=0.30$ does not form. The fields pass smoothly through the gear-ratio
  crossing near $\Rey=44$, showing that the spin changes sign without a bulk
  topological rearrangement.}
  \label{fig:vecRe22}
\end{figure*}

Figure~\ref{fig:reroute22} explains why the $G=0.22$ crossing is not the
experimental inertial route translated to a smaller gap. Across the spin
crossing, the flow remains organized by a single $-$side gap vortex: the
$+$side vortex never appears because this gap lies below the second-vortex
threshold in the horizontal sweep. The lone vortex moves inward and strengthens
monotonically [Fig.~\ref{fig:reroute22}(b)], but it is not born, absorbed, or
merged at the crossing. The streamline fields show the same continuity
directly (Fig.~\ref{fig:vecRe22}). The wall diagnostics are equally smooth. A
single gap-facing counterrotation arc persists at every Reynolds number, its
length drifting only from about $0.24$ to $0.21$, and its sign does not reverse
on the gap axis [Figs.~\ref{fig:reroute22}(c,d)]. Thus the terminal spin
changes sign while the dominant gap-facing traction remains counterrotating;
the net torque is tipped instead by weaker flanking stresses. The crossing is
therefore a change in integrated torque, not a topological rearrangement of the
gap vortex or a collapse of the inner shear zone.

The contrast with the experiment is therefore a contrast in boundary placement
and mechanism, not simply in the presence or absence of a Reynolds-driven
crossing. In the experimental map, the counter-to-corotation boundary at
$C=4.5$ lies nearly flat near $\Rey\approx55$ over a broad interval of gaps,
including both $G\approx0.22$ and $G\approx0.30$ [Fig.~2(b) of
Ref.~\cite{SmithRistrophZhang2026}]. In the planar map, the boundary is much
steeper: it crosses $G=0.22$ near $\Rey=44$ but has not reached $G=0.30$ by
$\Rey=400$. The two-dimensional boundary is therefore displaced toward smaller
gaps and steepened relative to the quasi-two-dimensional experiment.

The shear-zone diagnostic rules out identifying the planar crossing with the
experimental inertial route at the level of the wall traction. In the
experiment, reversal is accompanied by collapse and deflection of the
gap-facing inner shear zone. In the plane, $S$ drifts only weakly with $\Rey$:
from about $0.24$ to $0.21$ at $G=0.22$ and from about $0.24$ to $0.20$ at
$G=0.30$ [Figs.~\ref{fig:reroute22}(c) and~\ref{fig:reroute}(c)]. At
$G=0.22$, the gap-facing counterrotation arc persists through the spin crossing
without collapsing, rotating away from the gap axis, or changing sign. The
crossing therefore reproduces a reversal in the terminal spin, but not the
experimental shear-zone mechanism. It should be read as a Reynolds-driven
boundary of the planar model, not as a direct surface-diagnostic match to the
experimental inertial route.

Two caveats delimit this interpretation. First, a $G=0.30$ crossing could in
principle occur beyond the computed range; in the present calculations,
however, $\Gamma$ flattens toward zero from below rather than steepening toward
a crossing as $\Rey$ rises to $400$. Second, an unresolved unsteady terminal
state could obscure a crossing; the monitored passive-spin histories remain
steady throughout the computed range, and the cylinder-shedding control
discussed in Sec.~\ref{sec:compare} supports reading this steadiness as
genuine. The remaining discrepancy is therefore most
naturally interpreted as a shift of the high-Reynolds-number torque balance,
with finite-depth axial/end-wall motion and apparatus geometry as plausible
contributors.

\subsection{Phase diagrams at $C=3$, $4.5$, and $6$}\label{sec:maps}

\begin{figure}[!tbp]
  \centering
  \newcommand{\phasetoppad}{2mm}
  \newcommand{\wcol}[3][0pt]{\begin{minipage}[t]{0.33\linewidth}\centering
    {\footnotesize #3}\\[0pt]
    \vspace*{#1}\makebox[\linewidth][r]{\includegraphics[width=0.96\linewidth]{#2}}%
  \end{minipage}}
  \wcol[\phasetoppad]{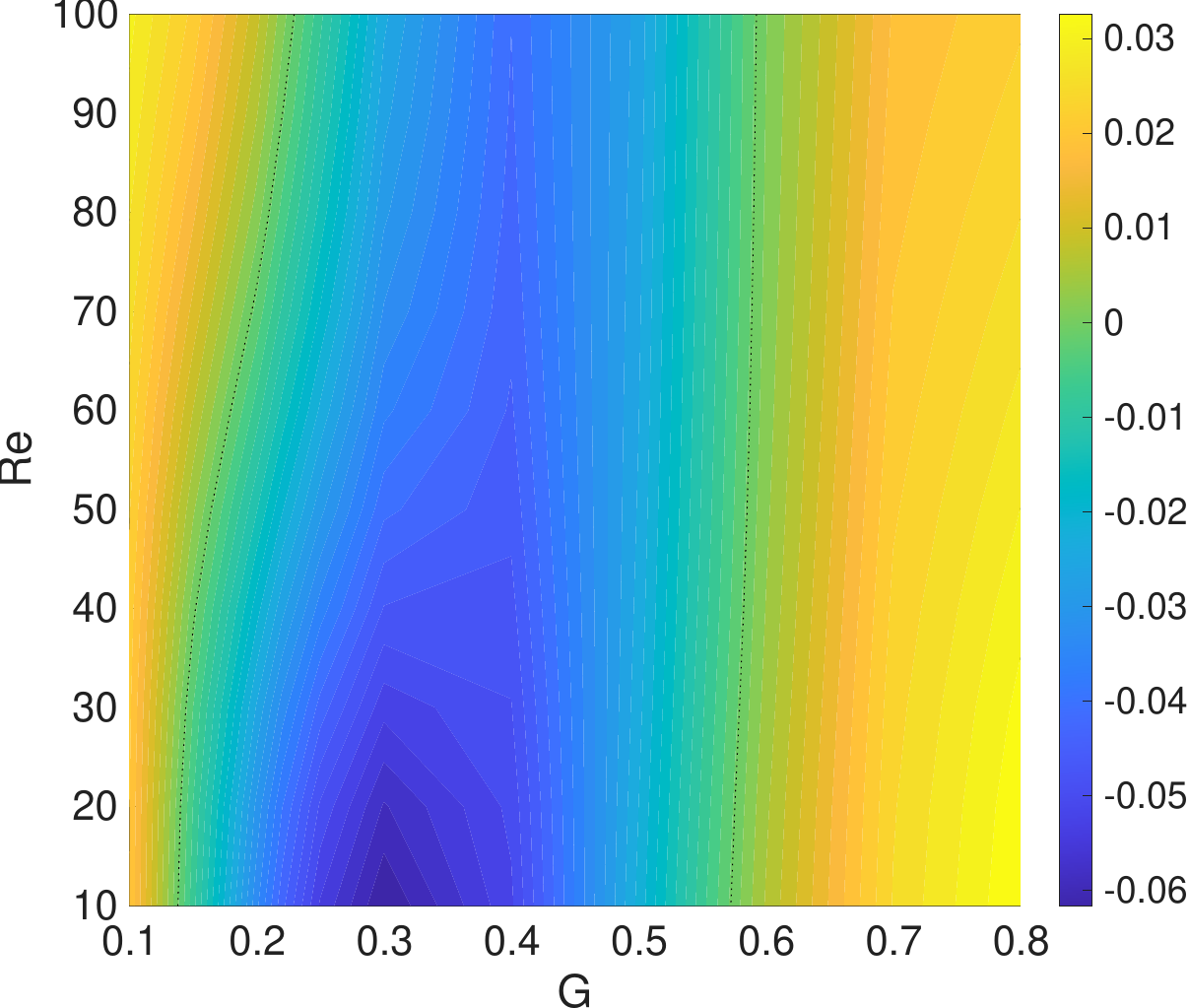}{(a)~$\Gamma(G,\Rey)$ at $C = 3$}\hfill
  \wcol[\phasetoppad]{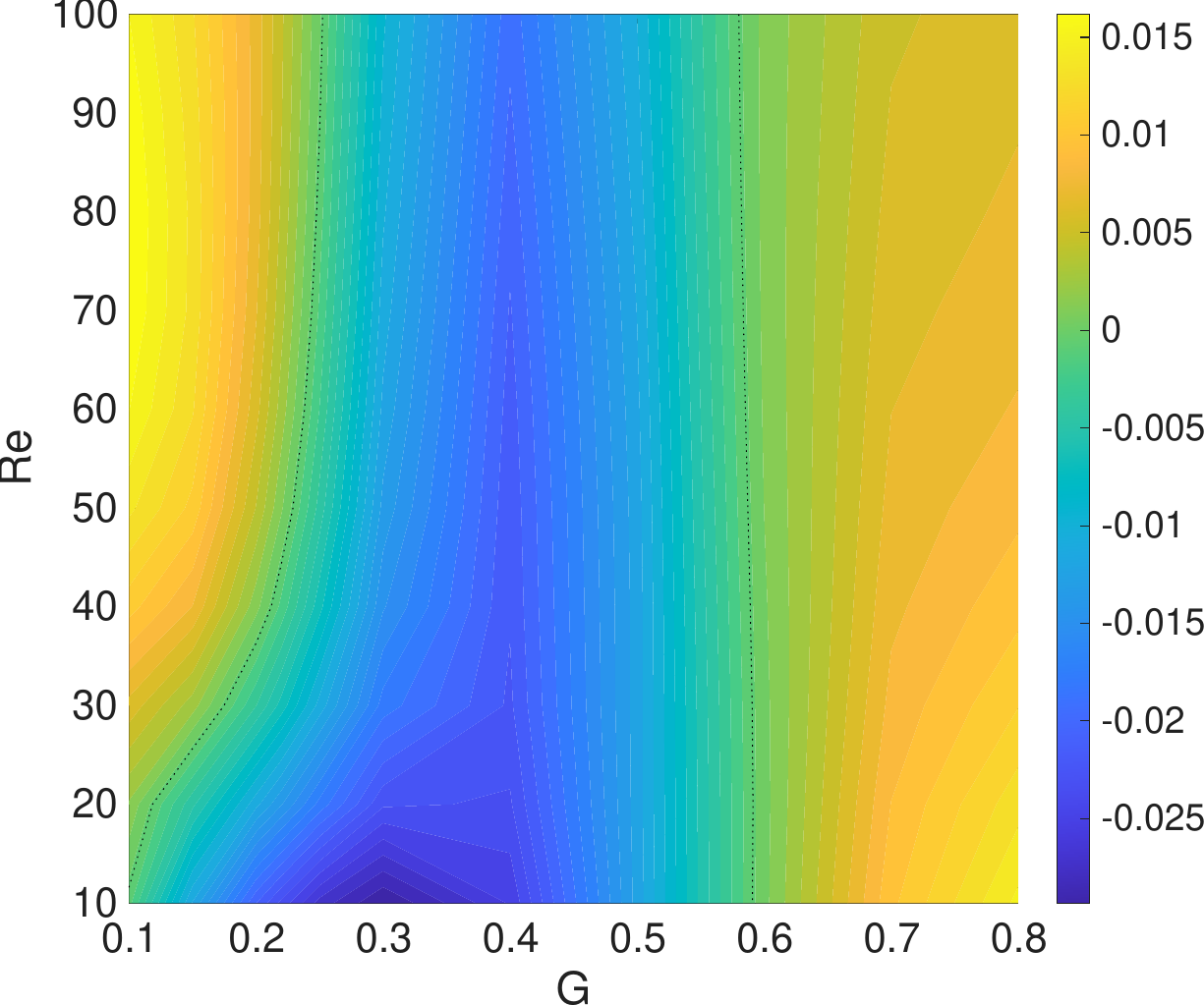}{(b)~$\Gamma(G,\Rey)$ at $C = 4.5$}\hfill
  \wcol{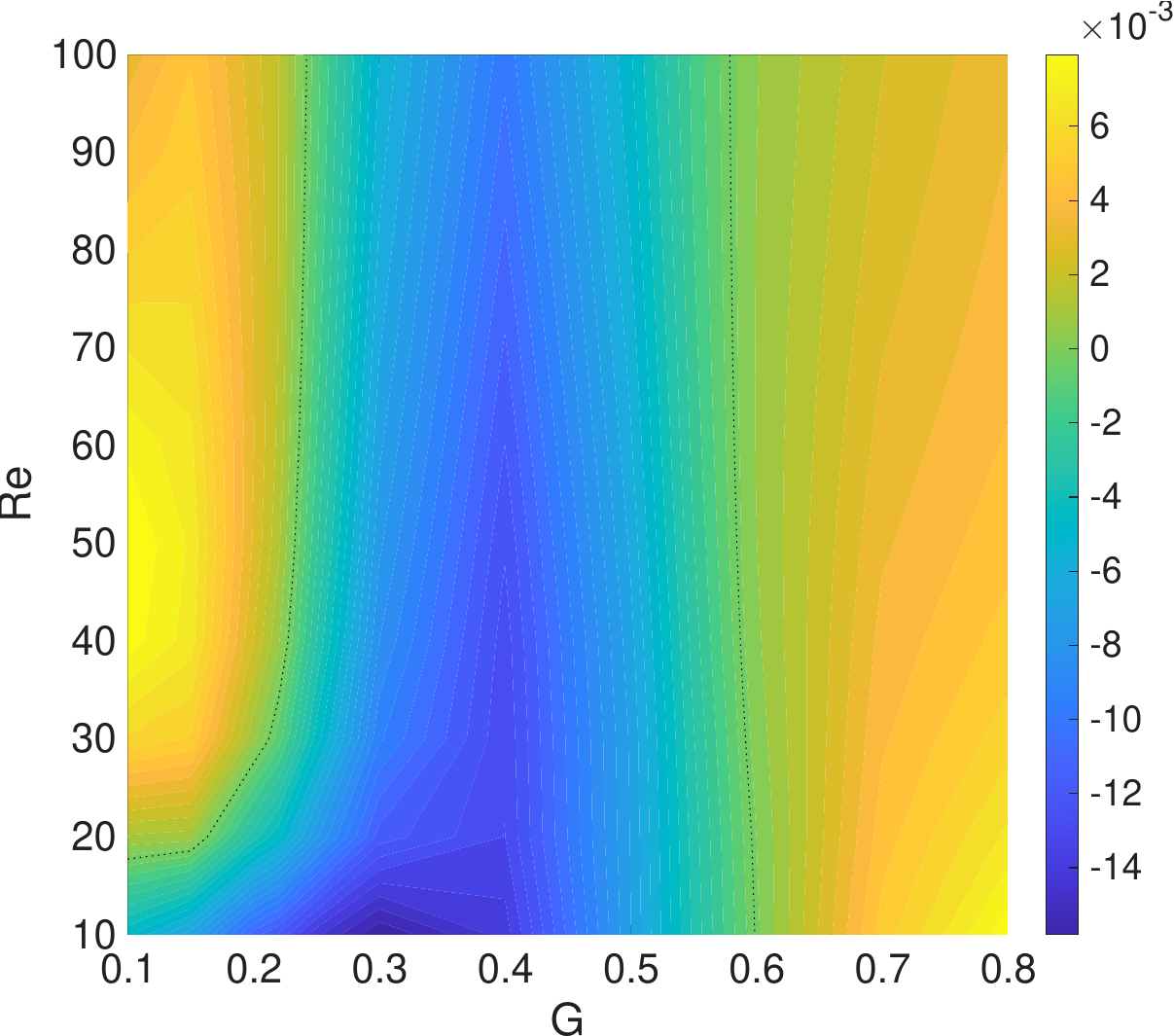}{(c)~$\Gamma(G,\Rey)$ at $C = 6$}
  \caption{Computed two-dimensional phase diagrams of the gear ratio
  $\Gamma$ in the $(G,\Rey)$ plane at (a)~$C=3$, (b)~$C=4.5$, and
  (c)~$C=6$. Blue and red denote counterrotation and corotation, respectively,
  on opposite sides of the dashed $\Gamma=0$ stagnation contour. The three
  panels share this sign convention, but each uses its own magnitude scale.
  Each map contains a reentrant-like gap structure: small-gap corotation, a
  connected intermediate counterrotation band, and wide-gap corotation. The
  left stagnation contour moves to larger gap as $\Rey$ rises, producing the
  fixed-gap Reynolds-driven crossing of Sec.~\ref{sec:inertial}. The coupling
  weakens as the corral grows, with extrema of order $10^{-2}$ throughout.}
  \label{fig:phase}
\end{figure}

We next assemble the fixed-gap and fixed-Reynolds-number cuts into phase maps
at three corral sizes, $C=3$, $4.5$, and $6$ (Fig.~\ref{fig:phase}). Each map
samples $G=0.1$ to $0.8$ and $\Rey=10$ to $100$, with about ninety terminal
states per corral. All three maps share the same reentrant-like gap structure:
small-gap corotation, an intermediate counterrotation band, and wide-gap
corotation. The counterrotation band is strongest near $G\approx0.3$ to $0.4$
at low Reynolds number, while the small-gap corotation strengthens as $\Rey$
increases.

The right stagnation contour, where the intermediate band gives way to
wide-gap corotation, lies between $G\approx0.57$ and $0.60$ at $\Rey=20$
for the three corrals, shifting only weakly as $C$ changes. In physical units, however, the
corresponding separation varies substantially because
$g/r=2(C-2)G$. Across the three corrals, the physical gap at this boundary
runs from about $1.1$ to $4.8$ rotor radii. Thus the transition correlates
more closely with the normalized gap $G$ than with the dimensional separation
$g/r$ over the range computed. A normalized gap of $G=0.6$ lies just inside
corotation at $C=4.5$ but returns to counterrotation at larger corral size,
consistent with the wide-corral experimental photograph at $\Rey=20$,
$G=0.6$ [Fig.~1(c) of Ref.~\cite{SmithRistrophZhang2026}].

The vortex rearrangements themselves do not follow the spin boundary in the
same way. Comparing $C=3$ and $C=4.5$ at $\Rey=20$, the attachment and
detachment of the passive-side vortex shift to smaller normalized gaps as the
corral tightens, while the merger and the wide-gap spin transition remain
near their $C=4.5$ values. Thus the rearrangement gaps are sensitive to
confinement and rotor eccentricity, whereas the wide-gap spin transition is
better organized by the normalized gap. The experiment already separates the
flow rearrangements from the spin transition at $C=4.5$; the planar sweep
shows that this separation persists as the corral size changes.

The small-gap corotation leg is also present in the maps, but its
identification with the experiment's close-range geometric state remains
unsettled. In the experiment that state is a narrow, nearly
Reynolds-independent strip near $G=0.02$, below the lower end of the present
gap sweep. In the planar maps, the same left $\Gamma=0$ contour both bounds
the small-gap corotation leg and produces the $G=0.22$ Reynolds-driven
crossing discussed above. The coupling magnitude remains small throughout,
of order $10^{-2}$, and weakens as the corral grows:
$|\Gamma|\lesssim0.062$ at $C=3$, $\lesssim0.029$ at $C=4.5$, and
$\lesssim0.016$ at $C=6$, matching the experimental trend that tighter
corrals produce stronger coupling.

\subsection{Transient response of the passive rotor}\label{sec:transient}

The phase diagrams above are based on terminal zero-torque states, but the
start-up histories contain an additional diagnostic: they show which part of the
flow reaches the passive rotor first. This distinction is useful because the
experiment shows the same qualitative feature.
There, belt-driven terminal states can first rotate in the gearlike direction
before reversing to their long-time sense; the proposed explanation is a
diffusive delay, in which the change in active speed reaches the gap-facing side
of the passive rotor before it reaches the outer side~\cite{SmithRistrophZhang2026}.
Our planar histories should therefore be read as mechanism checks, not as a new
criterion for assigning the phase: throughout the paper the reported gear ratio
remains the terminal value selected by zero hydrodynamic torque.

\begin{figure*}[!tp]
  \centering
  \newcommand{\vstreamt}[4][=]{\begin{minipage}[t]{0.19\linewidth}\centering
    {\footnotesize $t=#3$, $\omega#1#4$}\\[1pt]\includegraphics[width=\linewidth]{#2}\end{minipage}}
  {\footnotesize (a)~Narrow gap, $G=0.22$}\\[2pt]
  \vstreamt{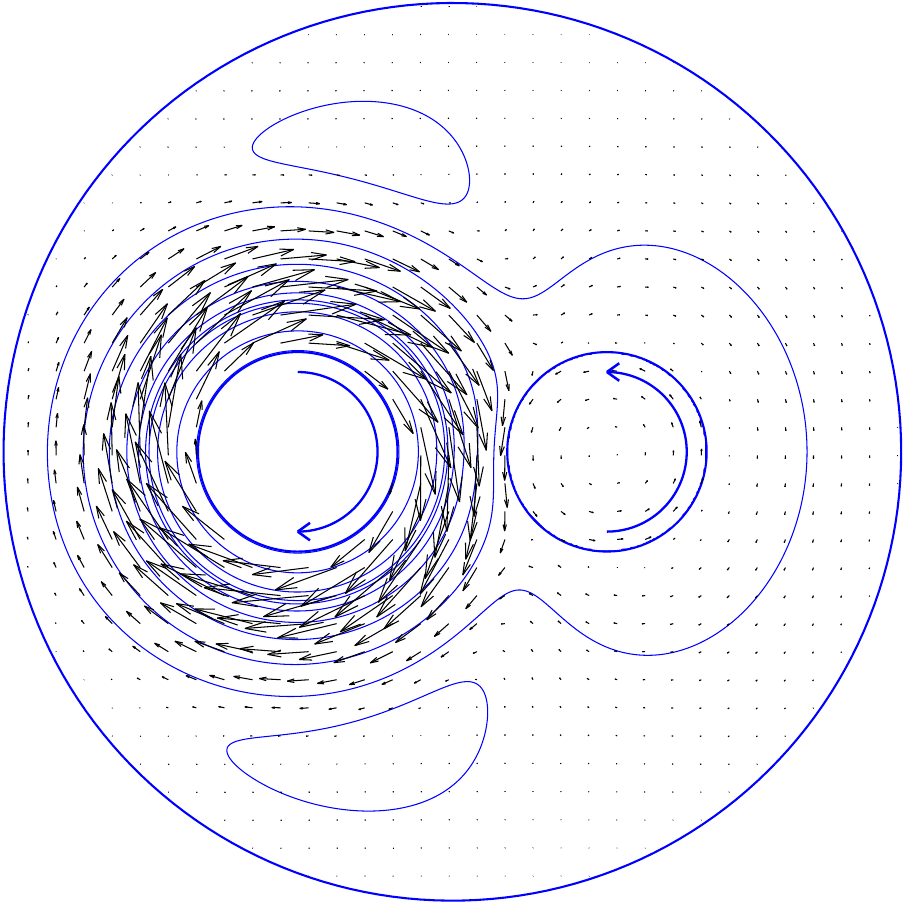}{0.33}{+0.50}\hfill
  \vstreamt{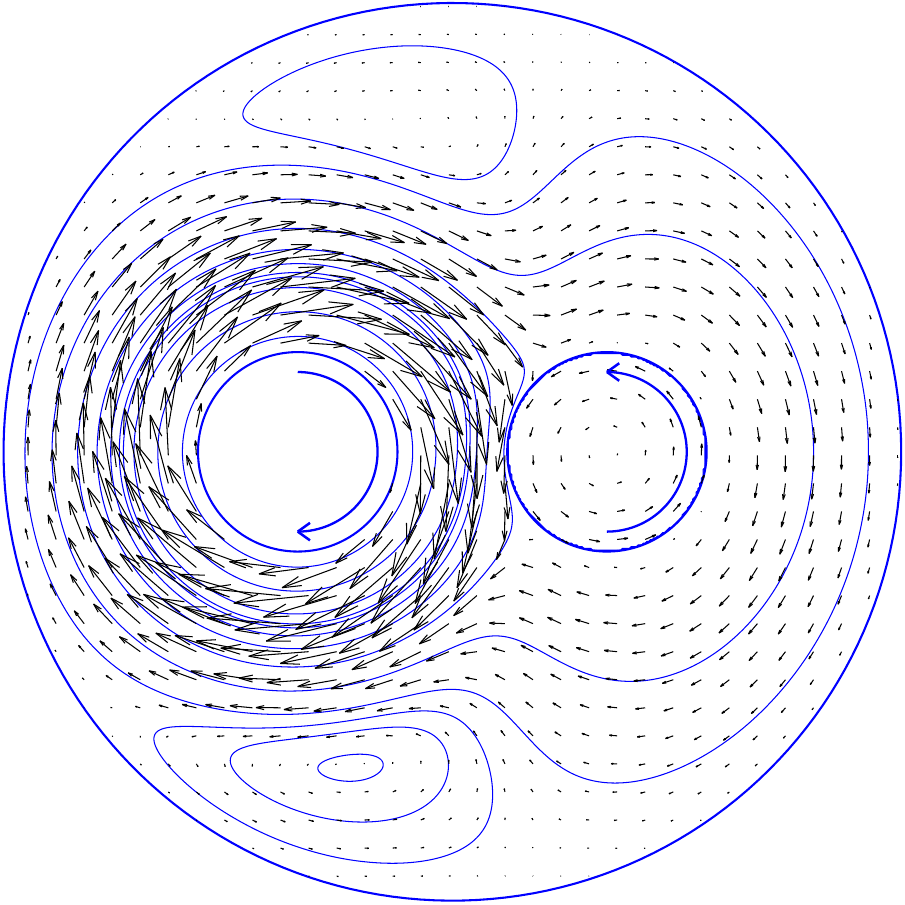}{0.69}{+0.98}\hfill
  \vstreamt{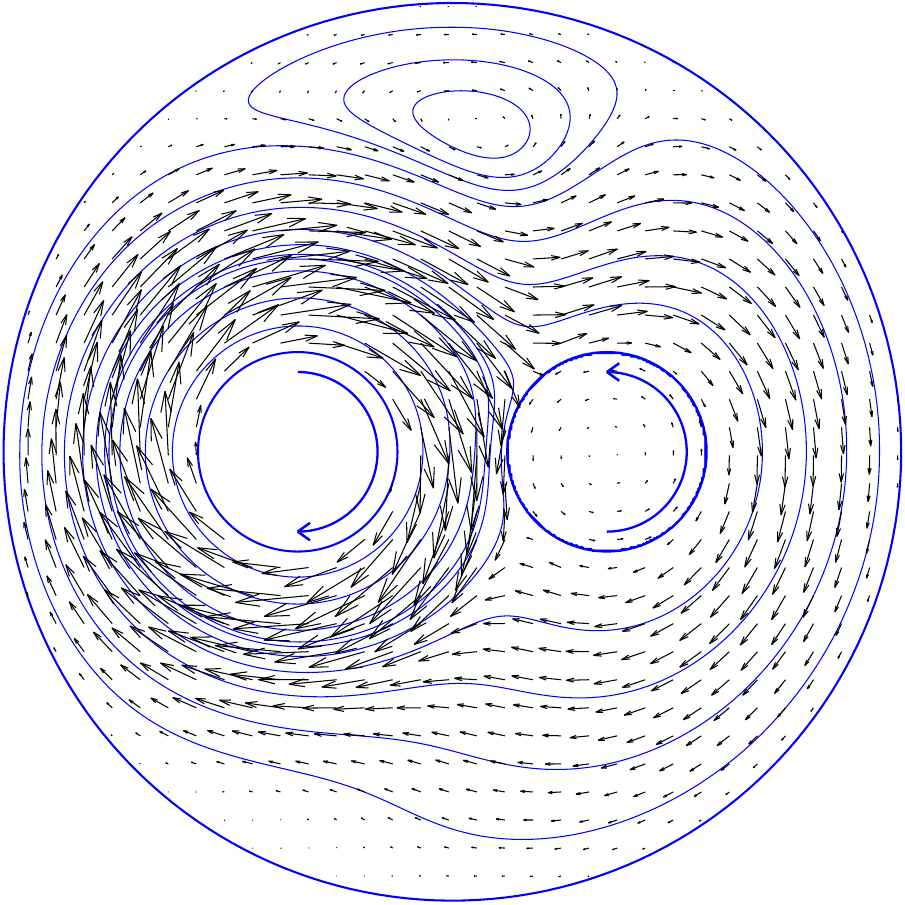}{1.31}{+0.50}\hfill
  \vstreamt[\approx]{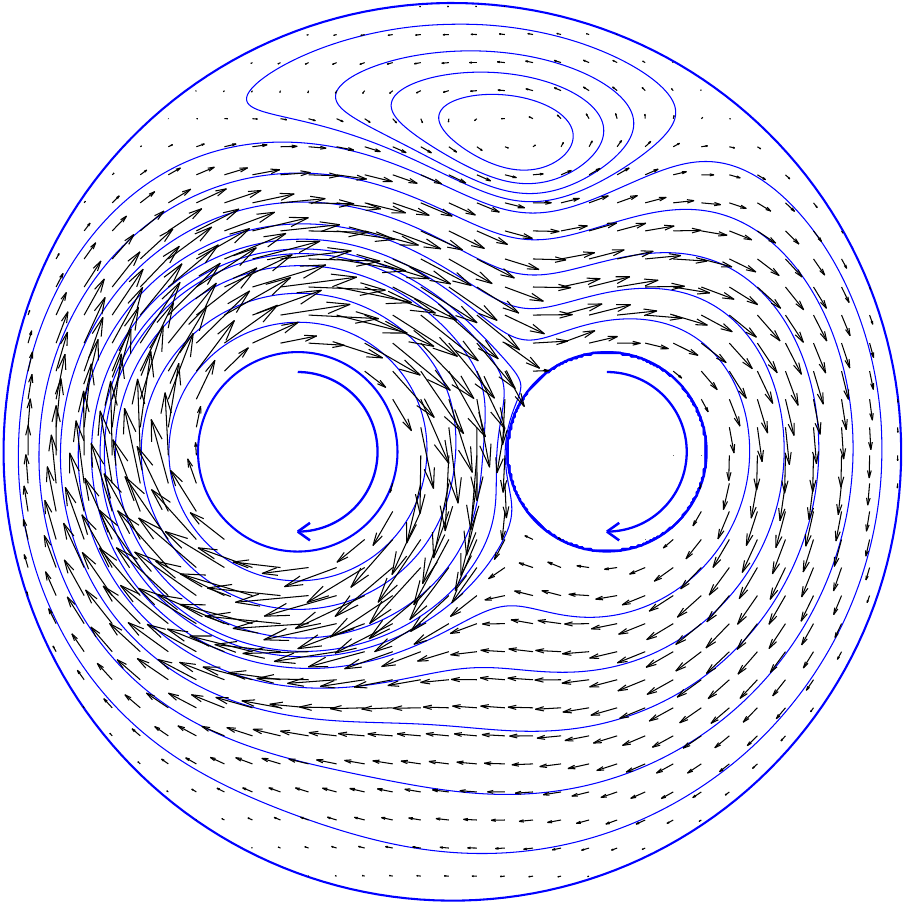}{2.19}{0.00}\hfill
  \vstreamt{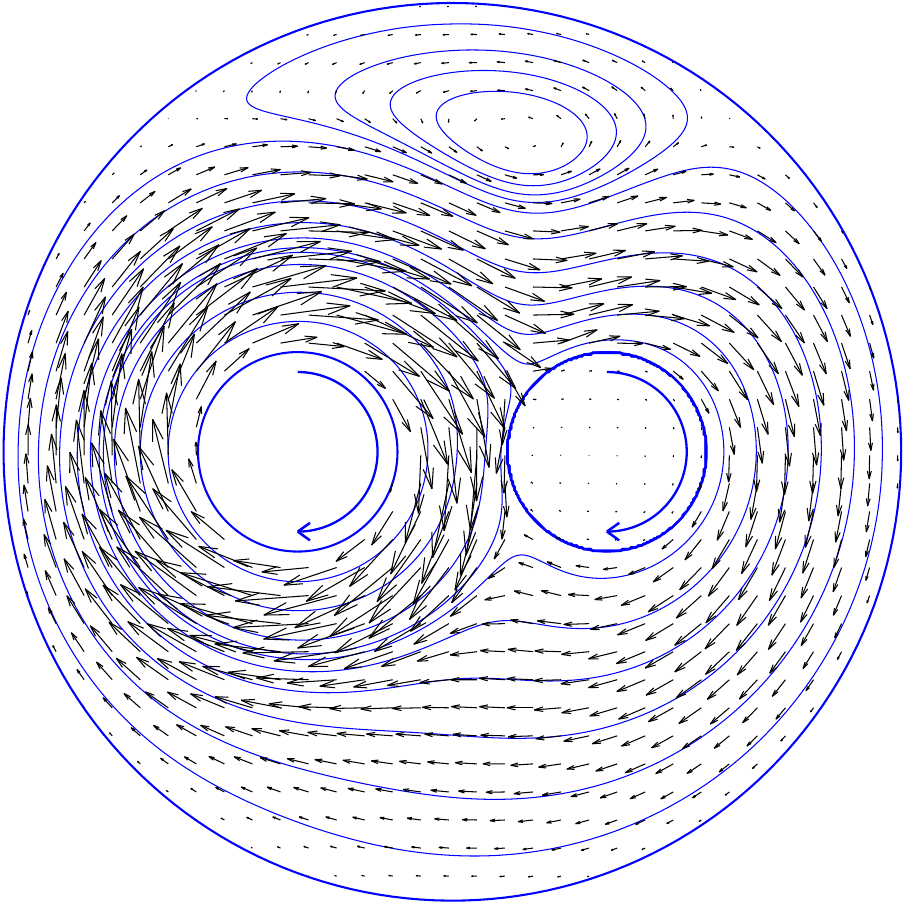}{8.81}{-0.15}

  \vspace{6pt}
  {\footnotesize (b)~Intermediate gap, $G=0.40$}\\[2pt]
  \vstreamt{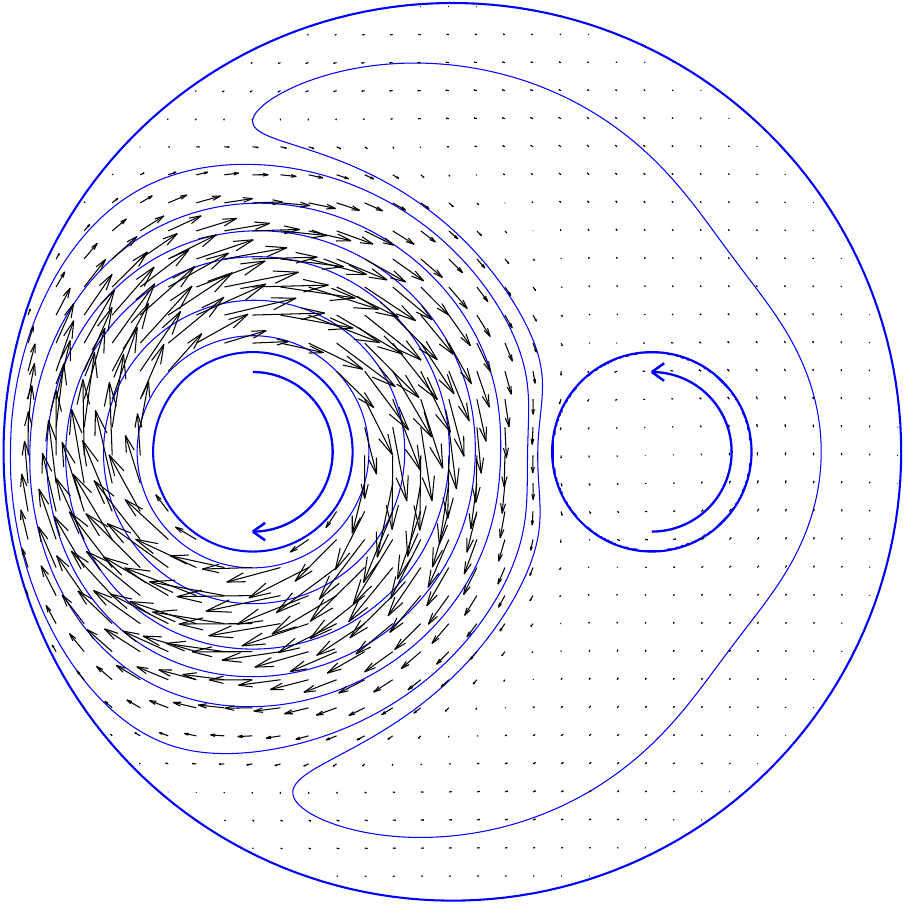}{0.66}{+0.20}\hfill
  \vstreamt{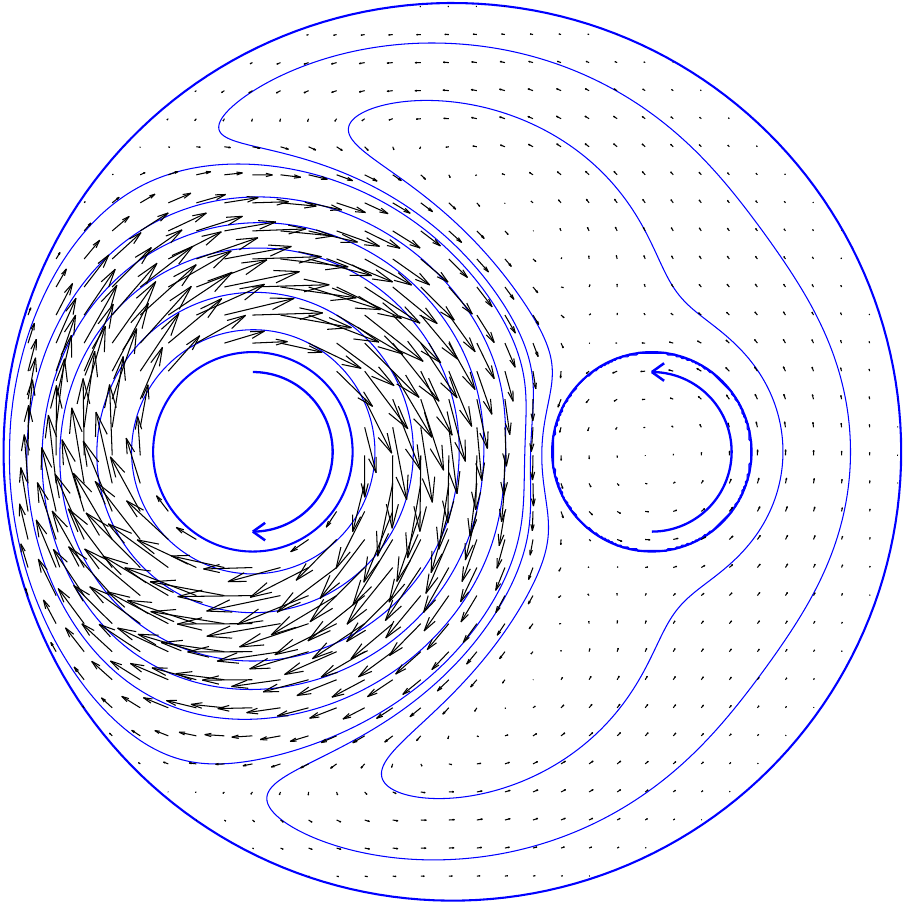}{0.92}{+0.40}\hfill
  \vstreamt{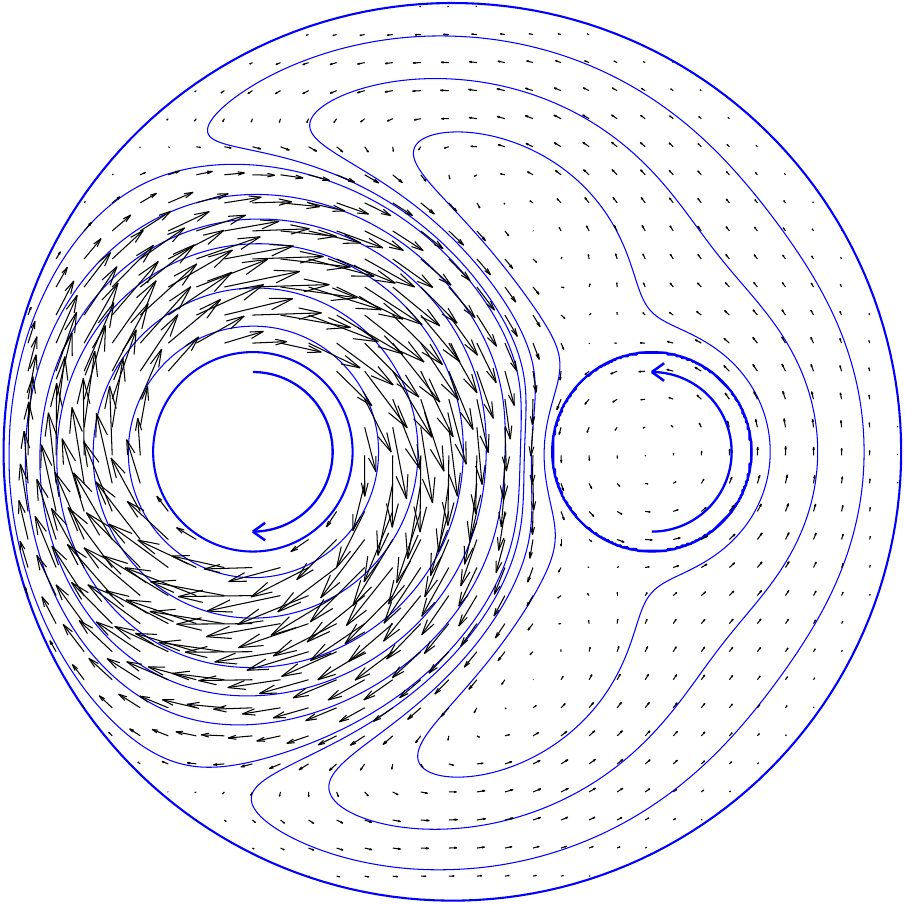}{1.23}{+0.60}\hfill
  \vstreamt{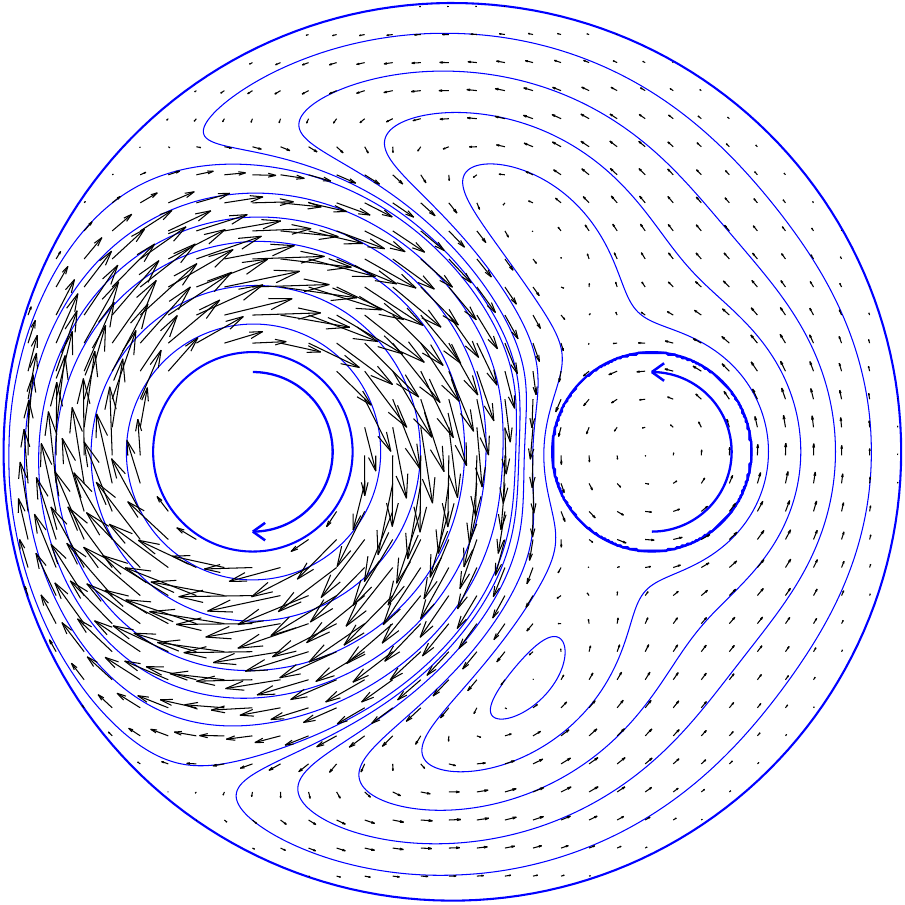}{1.68}{+0.80}\hfill
  \vstreamt{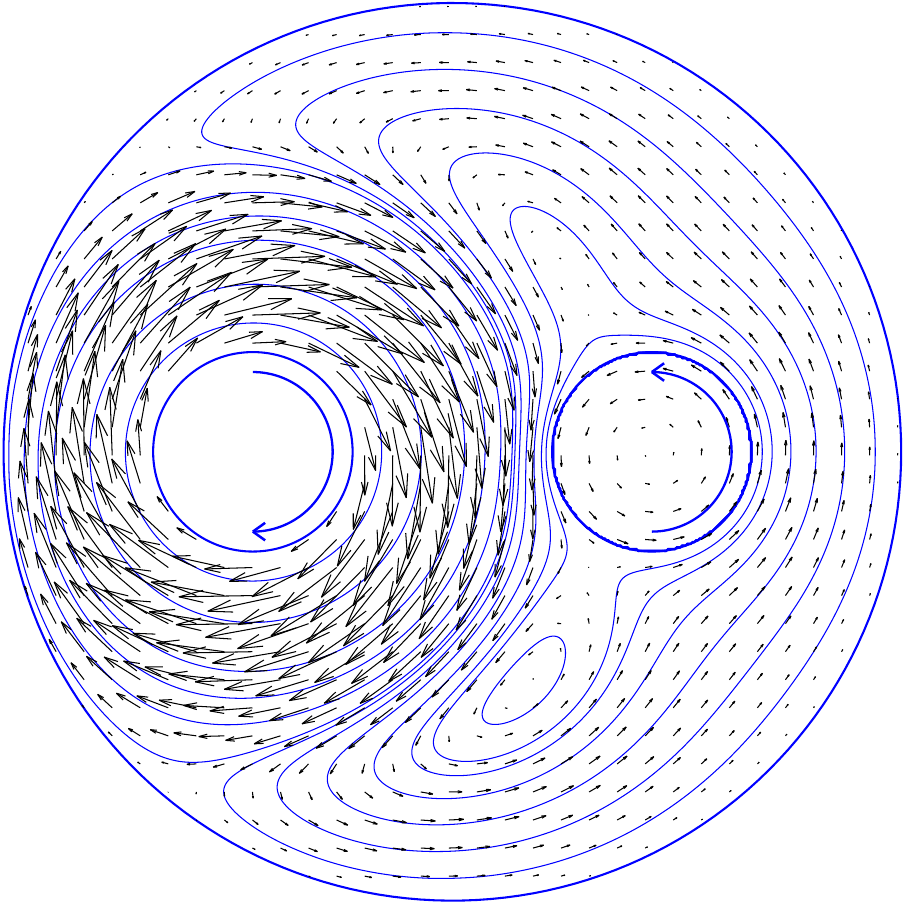}{7.74}{+1.01}

  \vspace{6pt}
  {\footnotesize (c)~Wide gap, $G=0.70$}\\[2pt]
  \vstreamt{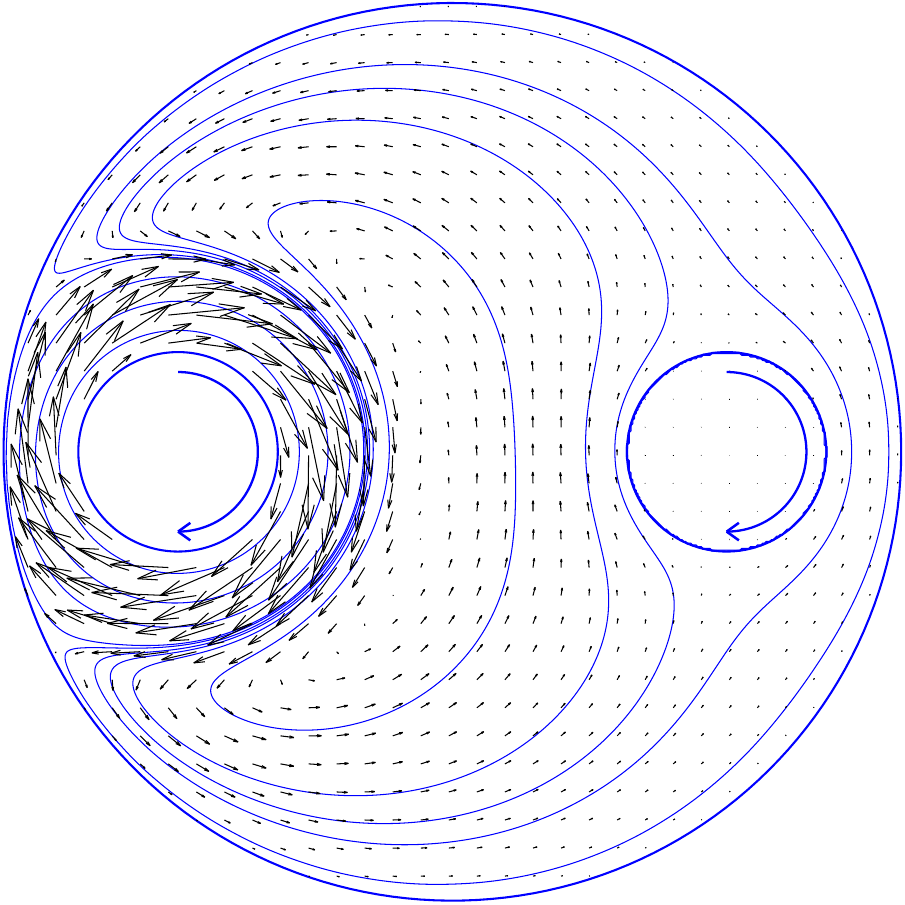}{0.39}{-0.01}\hfill
  \vstreamt{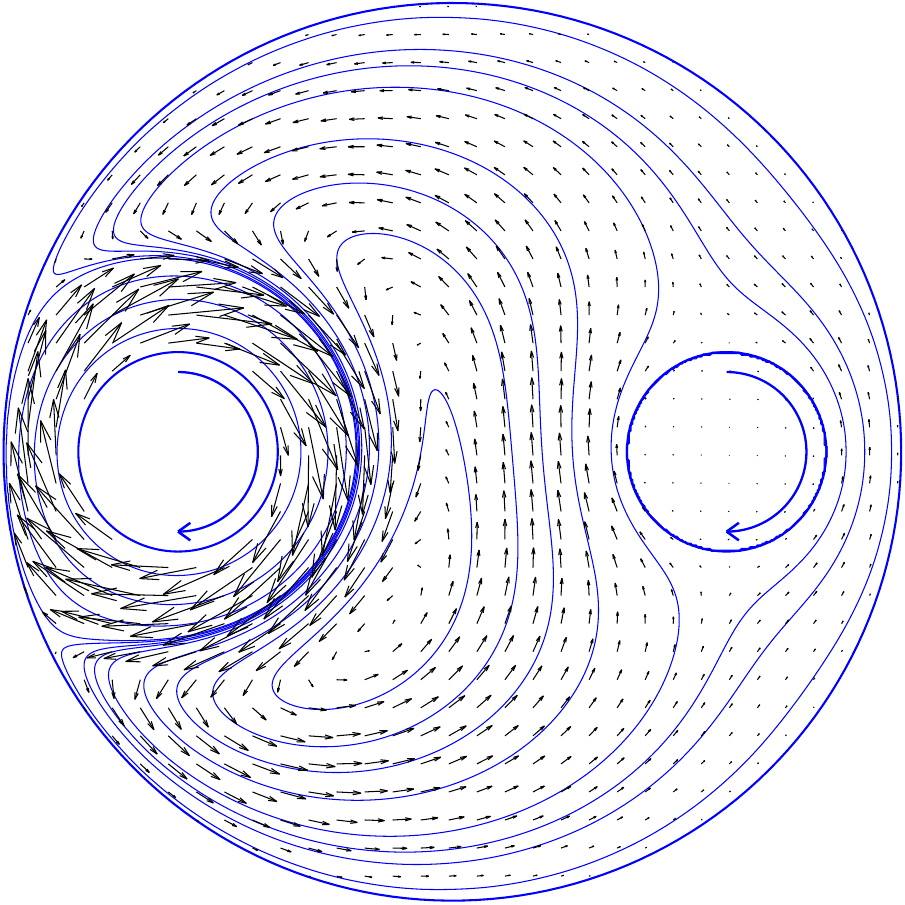}{0.77}{-0.10}\hfill
  \vstreamt{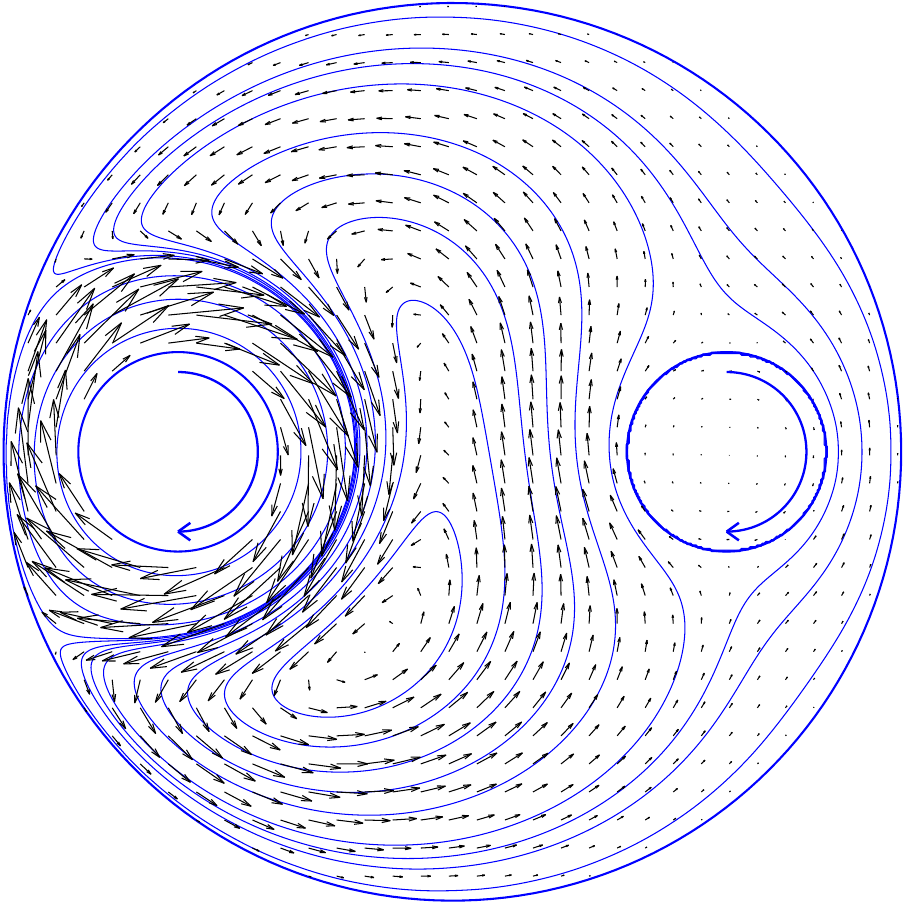}{1.10}{-0.20}\hfill
  \vstreamt{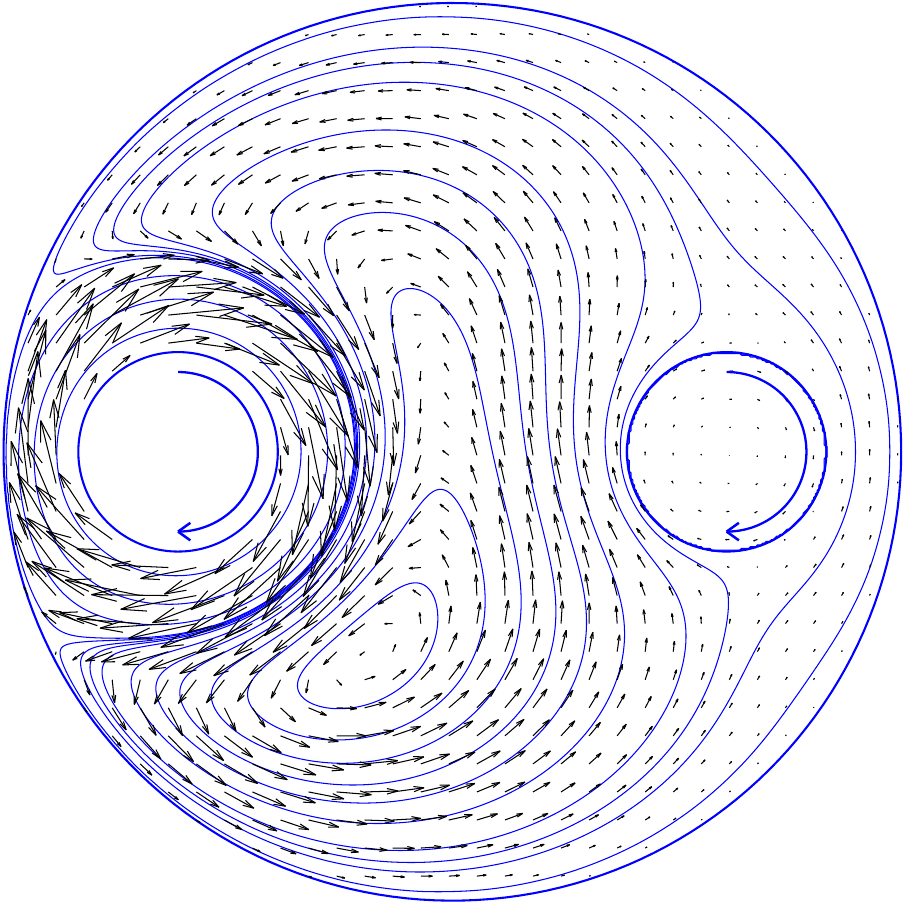}{1.69}{-0.30}\hfill
  \vstreamt{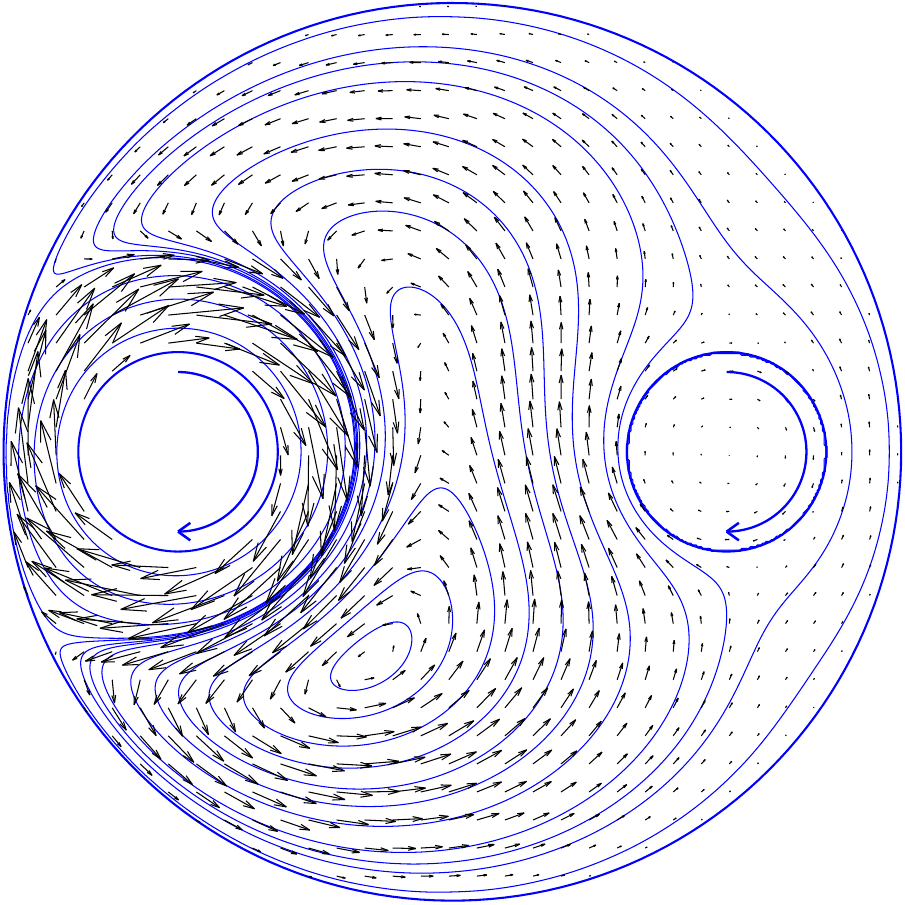}{7.55}{-0.34}
  \caption{Transient development of the passive spin at $C=4.5$ and
  $\Rey=60$ for three gaps.  The active rotor is set to its prescribed spin at
  $t=0$, the passive rotor starts from rest, and each frame shows the computed
  velocity field with streamlines, labeled by elapsed time $t$ and passive
  angular velocity $\omega$. Because $\Omega<0$, $\omega>0$ is counterrotation
  and $\omega<0$ is corotation. (a)~At $G=0.22$, a gap-localized vortex first
  drives a counterrotating overshoot; as momentum reaches the outer side of the
  passive rotor, the outer circulation cancels the inner drive near $t=2.2$ and
  leaves a weak corotating terminal state. (b)~At $G=0.40$, the developing
  recirculation drives the passive rotor counterrotating throughout the
  start-up, so
  the response is monotone. (c)~At $G=0.70$, the inner shear is too weak to
  produce a gearlike overshoot, and the corral-scale circulation drives
  corotation from the beginning. The comparison separates the short-time
  ordering of the shear signals from the terminal zero-torque state used in the
  phase diagrams.}
  \label{fig:transient}
\end{figure*}

Figure~\ref{fig:transient} gives three representative start-ups at $C=4.5$ and
$\Rey=60$. The passive rotor starts from rest, while the active rotor is set to
its prescribed clockwise spin at $t=0$. At the narrow gap $G=0.22$, the passive
rotor initially turns counterrotating, overshoots, crosses through rest, and
then settles to a weak corotating terminal state [Fig.~\ref{fig:history}(c)].
The flow fields show the sequence directly [Fig.~\ref{fig:transient}(a)]: a
gap-localized vortex first applies the inner, gearlike shear; only later does
the circulation extend around the outer side of the passive rotor and supply
the beltlike shear that determines the terminal spin. The relevant diffusion
estimate is $g^2/\nu\simeq1.5$ in the present nondimensional units for
$G=0.22$, $C=4.5$, and $\nu=0.01$, consistent with the time scale over which
the handoff occurs.

The two larger gaps show why this overshoot is selective rather than generic.
At $G=0.40$, the passive rotor climbs monotonically to counterrotation
[Fig.~\ref{fig:history}(b), Fig.~\ref{fig:transient}(b)]. The developing
recirculation already drives the passive rotor in the eventual terminal sense,
so the early and late torques do not compete in sign. At $G=0.70$, the
gap-facing shear is too weak to produce even a short-lived gearlike phase; the
corral-scale circulation controls the response from the start, and the passive
rotor relaxes monotonically to corotation [Fig.~\ref{fig:history}(d),
Fig.~\ref{fig:transient}(c)]. Thus the transient histories reinforce the
steady-state interpretation: at small gaps the inner shear can act first even
when it does not win the terminal torque balance, whereas at wider gaps the
large-scale circulation sets the sign immediately.

We use these histories only qualitatively. The absolute relaxation time depends
on passive-body inertia, numerical start-up protocol, and, in the experiment,
the imposed ramp in the active motor speed. The present fixed-axis, neutrally
buoyant calculations also do not address critical slowing near a stagnation
boundary. Their role is more
limited and more direct: they show that the same inner-first, outer-later
ordering invoked for the experimental transient response is already present in
the strictly planar calculation, while the terminal phase itself remains fixed
by the long-time zero-torque balance.

\subsection{Comparison with the quasi-two-dimensional experiment}
\label{sec:compare}

The comparison with the experiment is clearest when divided into three
questions. Does the planar model recover the time-averaged organization of the
phase diagram? Where does it place the Reynolds-driven corotation boundary? And
does it reproduce the high-$\Rey$ time dependence reported in the finite-depth
apparatus? The first answer is broadly positive. The planar maps recover the
gap-driven ordering seen in Figs.~2(b) and~2(c) of
Ref.~\cite{SmithRistrophZhang2026}: a close-gap corotation leg, an intermediate
counterrotation band, wide-gap corotation, gear ratios of order $10^{-2}$, and
weaker coupling in larger corrals. This agreement is notable because the
experiment has finite height, $H/r=8$, whereas the computation is strictly
two-dimensional.

A useful scale for the omitted end-wall motion is the corral Ekman number
$\Ek=\frac{\nu}{|\Omega|R^2}=\frac{1}{\Rey C^2}$, which gives the dimensionless
thickness $\sqrt{\Ek}$ of the Ekman layers on
the top and bottom walls~\cite{HouPanGlowinski2014}. This estimate is only a
scale comparison, not a rigorous error bound on the planar model. At the benchmark
condition $\Rey=20$ and $C=4.5$, it gives
$\Ek\simeq2.5\times10^{-3}$ and $\sqrt{\Ek}R\simeq0.22r$: small compared
with the corral radius, but not small compared with the rotor. Even with this
finite-depth scale present, the low- and moderate-$\Rey$ gap-route states
remain broadly consistent between the experiment and the planar calculation.
That consistency is also in line with the experimental vertical-plane
visualizations, which show Taylor-roll-like secondary motions above
$\Rey\approx60$ but interpret their action as primarily axial rather than
directly rotational.

The residual differences are organized rather than random. Quantitatively, the
planar calculation gives larger $|\Gamma|$, by up to about a factor of two at
$C=4.5$. It also differs in sign at the smallest gaps: the experiment retains
weak counterrotation near $G\approx0.05$ to $0.1$, including a thin
high-$\Rey$ strip, whereas the planar maps are corotating there. These
differences are plausibly connected to end-wall flow, finite-gap details in
near-contact configurations, and apparatus-specific geometry.

The more important discrepancy is the placement and topology of the
high-$\Rey$ boundary. In the experiment, the counterrotation band breaks into
isolated tongues as the corral size increases, already by $C=4.5$ and $5$
[Fig.~2(c) of Ref.~\cite{SmithRistrophZhang2026}]. In the planar maps, the band
narrows with $C$ but remains connected through $C=6$ on the present grid.
Cleaving that band would require corotation to invade the mid-gap region. In
the strictly planar calculation, however, the Reynolds-driven corotation
boundary reaches $G=0.22$ but not the band-centered transect near $G=0.30$.
This is the phase-map counterpart of the fixed-gap results above: the planar
model contains a Reynolds-driven spin reversal, but the boundary is displaced
to smaller gap and therefore does not reproduce the experimental tongue
topology.

The comparison becomes sharper when the high-$\Rey$ dynamics are included.
The experiment reports long-time unsteadiness of the passive spin under some
high-$\Rey$ conditions. The representative case $G=0.33$,
$C=4.5$ becomes oscillatory as $\Rey$ is increased from 60 to 70~\cite{SmithRistrophZhang2026}; more
generally, such oscillations are reported as most prevalent for
$\Rey\gtrsim70$ when the flow near the passive rotor is strong. The observed
amplitudes can be comparable to the mean and, in more unsteady cases, can
produce momentary reversals. Flow visualization attributes the motion to an
unsteady wake below the passive rotor, with crossing pathlines and an
oscillating separation point. Thus the experimental high-$\Rey$ state differs
from the planar model in both mean torque balance and long-time
dynamics.

The planar calculations do not show the corresponding long-time oscillation.
For the phase-map sweeps and the extended fixed-gap transects reported here,
the monitored passive-spin histories relax to steady terminal values, including
cases with $\Rey\le400$. This steadiness should not be interpreted simply as a
failure to support unsteady two-dimensional flow. As a separate
control, the same solver reproduces the two-dimensional vortex-shedding
transition for uniform flow past an isolated cylinder above
$Ud/\nu\approx47$, where that Reynolds number is based on the incoming speed
and cylinder diameter. Rather, within the confined two-rotor geometry studied
here, the strictly planar flow selects steady fixed points over the computed
range. The unsteady experimental wake is therefore an additional high-$\Rey$
feature of the finite-depth apparatus that the planar model does not reproduce.

This dynamical difference should be kept separate from the location of the
spin-reversal boundary. The experimental inertial transition near
$\Rey\approx55$ at $G\approx0.3$ occurs before the broadly reported
$\Rey\gtrsim70$ unsteady regime, and the planar model remains steady while
still producing a Reynolds-driven crossing at the narrower gap $G=0.22$.
Thus the central discrepancy is not merely that the experiment is unsteady and
the calculation is steady. It is that the experiment develops shear-zone
collapse, inner-zone deflection, outer-shear dominance, and, at higher $\Rey$,
long-time oscillations at the experimental mid-gap transect, whereas the
planar model develops a comparable terminal torque reversal only at smaller
gap.

The resulting picture is a localized, mechanism-level limitation of the planar
approximation. The gap-route rearrangements and wide-gap transition are
captured well. The surface shear-zone diagnostic is captured selectively: the
planar wall traction reproduces the lower and post-merger branches of $S(G)$
and the attachment jump, but not the full experimental bracket through
detachment and merger. At high Reynolds number, the planar model captures a
fixed-gap spin reversal, but not the experimental shear-zone collapse,
inner-zone deflection, tongue break-up, or long-time passive-spin oscillations.
The residual discrepancy is therefore not a generic failure of two-dimensional
modeling. It is localized to the placement, topology, and surface-stress
geometry of the high-$\Rey$ corotation boundary, together with the long-time
unsteadiness of the high-$\Rey$ state. End-wall motion
remains a plausible contributor, but Ekman scaling alone cannot determine the
corral-size trends: the rotor-scale thickness
$\sqrt{\nu/|\Omega|}/r\sim\Rey^{-1/2}$ is independent of $C$, whereas the
observed phase-boundary displacement depends on the corral geometry.

\section{Conclusions and outlook}
\label{sec:conclusion}

We have used a distributed Lagrange multiplier/\allowbreak{}fictitious domain
formulation to isolate the planar component of hydrodynamic spin coupling in a
two-rotor corral. The purpose was not to reproduce every feature of the
finite-depth experiment, but to determine which phase-diagram structures are
already contained in the horizontal-plane Navier--Stokes problem and which
require physics beyond that reduction. The implementation was validated against
an eccentric-Couette benchmark, and the two-rotor calculations were evolved to
terminal passive spins selected by near-zero hydrodynamic torque. The reported
gear ratio is therefore a property of the torque balance, not a prescribed
passive rotation rate.

The strongest agreement is the gap route. At $\Rey=20$ and $C=4.5$, the
planar model recovers the benchmark counterrotation state, gear-ratio
magnitudes of order $10^{-2}$, and the wide-gap transition to corotation. It
also reproduces the experimentally observed sequence of intervening flow
rearrangements: vortex attachment, detachment, and merger. The wall-traction
diagnostic explains why this agreement is meaningful but incomplete. A planar
surface shear arc follows the lower and post-merger branches of the experimental
shearing length and captures the attachment jump, whereas the larger
experimental bracket contains additional bulk-flow structure through detachment
and merger. Thus the two-dimensional model captures the horizontal vortex
architecture of the gap route, while the detailed experimental shearing length
contains information not reducible to a single surface-traction arc.

The principal mismatch is the high-$\Rey$ corotation boundary. The planar model
does contain a fixed-gap Reynolds-driven crossing: at $G=0.22$, the terminal
passive spin reverses near $\Rey\approx44$. Along the experimental mid-gap
transect near $G=0.3$, however, the planar gear ratio approaches zero from the
counterrotating side and remains negative through $\Rey=400$. The discrepancy
is therefore not whether inertial spin reversal can occur in two dimensions, but
where it occurs and by what surface-stress mechanism. In the experiment, the
inner shear zone shortens, deflects away from the gap axis, and loses dominance
to the outer shear. In the planar calculation at $G=0.3$, the corresponding
wall-traction arc narrows but remains gap-facing and counterrotating. Even at
$G=0.22$, where the terminal spin changes sign, the dominant gap-facing
traction persists; the reversal is produced by redistribution of the integrated
planar torque, not by the same shear-zone collapse inferred in the experiment.

Time dependence sharpens this distinction. The phase diagrams in this paper are
based on terminal states, and the monitored planar histories relax to steady
fixed points over the computed range, including extended transects up to
$\Rey=400$. By contrast, the experiment reports long-time passive-spin
oscillations and unsteady wake motion under some high-$\Rey$ conditions, with
a representative onset between $\Rey=60$ and 70 at $G=0.33$, $C=4.5$.
These oscillations should not be conflated with the mean spin-reversal
boundary: the experimental inertial transition occurs before the broadly
reported unsteady regime, and the planar model can reverse terminal spin while
remaining steady. The transient calculations nevertheless provide a useful
diagnostic. They show the early arrival of the gap-side shear and the later
establishment of the terminal torque balance, giving a planar analogue of the
experimental observation that start-up histories can initially move in the
gearlike direction before relaxing to the long-time state.

Taken together, the results support a boundary-displacement interpretation of
the planar approximation. The strictly two-dimensional model recovers the low-
and moderate-$\Rey$ gap route and the scale of the coupling. Its limitation is
localized to the high-$\Rey$ boundary: the calculation does not reproduce the
experimental tongue break-up, the mid-gap shear-zone deflection, or the
long-time unsteady wake. Finite-depth secondary motion, end-wall stresses, and
apparatus-specific geometry remain natural candidates for shifting the
surface-stress balance, but the present results also show that the missing
effect cannot be summarized by a simple ``two-dimensional versus
three-dimensional'' dichotomy. The planar problem contains inertial spin
reversal; it places that reversal in the wrong part of parameter space.

Several directions follow. A direct three-dimensional calculation with
controlled end-wall conditions would test whether axial inflow and Ekman-layer
transport supply the missing outward routing of the inner flow, or whether the
displacement is already determined by finer horizontal-plane structure.
Experiments that vary height, lid clearance, and corral construction would
provide the complementary test. The same numerical formulation also permits
passive-body freedoms that extend the present fixed-axis, circular-rotor
problem. Releasing the passive disk to translate would ask whether hydrodynamic
coupling selects a stable separation. Replacing the
disk by a freely rotating ellipse would ask whether shape can convert steady
spinning into stable flow-induced stalling at a preferred orientation. Finally,
many-body rotor arrays in a corral require no new governing equations in this
framework, but they may introduce non-pairwise flow paths, collective spin
states, and additional numerical stiffness. These extensions would use the
present two-rotor system as a calibrated benchmark for studying how confinement,
inertia, and passive-body design control hydrodynamic spin transmission.




\begin{thebibliography}{99}

\bibitem{Ristroph2008} L.~Ristroph and J.~Zhang,
  Anomalous hydrodynamic drafting of interacting flapping flags,
  Phys.\ Rev.\ Lett.\ \textbf{101}, 194502 (2008).
\bibitem{Newbolt2019} J.~W.~Newbolt, J.~Zhang, and L.~Ristroph,
  Flow interactions between uncoordinated flapping swimmers give rise to
  group cohesion,
  Proc.\ Natl.\ Acad.\ Sci.\ U.S.A.\ \textbf{116}, 2419 (2019).
\bibitem{Soni2019} V.~Soni, E.~S.~Bililign, S.~Magkiriadou, S.~Sacanna,
  D.~Bartolo, M.~J.~Shelley, and W.~T.~M.~Irvine,
  The odd free surface flows of a colloidal chiral fluid,
  Nat.\ Phys.\ \textbf{15}, 1188 (2019).
\bibitem{Bililign2022} E.~S.~Bililign, F.~Balboa Usabiaga, Y.~A.~Ganan,
  A.~Poncet, V.~Soni, S.~Magkiriadou, M.~J.~Shelley, D.~Bartolo, and
  W.~T.~M.~Irvine,
  Motile dislocations knead odd crystals into whorls,
  Nat.\ Phys.\ \textbf{18}, 212 (2022).
\bibitem{Brownstein2019} I.~D.~Brownstein, N.~J.~Wei, and J.~O.~Dabiri,
  Aerodynamically interacting vertical-axis wind turbines: Performance
  enhancement and three-dimensional flow,
  Energies \textbf{12}, 2724 (2019).
\bibitem{Chen2024} P.~Chen, S.~Weady, S.~Atis, T.~Matsuzawa,
  M.~J.~Shelley, and W.~T.~M.~Irvine,
  Self-propulsion, flocking and chiral active phases from particles
  spinning at intermediate Reynolds numbers,
  Nat.\ Phys.\ \textbf{21}, 146 (2025).
\bibitem{Kida1994} S.~Kida and M.~Takaoka,
  Vortex reconnection,
  Annu.\ Rev.\ Fluid Mech.\ \textbf{26}, 169 (1994).
\bibitem{Dritschel1995} D.~G.~Dritschel,
  A general theory for two-dimensional vortex interactions,
  J.\ Fluid Mech.\ \textbf{293}, 269 (1995).
\bibitem{Taylor1923} G.~I.~Taylor,
  Stability of a viscous liquid contained between two rotating cylinders,
  Phil.\ Trans.\ R.\ Soc.\ A \textbf{223}, 289 (1923).
\bibitem{Huisman2018} R.~Ezeta, S.~G.~Huisman, C.~Sun, and D.~Lohse,
  Turbulence strength in ultimate Taylor--Couette turbulence,
  J.\ Fluid Mech.\ \textbf{836}, 397 (2018).
\bibitem{Jana1994} S.~C.~Jana, G.~Metcalfe, and J.~M.~Ottino,
  Experimental and computational studies of mixing in complex Stokes
  flows: the vortex mixing flow and multicellular cavity flows,
  J.\ Fluid Mech.\ \textbf{269}, 199 (1994).
\bibitem{Hu2021} S.-Y.~Hu, J.-J.~Chu, M.~J.~Shelley, and J.~Zhang,
  L\'evy walks and path chaos in the dispersal of elongated structures
  moving across cellular vortical flows,
  Phys.\ Rev.\ Lett.\ \textbf{127}, 074503 (2021).

\bibitem{SmithRistrophZhang2026} J.~E.~Smith, L.~Ristroph, and J.~Zhang,
  Hydrodynamic spin-coupling of rotors,
  Phys.\ Rev.\ Lett.\ \textbf{136}, 024001 (2026).

\bibitem{GuoManZhu2024} H.~Guo, Y.~Man, and H.~Zhu,
  Hydrodynamic bound states of rotating microcylinders in a confining
  geometry,
  Phys.\ Rev.\ Fluids \textbf{9}, 014102 (2024).

\bibitem{Glowinski1999} R.~Glowinski, T.-W.~Pan, T.~I.~Hesla, and
  D.~D.~Joseph,
  A distributed Lagrange multiplier/\allowbreak{}fictitious domain method for
  particulate flows,
  Int.\ J.\ Multiphase Flow \textbf{25}, 755 (1999).
\bibitem{Glowinski2001} R.~Glowinski, T.-W.~Pan, T.~I.~Hesla,
  D.~D.~Joseph, and J.~P\'eriaux,
  A fictitious domain approach to the direct numerical simulation of
  incompressible viscous flow past moving rigid bodies: Application to
  particulate flow,
  J.\ Comput.\ Phys.\ \textbf{169}, 363 (2001).
\bibitem{PanGlowinskiHou2007} T.-W.~Pan, R.~Glowinski, and S.~C.~Hou,
  Direct numerical simulation of pattern formation in a rotating
  suspension of non-Brownian settling particles in a fully filled
  cylinder,
  Comput.\ Struct.\ \textbf{85}, 955 (2007).
\bibitem{HouPanGlowinski2014} S.~C.~Hou, T.-W.~Pan, and R.~Glowinski,
  Circular band formation for incompressible viscous fluid--rigid-particle
  mixtures in a rotating cylinder,
  Phys.\ Rev.\ E \textbf{89}, 023013 (2014).
\bibitem{PanGlowinski2002} T.-W.~Pan and R.~Glowinski,
  Direct simulation of the motion of neutrally buoyant circular cylinders in
  plane Poiseuille flow,
  J.\ Comput.\ Phys.\ \textbf{181}, 260 (2002).
\bibitem{PanGlowinski2005} T.-W.~Pan and R.~Glowinski,
  Direct simulation of the motion of neutrally buoyant balls in a
  three-dimensional Poiseuille flow,
  C.\ R.\ M\'ecanique \textbf{333}, 884 (2005).
\bibitem{Glowinski2003} R.~Glowinski, \textit{Finite Element Methods for
  Incompressible Viscous Flow}, Handbook of Numerical Analysis, Vol.~IX,
  edited by P.~G.~Ciarlet and J.~L.~Lions (North-Holland, Amsterdam,
  2003), pp.~3--1176.
\bibitem{DeanGlowinski1997} E.~J.~Dean and R.~Glowinski,
  A wave equation approach to the numerical solution of the Navier-Stokes
  equations for incompressible viscous flow,
  C.\ R.\ Acad.\ Sci.\ Paris, S\'er.\ I \textbf{325}, 783 (1997).
\bibitem{Chorin1978} A.~J.~Chorin, T.~J.~R.~Hughes, M.~F.~McCracken, and
  J.~E.~Marsden,
  Product formulas and numerical algorithms,
  Commun.\ Pure Appl.\ Math.\ \textbf{31}, 205 (1978).
\bibitem{Wannier1950} G.~H.~Wannier,
  A contribution to the hydrodynamics of lubrication,
  Q.\ Appl.\ Math.\ \textbf{8}, 1 (1950).

\end{thebibliography}
\end{document}